\documentclass[aps,prd,nofootinbib,floatfix,superscriptaddress,secnumarabic,preprintnumbers,longbibliography,10pt]{revtex4-2}
\usepackage{amsmath}
\usepackage{graphicx}
\usepackage[dvipsnames]{xcolor}
\usepackage{amsfonts}
\usepackage{hyperref}
\usepackage{bm}
\usepackage[normalem]{ulem}

\newcommand{\MSbar}{{\overline{\rm MS}}}

\newcommand{\be}{\begin{equation}}
\newcommand{\ee}{\end{equation}}
\newcommand{\bea}{\begin{eqnarray}}
\newcommand{\eea}{\end{eqnarray}}
\def\lsim{\mathrel{\rlap{\lower4pt\hbox{\hskip1pt$\sim$}}
    \raise1pt\hbox{$<$}}}                
\def\slashed{{/}\mskip-10.0mu}

\newcommand{\bmb}[1]{{\color{Blue} \bm{#1}}}
\newcommand{\bmp}[1]{{\color{Plum} \bm{#1}}}
\newcommand{\bmr}[1]{{\color{Red} \bm{#1}}}

\begin{document}

\title{Gauge-invariant renormalization of four-quark operators}

\author{M.~Constantinou}
\email[]{marthac@temple.edu}
\affiliation{Department of Physics, Temple University, Philadelphia, PA 19122 - 1801, USA}

\author{M.~Costa}
\email[]{kosta.marios@ucy.ac.cy}
\affiliation{Department of Physics, University of Cyprus, Nicosia, CY-1678, Cyprus}
\affiliation{Rinnoco Ltd, Limassol, CY-3047, Cyprus}
\affiliation{Department of Chemical Engineering, Cyprus University of Technology, Limassol, CY-3036, Cyprus\bigskip}

\author{H.~Herodotou}
\email[]{herodotos.herodotou@ucy.ac.cy}
\affiliation{Department of Physics, University of Cyprus, Nicosia, CY-1678, Cyprus}

\author{H.~Panagopoulos}
\email[]{panagopoulos.haris@ucy.ac.cy}
\affiliation{Department of Physics, University of Cyprus, Nicosia, CY-1678, Cyprus}

\author{G.~Spanoudes}
\email[]{spanoudes.gregoris@ucy.ac.cy}
\affiliation{Department of Physics, University of Cyprus, Nicosia, CY-1678, Cyprus}

\begin{abstract}

We study the renormalization of four-quark operators in one-loop perturbation theory. We employ a coordinate-space Gauge-Invariant Renormalization Scheme (GIRS), which can be advantageous compared to other schemes, especially in nonperturbative lattice investigations. From our perturbative calculations, we extract the conversion factors between GIRS and the modified Minimal Subtraction scheme ($\MSbar$) at the next-to-leading order. As a by-product, we also obtain the relevant anomalous dimensions in the GIRS scheme. A formidable issue in the study of the four-quark operators is that operators with different Dirac matrices mix among themselves upon renormalization. We focus on both parity-conserving and parity-violating four-quark operators, which change flavor numbers by two units ($\Delta F = 2$).  The extraction of the conversion factors entails the calculation of two-point Green's functions involving products of two four-quark operators, as well as three-point Green's functions with one four-quark and two bilinear operators. The significance of our results lies in their potential to refine our understanding of QCD phenomena, offering insights into the precision of Cabibbo-Kobayashi-Maskawa (CKM) matrix elements and shedding light on the nonperturbative treatment of complex mixing patterns associated with four-quark operators.
\end{abstract}

\maketitle

\section{Introduction}

A crucial feature of the Standard Model (SM), which is remarkably successful in accurately describing electroweak and strong interactions at the fundamental level, is the fact that the SM Lagrangian incorporates all pertinent operators with dimensions $D \leq 4$, and is a renormalizable Quantum Field Theory.
These operators are constructed using the elementary particle fields that have already been observed and adhere to the principles of Lorentz invariance and gauge symmetry. The potential influence of higher-dimensional ($D > 4$) effective operators, not encompassed within the SM Lagrangian, is anticipated to be inherently small. This is due to their suppression by negative powers of the high-energy scale $M$, characterizing physics beyond the SM, expressed as $M^{4-D}$, with allowances for logarithmic correction. This suppression is rooted in the principles of effective field theory, where new physics at a higher scale in $M$ can be systematically integrated out, leaving behind a series of higher-dimensional operators in the low-energy effective theory; see \cite{Buchalla:1995vs} for a classic review, and \cite{USQCD:2019hyg,FlavourLatticeAveragingGroupFLAG:2021npn} for more recent reviews. In this framework, operators with a dimension of $D = 6$, such as the four-quark operators, assume a particular significance; their impact is suppressed by $M^{-2}$, making them (along with the contributions of $D=5$ operators, e.g., the Weinberg operator) prime candidates for corrections to SM contributions. Thus, these operators can contribute to processes that are either forbidden or highly suppressed in the SM, offering a window into potential new physics phenomena.

In the context of lattice simulations of pure QCD, scalar and pseudoscalar four-quark operators naturally incorporate weak interaction effects. Furthermore, by the high precision achieved in experimental CKM matrix element measurements, the study of four-quark operators becomes even more pertinent in the context of potential discoveries at the Large Hadron Collider (LHC), such as new tetraquarks, which have been of significant interest in recent experimental results from the LHCb collaboration~\cite{LHCb:2022aki,LHCb:2022dvn,LHCb:2022sfr,LHCb:2022lzp}. Thus, it is important to explore their properties numerically on the lattice; this calls for a detailed investigation of the corresponding four-quark operators. Calculating matrix elements of four-quark operators in lattice QCD offers insights into a wide range of phenomena, including electroweak decays of hadrons, and beyond the Standard Model physics; see, e.g., Refs.~\cite{Ishikawa:2011dd,DiPierro:1998ty,DiPierro:1999tb,Gimenez:1998mw,Ishizuka:2018qbn}. In particular, phenomenological bag parameters~\cite{Carrasco:2015pra} are important lattice quantities associated with four-quark operators~\cite{Carrasco:2015pra}, which, among others, are relevant in hadronic decays and CP violation. One of the most widely studied bag parameters is the so-called $B_K$ parameter that parametrizes the neutral $K^0 - \overline{K}^0$ meson oscillations~\cite{Gupta:1992bd,Becirevic:2001re,Aoki:2010pe,ETM:2010ubf,Aebischer:2020dsw,Suzuki:2020zue}. Other related studies can be found in Refs.~\cite{Neubert:1996we,Lenz:2014jha,Gabbiani:1996hi}.

One of the practical difficulties in calculating physical matrix elements of four-quark operators is the mixing encountered under renormalization, which is generally complex: Depending on the flavor content of the four-quark operators, mixing is allowed among operators of equal and/or lower dimensions, which have the same transformation properties under symmetries of the action. Such mixing is especially challenging for lattice QCD calculations, where symmetries are broken (e.g., Lorentz symmetry is reduced to hypercubic symmetry, and chiral symmetry is fully broken for Wilson-like fermion actions) and which may further complicate the renormalization procedure. In the simplest case of scalar and pseudoscalar four-quark operators that change flavor number by two units ($\Delta F=2$), the calculation of four $5\times5$ mixing matrices is required in order to address the mixing among a complete basis of 20 four-quark operators as classified by discrete symmetries (parity and flavor switching symmetries; see section~\ref{def+symm}). This leads to the necessity of determining $4 \times 25 = 100$ nontrivial renormalization coefficients (some of them are zero according to the discrete symmetries). Note that by employing chiral symmetry, the mixing pattern is further simplified leading to additional zero elements in the mixing matrices for parity-even operators, as happened in dimensional regularization, in the $\MSbar$ scheme. However, in the general case of a regularization-independent scheme which does not impose chirally invariant conditions, applicable in both continuum and lattice regularizations, the mixing sets are not decreased. For this purpose, the current study does not employ chiral symmetry in determining mixing patterns.

The renormalization and mixing of four-quark operators with different flavor content have been investigated before in various perturbative and nonperturbative studies using $\MSbar$~\cite{Ciuchini:1997bw, Capitani:2000da,Nakamura:2006zx,Taniguchi:2012xm}, RI$'$/MOM~\cite{Ciuchini:1997bw, Donini:1999sf,Becirevic:2002qr,ETM:2010ubf,Constantinou:2010zs,Carrasco:2015pra}, RI$'$/SMOM~\cite{Aoki:2010pe,Lehner:2011fz,Boyle:2017skn,Garron:2018tst,Boyle:2024gge} and Schr\"odinger Functional~\cite{Guagnelli:2002rw,Guagnelli:2005zc,Palombi:2005zd,Dimopoulos:2006ma,Palombi:2006pu,Palombi:2007dr,Dimopoulos:2007ht,Papinutto:2016xpq,DallaBrida:2016zxs,Dimopoulos:2018zef,Marinelli:2023} schemes. In this work, we revisit the renormalization of the four-quark operators by employing a gauge-invariant renormalization scheme (GIRS)~\cite{Costa:2021iyv}, which involves Green’s functions of gauge-invariant operators in coordinate space. A similar recent study using coordinate-space renormalization prescription has been implemented in~\cite{Lin:2024mws}. GIRS is a promising renormalization prescription that does not encounter issues in lattice studies related to gauge fixing. Our goal is to provide appropriate renormalization conditions, which address the mixing of the four-quark operators and which are applicable in nonperturbative calculations on the lattice, as well as to provide the conversion factors between our proposed prescription and the $\MSbar$ scheme (typically used in phenomenology). The conversion factors are regularization-independent, and thus, one can compute them in dimensional regularization (DR), where perturbative computation can be performed more readily and in higher-loop order. To this end, we calculate the first quantum corrections for appropriate two-point and three-point Green’s functions in coordinate space using DR. We focus on the renormalization of four-quark operators, which are involved in flavor-changing $\Delta F=2$ processes. These are categorized into 2 sets of parity-conserving and 2 sets of parity-violating operators. The two-point functions involve the product of two four-quark operators positioned at a nonzero distance in coordinate space. The three-point functions include the product of one four-quark operator and two quark bilinear operators located at three different spacetime points. For determining all renormalization coefficients, a number of three-point functions are involved in the renormalization conditions.

The paper is organized as follows: Section~\ref{formula} outlines the formulation of our calculation, including the definitions of the operators under study and their symmetry properties. Also, a detailed description of the calculated Green's functions is provided, along with general renormalization conditions in GIRS and the definition of the conversion matrices between GIRS and $\MSbar$ schemes. In Sec.~\ref{CalculationProcedure}, we collect a number of prototype Feynman integrals appearing in our calculation for both two-point and three-point Green's functions, and we give some details regarding the methodology that we follow for their computation. Section~\ref{results} presents our main results for the $\MSbar$-renormalized Green's functions, along with the corresponding mixing matrices. A detailed discussion follows regarding the selection of a suitable set of renormalization conditions in GIRS among a large number of acceptable GIRS variants. The conversion matrices between the selected version of GIRS and the $\MSbar$ scheme are presented; additional information for extracting conversion matrices for any other version of GIRS is also provided. As a by-product of our calculation, we also provide the next-to-leading order anomalous dimensions of the operators under study in the GIRS scheme. To make our results easily accessible to the reader, we include a supplemental Mathematica file containing the one-loop expressions of the $\MSbar$-renormalized Green's functions under study in electronic form. Finally, in Section~\ref{conclusions}, we summarize our findings and suggest potential extensions for future research.

\section{Formulation}
\label{formula}

In this section, we briefly introduce the formulation of our study, along with the notation utilized throughout the paper. We provide definitions of the four-quark operators, as well as their transformation properties under parity, charge conjugation, and flavor exchange symmetries. These symmetries allow mixing between specific groups of operators, which arise at the quantum level. A gauge-invariant renormalization scheme (GIRS) is outlined for the renormalization of quark-operators, which is constructed to be regularization-independent and thus also applicable in lattice regularizations that break chiral symmetry. We describe appropriate Green's functions for studying the renormalization of four-quark operators in GIRS, the implied renormalization conditions, and we define the conversion matrices between GIRS and $\MSbar$ scheme. There are multiple possibilities for defining GIRS, each leading to different conversion matrices; we present one of them in Sec.~\ref{results} while further options can be extracted by our results.

Our calculations are performed within the framework of Quantum Chromodynamics (QCD). The action of QCD in Euclidean spacetime is given by: 
\begin{equation}
S_{\text{QCD}} = \int d^4x \left[\frac{1}{4} F_{\mu\nu}^a F^a_{\mu\nu} + \sum_f \bar{\psi}_f (\gamma_\mu D_\mu + m_f) \psi_f \right],
\end{equation}
where $F_{\mu\nu}^a$ represents the gluon field-strength tensor, $\psi$ denotes the quark field of flavor $f$, and $D_\mu$ is the covariant derivative, which accounts for the interaction of quarks with the gluon ($A_\mu$) fields: $D_\mu \psi= \partial_{\mu}\psi + i g A_\mu \psi$\,; $g$ is the coupling constant. 

Note that we use a mass-independent scheme, and thus, the masses $m_f$ are set to zero.

\subsection{Definition of the four-quark operators and their symmetry properties}
\label{def+symm}

We investigate four-quark composite operators of the form:
\begin{equation}
{\cal O}_{\Gamma \Tilde{\Gamma}} \equiv (\bar \psi_{f_1}\,\Gamma\,\psi_{f_3})(\bar \psi_{f_2}\,\Tilde{\Gamma}\,\psi_{f_4}) \equiv \Big(\sum_{a} \sum_{\alpha,\beta} {\bar \psi_{\alpha, f_1}^a}(x)  \,\Gamma_{\alpha \beta} \,  {\psi_{\beta, f_3}^a}(x) \Big) \Big(\sum_{a'} \sum_{\alpha',\beta'} {\bar \psi_{\alpha',f_2}^{a'}}(x) \,\Tilde{\Gamma}_{\alpha' \beta'} \, {\psi_{\beta',f_4}^{a'}}(x) \Big),
\label{fourQ}
\end{equation}
where $\Gamma$ and $\Tilde{\Gamma}$ denote products of Dirac matrices:
\begin{eqnarray}
\Gamma, \Tilde{\Gamma} \in \{\openone,\, \gamma_5,\, \gamma_\mu,\, \gamma_\mu \gamma_5,
  \,\sigma_{\mu\nu}, \,\gamma_5\sigma_{\mu\nu}\} \equiv
  \{S,P,V,A,T,\tilde T \},
\label{Gamma}
\end{eqnarray}
and $\sigma_{\mu\nu}=\frac{1}{2}[\gamma_\mu,\gamma_\nu]$; spinor indices are denoted by Greek letters ($\alpha, \beta, \alpha', \beta'$), while color and flavor indices denoted by Latin letters ($a, a'$) and $f_i$, respectively. In our study, we focus on four-quark operators with $\Gamma = \Tilde{\Gamma}$ and $\Gamma = \Tilde{\Gamma} \gamma_5$ (repeated Lorentz indices are summed over), which are scalar or pseudoscalar quantities under rotational symmetry.

One complication in the study of these operators is that mixing is allowed among four-quark operators with different Dirac matrices under renormalization, as dictated by symmetries. In order to study the mixing of the four-quark operators at the quantum level, it is convenient to construct operators with exchanged flavors of their quark fields [cf. Eqs. (\ref{fourQ}, \ref{O^F_XY})], which are related to the original operators through the Fierz–Pauli–Kofink identity (the superscript letter $F$ stands for Fierz):
\begin{equation}
    {\cal O}^F_{\Gamma \Tilde{\Gamma}} \equiv (\bar \psi_{f_1}\,\Gamma\,\psi_{f_4})(\bar \psi_{f_2}\,\Tilde{\Gamma}\,\psi_{f_3}), \label{O^F_XY}
\end{equation}
where color and spinor indices are implied. 

We considered the symmetries of the QCD action: Parity $\mathcal{P}$, Charge conjugation $\mathcal{C}$, Flavor exchange symmetry $\mathcal{S} {\equiv} (f_3 \leftrightarrow f_4)$, Flavor Switching symmetries $\mathcal{S}' {\equiv} (f_1 \leftrightarrow f_3 , f_2 \leftrightarrow f_4)$ and
$\mathcal{S}'' {\equiv} (f_1 \leftrightarrow f_4 , f_3 \leftrightarrow f_2)$), with 4 mass-degenerate quarks \cite{Frezzotti:2004wz}. Chiral symmetry can be violated in some regularizations and thus, it is not considered in the present study for identifying the mixing pattern. In particular, the operator-mixing setup which follows is also applicable in lattice regularizations that break chiral symmetry (such as Wilson fermions). Operators with the same transformation properties under these symmetries can and will mix. The parity $\mathcal{P}$ and charge conjugation $\mathcal{C}$ transformations on quarks and antiquarks are defined below:
\begin{eqnarray}
\rm{Parity:}
&&\begin{cases}
    \mathcal{P}\psi_f(x) & = \gamma_4 \ \psi_f(x_P) \\
    \mathcal{P}\bar{\psi}_f(x) &= \bar{\psi}_f(x_P) \ \gamma_4,
\end{cases} \\
\rm{Charge\,conjugation:}
&&\begin{cases}
    \mathcal{C}\psi_f(x) &= -C \ \bar{\psi}_f^T(x) \\
    \mathcal{C}\bar{\psi}_f(x) &= \psi_f^T(x) \ C,
\end{cases}
\end{eqnarray}
where $x_P = (-\textbf{x},t)$, $^{\,T}$ means transpose and the matrix $C$ satisfies: $(C \gamma_{\mu})^{T}= C \gamma_{\mu}$, $C^T=-C$ and $C^{\dagger} C=1$. 
The transformations of the four-quark operators of Eqs. (\ref{fourQ}, \ref{O^F_XY}) are shown in Table~\ref{tb1}. Note that different combinations of the original operators are considered, which are odd/even under the $\mathcal{P}$, $\mathcal{CS'}$, $\mathcal{CS''}$, $\mathcal{CPS'}$ and $\mathcal{CPS''}$ transformations.
\begin{table}[ht!]
    \centering
    \begin{tabular}{c|c|c|c|c|c}
      &  \, \,$\mathcal{P}$ \,\,& \, \,$\mathcal{C S'}$\,\, & \, \,$\mathcal{C S''}$\,\, &\, \,$\mathcal{C P S'}$\,\, &\, \,$\mathcal{C P S''}$\,\, \\
         \hline
           \hline
    ${\cal O}_{VV}$ & $+$ & $+$ & $+$ & $+$ & $+$\\  \hline
    ${\cal O}_{AA}$ & $+$ & $+$ & $+$ & $+$ & $+$ \\  \hline
    ${\cal O}_{PP}$ & $+$ & $+$ & $+$ & $+$ & $+$ \\  \hline
    ${\cal O}_{SS}$ & $+$ & $+$ & $+$ & $+$ & $+$ \\  \hline
    ${\cal O}_{TT}$ & $+$ & $+$& $+$ & $+$ & $+$ \\  \hline
    ${\cal O}_{[VA+AV]}$ & $-$ & $-$ & $-$ & $+$ & $+$\\  \hline
    ${\cal O}_{[VA-AV]}$ & $-$ & $-$ & $+$ & $+$ & $-$\\  \hline
    ${\cal O}_{[SP-PS]}$ & $-$ &  $+$ & $-$ & $-$ & $+$\\  \hline
    ${\cal O}_{[SP+PS]}$ & $-$ & $+$ & $+$ & $-$ & $-$ \\  \hline
   ${\cal O}_{T \tilde T}$ & $-$ & $+$ & $+$ & $-$ & $-$ \\  \hline
    \end{tabular}
    \caption{Transformations of the four-quark operators ${\cal O}_{\Gamma \Tilde{\Gamma}}$ under $\mathcal{P}$, $\mathcal{CS'}$, $\mathcal{CS''}$, $\mathcal{CPS'}$ and $\mathcal{CPS''}$ are noted. The operators ${\cal O}_{\tilde T T}$ and ${\cal O}_{\tilde T \tilde T}$ are not explicitly shown in the above matrix, as they coincide with ${\cal O}_{T \tilde T}$ and ${\cal O}_{T T}$, respectively. For the Fierz four-quark operators ${\cal O}^F_{\Gamma \Tilde{\Gamma}}$, we must exchange the columns $\mathcal{CS'} \rightarrow \mathcal{CS''}$ and $\mathcal{CPS'} \rightarrow \mathcal{CPS''}$.}
\label{tb1}
\end{table}

The new basis of operators can be further decomposed into smaller independent bases according to the discrete symmetries $P,\,S,\,CPS',\,CPS''$. Following the notation of Ref.~\cite{Donini:1999sf}, the 20 operators of Table \ref{tb1} (including the Fierz operators) are classified into 4 categories: 
\begin{itemize}
    \item[(a)] Parity Conserving ($P=+1$) operators with $S=+1$: $Q_i^{S=+1}$, \qquad $(i=1,2,\ldots,5)$,
    \item[(b)] Parity Conserving ($P=+1$) operators with $S=-1$: $Q_i^{S=-1}$, \qquad $(i=1,2,\ldots,5)$,
    \item[(c)] Parity Violating \ \ ($P=-1$) operators with $S=+1$: $\mathcal{Q}_i^{S=+1}$, \qquad $(i=1,2,\ldots,5)$,
    \item[(d)] Parity Violating \ \ ($P=-1$) operators with $S=-1$: $\mathcal{Q}_i^{S=-1}$, \qquad $(i=1,2,\ldots,5)$,
\end{itemize}
which are explicitly given below: 

\be
\begin{split}
\begin{cases}
Q_1^{S=\pm 1}\equiv \frac{1}{2}\left[{\cal O}_{VV} \pm {\cal O}^F_{VV}\right]+\frac{1}{2}\left[{\cal O}_{AA} \pm {\cal O}^F_{AA}\right]\\[0.4ex]
Q_2^{S=\pm 1}\equiv \frac{1}{2}\left[{\cal O}_{VV} \pm {\cal O}^F_{VV}\right]-\frac{1}{2}\left[{\cal O}_{AA} \pm {\cal O}^F_{AA}\right]\\[0.4ex]
Q_3^{S=\pm 1}\equiv \frac{1}{2}\left[{\cal O}_{SS} \pm {\cal O}^F_{SS}\right]-\frac{1}{2}\left[{\cal O}_{PP} \pm {\cal O}^F_{PP}\right]\\[0.4ex]
Q_4^{S=\pm 1}\equiv \frac{1}{2}\left[{\cal O}_{SS} \pm {\cal O}^F_{SS}\right]+\frac{1}{2}\left[{\cal O}_{PP} \pm {\cal O}^F_{PP}\right]\\[0.4ex]
Q_5^{S=\pm 1}\equiv \frac{1}{2}\left[{\cal O}_{TT} \pm {\cal O}^F_{TT}\right],
\end{cases}
\end{split}
\qquad \quad 
\begin{split}
&\begin{cases}
{\cal Q}_1^{S=\pm 1}\equiv \frac{1}{2}\left[{\cal O}_{VA} \pm {\cal O}^F_{VA}\right]+\frac{1}{2}\left[{\cal O}_{AV} \pm {\cal O}^F_{AV}\right],
\end{cases}\\
&\begin{cases}
{\cal Q}_2^{S=\pm 1}\equiv \frac{1}{2}\left[{\cal O}_{VA} \pm {\cal O}^F_{VA}\right]-\frac{1}{2}\left[{\cal O}_{AV} \pm {\cal O}^F_{AV}\right]\\
{\cal Q}_3^{S=\pm 1}\equiv \frac{1}{2}\left[{\cal O}_{PS} \pm {\cal O}^F_{PS}\right]-\frac{1}{2}\left[{\cal O}_{SP} \pm {\cal O}^F_{SP}\right],
\end{cases}\\
&\begin{cases}
{\cal Q}_4^{S=\pm 1}\equiv \frac{1}{2}\left[{\cal O}_{PS} \pm {\cal O}^F_{PS}\right]+\frac{1}{2}\left[{\cal O}_{SP} \pm {\cal O}^F_{SP}\right]\\
{\cal Q}_5^{S=\pm 1}\equiv \frac{1}{2}\left[{\cal O}_{T\tilde T} \pm {\cal O}^F_{T\tilde T}\right].
\end{cases}    
\end{split}
\label{Q_definitions}
\ee

\noindent Note that a summation over all independent Lorentz indices (if any) of the Dirac matrices is understood. The operators of Eq. \eqref{Q_definitions} are grouped together according to their mixing pattern. Therefore, the mixing matrices $Z^{S=\pm 1}$ (${\cal Z}^{S=\pm 1}$), which renormalize the Parity Conserving (Violating) operators, take the following form:

{\small{
\be
Z^{S=\pm 1}
=
\left(\begin{array}{rrrrr}
Z_{11}\,\, & Z_{12}\,\, & Z_{13}\,\, & Z_{14}\,\, & Z_{15} \\
Z_{21}\,\, & Z_{22}\,\, & Z_{23}\,\, & Z_{24}\,\, & Z_{25} \\
Z_{31}\,\, & Z_{32}\,\, & Z_{33}\,\, & Z_{34}\,\, & Z_{35} \\
Z_{41}\,\, & Z_{42}\,\, & Z_{43}\,\, & Z_{44}\,\, & Z_{45} \\
Z_{51}\,\, & Z_{52}\,\, & Z_{53}\,\, & Z_{54}\,\, & Z_{55} 
\end{array}\right)^{S=\pm 1},
\quad
{\cal Z}^{S=\pm 1}
=
\left(\begin{array}{rrrrr}
 {\cal Z}_{11}  &0\,\,         &0\,\,         &0\,\,        &0\,\,  \\
 0\,\,         &{\cal Z}_{22}  &{\cal Z}_{23}  &0\,\,        &0\,\,  \\
 0\,\,         &{\cal Z}_{32}  &{\cal Z}_{33}  &0\,\,        &0\,\,  \\
 0\,\,         &0\,\,         &0\,\,         &{\cal Z}_{44}  &{\cal Z}_{45} \\
 0\,\,         &0\,\,         &0\,\,         &{\cal Z}_{54}  &{\cal Z}_{55}
\end{array}\right)^{S=\pm 1}.
\label{MixingMatrix}
\ee
}}

The renormalized Parity Conserving (Violating) operators,
$\hat{Q}^{S=\pm 1}$ ($\hat{\cal Q}^{S=\pm 1}$), are defined via the
equations:
\be
{\hat Q}_l^{S=\pm 1} = Z^{S=\pm 1}_{lm} \cdot Q^{S=\pm 1}_{m} ,\quad
\hat{\cal Q}^{S=\pm 1}_l = {\cal Z}^{S=\pm 1}_{lm} \cdot {\cal Q}^{S=\pm 1}_m,
\ee
where $l,m = 1,\dots ,5$ (a sum over $m$ is implied). 

Note that the mixing pattern of the Parity Conserving operators is reduced when chiral symmetry is considered leading to smaller mixing blocks similar to those of Parity Violating operators. Therefore, it is expected that the renormalization matrices for the two partity sectors take a block-diagonal form in dimensional regularization (DR), in the $\MSbar$ scheme. However, this is not true for our proposed gauge-invariant scheme which does not impose chiral symmetry. As a result, the renormalization matrix in (DR,GIRS) for the Parity-Conserving operators does not preserve the block-diagonal form, even though the regulator (DR) preserves chiral symmetry. This means that an additional finite mixing will be obtained in (DR, GIRS) among operators which can behave differently under chiral transformations.

In principle, the four-quark operators can also mix with a number of lower-dimensional operators, which have the same symmetry properties. However, in this work, we focus on operators with $\Delta F = 2$; thus, $f_1 \notin \{f_3, f_4\} $ and $f_2 \notin \{f_3, f_4\}$, which forbid such additional mixing. 

\subsection{Renormalization in GIRS}

In this work, the GIRS scheme~\cite{Costa:2021iyv} is employed for extracting the renormalization matrices $Z^{S=\pm 1}$ and ${\cal Z}^{S=\pm 1}$. GIRS is an extension of the X-space scheme~\cite{Gimenez:2004me,Chetyrkin:2010dx,Cichy:2012is,Tomii:2018zix}, in which Green’s functions of products of gauge-invariant
operators at different spacetime points are considered. In the case of a multiplicatively renormalizable operator, ${\cal O}$, a typical condition in GIRS has the following form:
 \begin{equation}
(Z_{\cal O}^{\rm GIRS} )^2 \langle{\cal O}(x) \,  {\cal O}^\dagger (y) \rangle \big{|}_{x-y=\bar z}=  \langle {\cal O}(x)  \, {\cal O}^\dagger (y)\rangle ^{\rm tree} \Big{|}_{x-y= \bar z},
\label{GIRScond}
 \end{equation}
where $\bar z$ is a nonzero renormalization 4-vector scale. The Green's function $\langle{\cal O}(x) \,  {\cal O}^\dagger (y) \rangle$ is gauge independent and thus, a nonperturbative implementation of such a scheme on the lattice avoids gauge fixing altogether and the numerical simulation becomes more straightforward and statistically robust without the issue of Gribov copies. [Note that GIRS is not a unique gauge-independent scheme. While the $\MSbar$ and RI$'$ families of schemes do require gauge fixing, there are other nonperturbative schemes, such as the Schr{\"o}dinger Functional family, which do not.] When operator mixing occurs, one needs to consider a set of conditions involving more than one Green's functions of two or more gauge-invariant operators, each of which has a similar form to Eq.\eqref{GIRScond}, i.e., the renormalized Green's functions are set to their tree-level values when the operators' space-time separations equal to specific reference scales. 

In our study, the determination of the $5 \times 5$ mixing matrices of Eq. \eqref{MixingMatrix} requires the calculation of (i) two-point Green's functions with two four-quark operators and (ii) three-point Green's functions with one four-quark operator and two lower dimensional operators, e.g., quark bilinear operators ${\cal O}_{\Gamma}(x) = \bar{\psi}_{f_1}(x) \Gamma \psi_{f_2}(x)$. All operators are placed at different spacetime points, in a way as to avoid potential contact singularities: 
\begin{eqnarray}
G^{\rm 2pt}_{\mathcal{O}_{\Gamma \Tilde{\Gamma}};\, \mathcal{O}_{\Gamma' \Tilde{\Gamma'}}} (z) &\equiv& \langle {\cal O}_{\Gamma \Tilde{\Gamma}} (x) \,  {\cal O}^\dagger_{\Gamma' \Tilde{\Gamma'}} (y) \rangle, \qquad \qquad \ \, z \equiv x-y, \ x \neq y, \label{4Q4Q}
\\
G^{\rm 3pt}_{\mathcal{O}_{\Gamma'};\, \mathcal{O}_{\Gamma\Tilde{\Gamma}};\, \mathcal{O}_{\Gamma''}} (z,z') &\equiv& \langle {\cal O}_{\Gamma'} (x) \, {\cal O}_{\Gamma\Tilde{\Gamma}} (y) {\cal O}_{\Gamma''} (w) \rangle, \qquad z \equiv x-y, \ z' \equiv y-w, \ x \neq y \neq w \neq x.
\label{4Q2Q2Q}
\end{eqnarray}
Two-point Green's functions with one four-quark operator and one bilinear operator are not considered since they vanish when $\Delta F=2$. In principle, the perturbative calculation of the Green's functions of Eqs.~(\ref{4Q4Q} -- \ref{4Q2Q2Q}) can be performed for generic Dirac matrices $\Gamma, \Tilde{\Gamma}, \Gamma',\Tilde{\Gamma}',\Gamma''$, which do not lead to vanishing result. However, when constructing the renormalization conditions, we specify the Dirac matrices for both four-quark ($Q_i$ and $\mathcal{Q}_i$ combinations) and bilinear operators (see section \ref{conditions}). 

In order to determine a consistent and solvable set of nonperturbative renormalization conditions, we need to examine multiple choices of three-point functions with different bilinear operators. Also, since there is no unique way of selecting solvable conditions in GIRS, a perturbative calculation of all possible Green’s functions will be useful for determining conversion factors from different variants of GIRS to the $\MSbar$ scheme. To this end, we calculate the Green's functions of Eqs.~(\ref{4Q4Q} -- \ref{4Q2Q2Q}), up to one-loop order and in DR. 

\begin{figure}[ht!]
\includegraphics[scale=1]{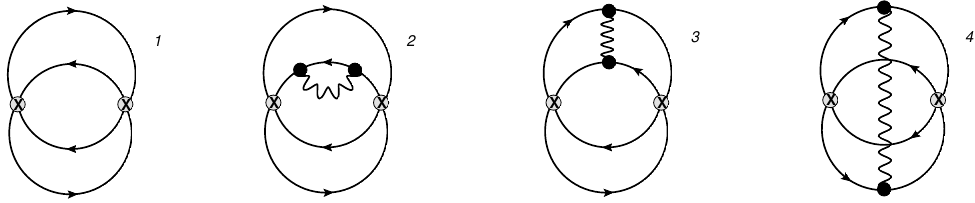} 
\caption{Feynman diagrams contributing to $\langle {\cal O}_{\Gamma \Tilde{\Gamma}} (x) \,  {\cal O}^\dagger_{\Gamma' \Tilde{\Gamma'}} (y) \rangle$, to order $\mathcal{O} (g^0)$ (diagram 1) and $\mathcal{O} (g^2)$ (the remaining diagrams). Wavy (solid) lines represent gluons (quarks). Diagrams 2 and 4 have also mirror variants.}
\label{fig4Q4Q:2pt}
\end{figure}

\begin{figure}[ht!]
\includegraphics[scale=0.95]{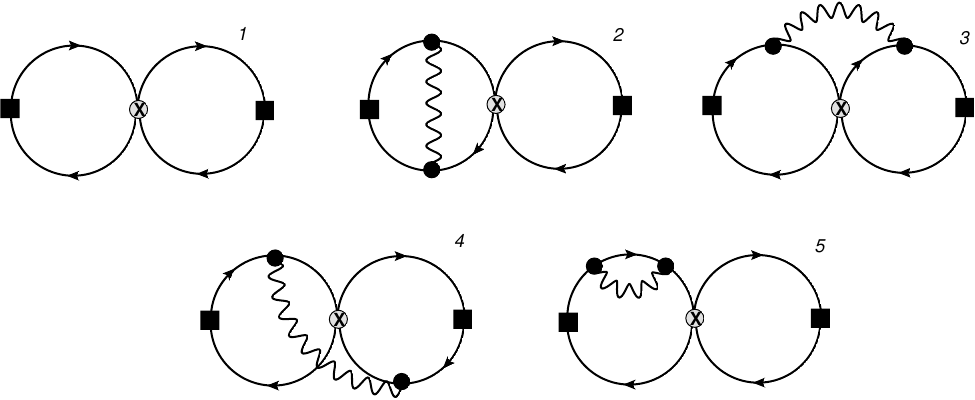} 
\caption{Feynman diagrams contributing to $\langle {\cal O}_{\Gamma'} (x) \, {\cal O}_{\Gamma \Tilde{\Gamma}} (0) {\cal O}_{\Gamma''} (y) \rangle$, to order $\mathcal{O} (g^0)$ (diagram 1) and $\mathcal{O} (g^2)$ (the remaining diagrams). Wavy (solid) lines represent gluons (quarks). A circled cross denotes the insertion of the four-quark operator, and the solid squares denote the quark bilinear operators. Diagrams 2-5 have also mirror variants. 
}
\label{fig4Q2Q2Q:3pt}
\end{figure}

The Feynman diagrams contributing to the two-point functions of the four-quark operators, to order ${\cal O}(g^0)$ (diagram 1) and ${\cal O} (g^2)$ (the remaining diagrams), are shown in Fig.~\ref{fig4Q4Q:2pt}. Likewise, the Feynman diagrams contributing to the three-point Green's functions of the product of one four-quark operator and two quark bilinear operators are shown in Fig.~\ref{fig4Q2Q2Q:3pt}. For simplicity, we have not drawn separate diagrams to specify which quark/antiquark appearing in the definition of the four-quark operators is contracted in each fermion propagator. Thus, in each diagram, it is understood that all possible ways of contracting the quark/antiquark fields of the operators are summed over. The $\mathcal{O} (g^2)$ (and $\mathcal{O} (g^0)$ ) contributions for each Green's function are gauge-independent; indeed, terms dependent on the gauge parameter cancel out upon the summation of the Feynman diagrams.

There are numerous possible variants of GIRS, depending on which Green's functions and which renormalization four-vectors are selected for imposing renormalization conditions. A variant of choice, which is expected to result in reduced statistical noise in lattice simulations, includes integration (summation on the lattice) over time slices of the operator-insertion points in all Green’s functions:
\bea
\Tilde{G}^{\rm 2pt}_{\mathcal{O}_{\Gamma \Tilde{\Gamma}};\mathcal{O}_{\Gamma' \Tilde{\Gamma'}}} (z_4) &\equiv& \int d^3 \vec{z} \ G^{\rm 2pt}_{\mathcal{O}_{\Gamma \Tilde{\Gamma}};\mathcal{O}_{\Gamma' \Tilde{\Gamma'}}} (\vec{z},z_4), \qquad \qquad \qquad \qquad \qquad \ z_4 > 0, \label{2pt_int} \\
\Tilde{G}^{\rm 3pt}_{\mathcal{O}_{\Gamma'}; \mathcal{O}_{\Gamma\Tilde{\Gamma}}; \mathcal{O}_{\Gamma''}} (z_4, z_4') &\equiv& \int d^3 \vec{z} \int d^3 \vec{z'} \ G^{\rm 3pt}_{\mathcal{O}_{\Gamma'};\mathcal{O}_{\Gamma\Tilde{\Gamma}};\mathcal{O}_{\Gamma''}} ((\vec{z},z_4),(\vec{z'},z_4')), \qquad z_4 >0, \ z_4'>0. \label{3pt_int}
\eea
We have employed this variant in a number of previous studies of our group regarding the renormalization of fermion bilinear operators~\cite{Costa:2021iyv}, the study of mixing between the gluon and quark energy-momentum tensor operators~\cite{Costa:2021iyv}, as well as the renormalization of supersymmetric operators, such as gluino-glue~\cite{Costa:2021pfu} and supercurrent~\cite{Bergner:2022see} in $\mathcal{N} =1$ Supersymmetric Yang-Mills theory (SYM).

\subsection{Renormalization conditions}
\label{conditions}

In the case of the Parity Conserving operators ($Q_i$), the mixing matrix is $5\times5$ for both $S=+1$ and $S=-1$. Therefore, for each case, we need 25 conditions to obtain these mixing coefficients. Computing the relevant two-point Green's functions, we extract 15 conditions, and therefore, we need another 10 conditions that will be extracted from the relevant three-point Green's functions. The 15 conditions in GIRS, which include two-point Green's functions, are the following: 
\be
[\Tilde{G}^{\rm 2pt}_{Q_i^{S=\pm 1}; Q_j^{S=\pm 1}} (t)]^{\rm GIRS} \equiv \sum_{k,l = 1}^5 (Z_{ik}^{S \pm 1})^{\rm GIRS} (Z_{jl}^{S \pm 1})^{\rm GIRS} \ \Tilde{G}^{\rm 2pt}_{Q_k^{S=\pm 1}; Q_l^{S=\pm 1}} (t) = [\Tilde{G}^{\rm 2pt}_{Q_i^{S=\pm 1};Q_j^{S=\pm 1}} (t)]^{\rm tree},
\label{cond2pt}
\ee
where $i, j$ runs from 1 to 5 and $i \leq j$; $z_4:=t$ is the GIRS renormalization scale. We have a variety of options for selecting the remaining conditions involving three-point Green's functions:
\be
[\Tilde{G}^{\rm 3pt}_{\mathcal{O}_\Gamma;Q_i^{S=\pm 1}; \mathcal{O}_\Gamma} (t,t')]^{\rm GIRS} \equiv (Z_{\mathcal{O}_\Gamma}^{\rm GIRS})^2 \ \sum_{k=1}^5 \ (Z_{ik}^{S \pm 1})^{\rm GIRS} \ \Tilde{G}^{\rm 3pt}_{\mathcal{O}_\Gamma; Q_k^{S=\pm 1}; \mathcal{O}_\Gamma} (t,t') = [\Tilde{G}^{\rm 3pt}_{\mathcal{O}_\Gamma;Q_i^{S=\pm 1}; \mathcal{O}_\Gamma} (t,t')]^{\rm tree},
\label{cond3pconserving}
\ee
where $i \in [1,5]$, $\Gamma \in \{\openone,\, \gamma_5,\, \gamma_\mu,\, \gamma_\mu \gamma_5,\, \sigma_{\mu\nu} \}$, and $z_4:=t$, $z'_4 :=t'$ are GIRS renormalization scales. In this case, the two bilinears must be the same in order to obtain a nonzero Green's function. $Z_{\mathcal{O}_\Gamma}^{\rm GIRS}$ is the renormalization factor of the bilinear operator $\mathcal{O}_\Gamma$ calculated in Ref.~\cite{Costa:2021iyv}. 

In order to avoid having more than one renormalization scale, a natural choice is to set $t'=t$; in this way, the original set of two four-vector renormalization scales (given by specific values of $z$ and $z'$, cf. Eqs.~\ref{4Q4Q}-\ref{4Q2Q2Q}), following integration over time slices and setting $t'=t$, is reduced to just one real variable. Changing the values of $t$ and/or $t'$ would obviously affect the results for the nonperturbative Green's functions in Eqs.~\ref{cond2pt}-\ref{cond3pconserving}; nevertheless, after multiplication by the appropriate conversion factors, one should arrive at the {\it{same}} $\MSbar$-renormalized Green's functions, independently of $t$ and $t'$ (assuming that various standard sources of systematic error are under control). This then provides a powerful consistency check for the renormalization of four-quark operators. 

As we conclude, by doing the perturbative calculation, not all sets of conditions given in Eqs.~\ref{cond2pt}-\ref{cond3pconserving} can lead to viable solutions. We will provide a feasible choice in section~\ref{results}. In practice, one can choose the specific conditions that provide a more stable signal in numerical simulations. 

In the case of the Parity Violating operators ($\mathcal{Q}_i)$, the $5\times5$ mixing matrix is block diagonal for both $S=+1$ and $S=-1$, as dictated by symmetries. In particular, there are three mixing subsets: $\{\mathcal{Q}_1\}$, $\{ \mathcal{Q}_2, \mathcal{Q}_3 \}$ and $\{\mathcal{Q}_4, \mathcal{Q}_5 \}$, for each $S$. The first subset includes only 1 operator, which is multiplicatively renormalizable; thus, only one condition is needed and can be obtained from the two-point functions. The second and third subsets include two operators, and thus, 4 conditions are needed for each subset to obtain the mixing coefficients. Three of them will be extracted from the two-point functions, while the remaining 1 condition requires the calculation of three-point Green's functions. Thus, in total, we need 9 conditions for each $S$: 7 will be extracted from two-point Green's functions, and 2 will be extracted from three-point Green's functions. The seven conditions that include two-point Green's functions are the following: 
\bea
[\Tilde{G}^{\rm 2pt}_{\mathcal{Q}_1^{S=\pm 1}; \mathcal{Q}_1^{S=\pm 1}} (t)]^{\rm GIRS} &\equiv& [(\mathcal{Z}_{11}^{S \pm 1})^{\rm GIRS}]^2 \ \Tilde{G}^{\rm 2pt}_{\mathcal{Q}_1^{S=\pm 1}; \mathcal{Q}_1^{S=\pm 1}} (t) = [\Tilde{G}^{\rm 2pt}_{\mathcal{Q}_1^{S=\pm 1}; \mathcal{Q}_1^{S=\pm 1}} (t)]^{\rm tree}, \\
{[\Tilde{G}^{\rm 2pt}_{\mathcal{Q}_i^{S=\pm 1}; \mathcal{Q}_j^{S=\pm 1}} (t)]}^{\rm GIRS} &\equiv& \sum_{k,l=2}^3 \ (\mathcal{Z}_{ik}^{S \pm 1})^{\rm GIRS} (\mathcal{Z}_{jl}^{S \pm 1})^{\rm GIRS} \ \Tilde{G}^{\rm 2pt}_{\mathcal{Q}_k^{S=\pm 1}; \mathcal{Q}_l^{S=\pm 1}} (t) = [\Tilde{G}^{\rm 2pt}_{\mathcal{Q}_i^{S=\pm 1}; \mathcal{Q}_j^{S=\pm 1}} (t)]^{\rm tree}, \ (i,j = 2,3), \qquad \\
{[\Tilde{G}^{\rm 2pt}_{\mathcal{Q}_i^{S=\pm 1}; \mathcal{Q}_j^{S=\pm 1}} (t)]}^{\rm GIRS} &\equiv& \sum_{k,l = 4}^5 \ (\mathcal{Z}_{ik}^{S \pm 1})^{\rm GIRS} (\mathcal{Z}_{jl}^{S \pm 1})^{\rm GIRS} \ \Tilde{G}^{\rm 2pt}_{\mathcal{Q}_k^{S=\pm 1}; \mathcal{Q}_l^{S=\pm 1}} (t) = [\Tilde{G}^{\rm 2pt}_{\mathcal{Q}_i^{S=\pm 1}; \mathcal{Q}_j^{S=\pm 1}} (t)]^{\rm tree}, \ (i,j = 4, 5). \qquad
\eea
Note that in the above equations $i \leq j$. The two conditions that include three-point functions can be:
\bea
[\Tilde{G}^{\rm 3pt}_{\mathcal{O}_\Gamma;\mathcal{Q}_i^{S=\pm 1}; \mathcal{O}_{\Gamma \gamma_5}} (t,t')]^{\rm GIRS} &\equiv& Z_{\mathcal{O}_\Gamma}^{\rm GIRS} \ Z_{\mathcal{O}_\Gamma \gamma_5}^{\rm GIRS} \ \sum_{k=2}^3 \ (\mathcal{Z}_{ik}^{S \pm 1})^{\rm GIRS} \ \Tilde{G}^{\rm 3pt}_{\mathcal{O}_\Gamma; \mathcal{Q}_k^{S=\pm 1}; \mathcal{O}_{\Gamma \gamma_5}} (t,t') \nonumber \\
&=& [\Tilde{G}^{\rm 3pt}_{\mathcal{O}_\Gamma;\mathcal{Q}_i^{S=\pm 1}; \mathcal{O}_{\Gamma \gamma_5}} (t,t')]^{\rm tree}, \ (i=2 \ {\rm or} \ 3), \\
{[\Tilde{G}^{\rm 3pt}_{\mathcal{O}_\Gamma;\mathcal{Q}_i^{S=\pm 1}; \mathcal{O}_{\Gamma \gamma_5}} (t,t')]}^{\rm GIRS} &\equiv& Z_{\mathcal{O}_\Gamma}^{\rm GIRS} \ Z_{\mathcal{O}_\Gamma \gamma_5}^{\rm GIRS} \ \sum_{k=4}^5 \ (\mathcal{Z}_{ik}^{S \pm 1})^{\rm GIRS} \ \Tilde{G}^{\rm 3pt}_{\mathcal{O}_\Gamma; \mathcal{Q}_k^{S=\pm 1}; \mathcal{O}_{\Gamma \gamma_5}} (t,t') \nonumber \\
&=& [\Tilde{G}^{\rm 3pt}_{\mathcal{O}_\Gamma;\mathcal{Q}_i^{S=\pm 1}; \mathcal{O}_{\Gamma \gamma_5}} (t,t')]^{\rm tree}, \ (i=4 \ {\rm or} \ 5),
\label{cond3pt}
\eea
where $\Gamma \in \{\openone,\, \gamma_\mu,\, \sigma_{\mu\nu} \}$. In this case, the two bilinears must differ by $\gamma_5$ in order to obtain a nonzero Green's function. As in the Parity Conserving operators, we simplify the conditions by setting $t' = t$. It is not guaranteed that all possible choices can give a solution to the system of conditions. We test all options, and we provide the choices that can work in section~\ref{results}. 

\subsection{Conversion matrices and anomalous dimensions}

In order to arrive at the renormalized four-quark operators in the more standard $\MSbar$ scheme (which is the typical scheme used in phenomenological work), the conversion matrices $(C^{S=\pm 1})^{\MSbar, \rm{GIRS}}$ and $(\Tilde{C}^{S=\pm 1})^{\MSbar, \rm{GIRS}}$ between GIRS and $\MSbar$ schemes are necessary:
\begin{equation}
(Z^{S=\pm 1})^\MSbar  = (C^{S=\pm 1})^{\MSbar, \rm{GIRS}} (Z^{S=\pm 1})^{\rm{GIRS}},  \qquad (\mathcal{Z}^{S=\pm 1})^\MSbar  = (\Tilde{C}^{S=\pm 1})^{\MSbar, \rm{GIRS}} (\mathcal{Z}^{S=\pm 1})^{\rm{GIRS}}.
\label{Cdef}
\end{equation}
These can be computed only perturbatively due to the very nature of $\MSbar$. Being regularization-independent, they are evaluated more easily in DR. The one-loop expressions of the conversion matrices for different variants of GIRS are extracted from our calculations and are given in section~\ref{conv_matrices} for a selected version of GIRS. 
The conversion matrices, along with the lattice mixing matrices in GIRS, calculated nonperturbatively, allow the extraction of the lattice mixing matrices in the $\MSbar$ scheme.

A by-product of our calculation is the determination of the next-to-leading (NLO) order anomalous dimensions of the four-quark operators in GIRS. Using standard notation, we define (in terms of the renormalization scale $\mu$) the Callan-Symanzik equations satisfied by the renormalized gauge coupling $g^R$ and the renormalized four-quark operators in an arbitrary renormalization scheme, denoted by ``R":
\begin{eqnarray}
    \mu \frac{d}{d\mu} g^R (\mu) &=& \beta^R (g^R(\mu)), \label{beta}\\
    \mu \frac{d}{d\mu} \hat{Q}_i^{S=\pm1,R} &=& \sum_j \gamma_{ij}^{\pm,R} (g^R(\mu)) \ \hat{Q}_j^{S=\pm1,R}, \\
    \mu \frac{d}{d\mu} \hat{\cal Q}_i^{S=\pm1,R} &=& \sum_j {\tilde{\gamma}}_{ij}^{\pm,R} (g^R(\mu)) \ \hat{Q}_j^{S=\pm1,R}.
\end{eqnarray}
Also, we define the perturbative expansions of the $\beta$ function, the anomalous dimensions and the conversion matrices (between two schemes) of the four-quark operators, as follows:
\begin{eqnarray}
    \beta (g) &=& -g^3 (b_0 + b_1 g^2 + \ldots), \\
    \gamma^{\pm} (g) &=& -g^2 (\gamma_0^{\pm} + \gamma_1^{\pm} g^2 + \ldots), \\
    \tilde{\gamma}^{\pm} (g) &=& -g^2 (\tilde{\gamma}_0^{\pm} + \tilde{\gamma}_1^{\pm} g^2 + \ldots), \\
    C^{S=\pm 1} (g) &=& (\openone +  c^{\pm}_1 g^2 + \ldots), \\
    \tilde{C}^{S=\pm 1} (g) &=& (\openone +  \tilde{c}^{\pm}_1 g^2 + \ldots). \label{pertexpansion}
\end{eqnarray}
By considering Eqs. (\ref{beta} -- \ref{pertexpansion}) in both $\MSbar$ and GIRS, one can extract the following scheme-conversion formula of NLO coefficients ($g^4$) for the operator anomalous dimensions~\cite{Papinutto:2016xpq}:   
\begin{eqnarray}
    \gamma_1^{\pm, {\rm GIRS}} &=& \gamma_1^{\pm, \MSbar} - 2 b_0 c_1^{\pm} + [\gamma_0^\pm, c_1^{\pm}], \label{gamma_cons} \\
        \tilde{\gamma}_1^{\pm, {\rm GIRS}} &=& \tilde{\gamma}_1^{\pm, \MSbar} - 2 b_0 \tilde{c}_1^{\pm} + [\tilde{\gamma}_0^\pm, \tilde{c}_1^{\pm}]. \label{gamma_viol}
\end{eqnarray}
In the above relations, we set the GIRS coupling equal to the $\MSbar$ coupling, at least up to the NLO order, and thus, an additional term containing the conversion factor of the coupling constant is not present. Also, as is standard practice, we relate the renormalization scales in the two schemes: thus, we set the GIRS scales $t,t'$ proportional to $1/\bar{\mu}$. The coefficients $b_0$, $\gamma_0^\pm$, $\gamma_1^{\pm,\MSbar}$, $\tilde{\gamma}_0^\pm$, $\tilde{\gamma}_1^{\pm,\MSbar}$ are well-known in the literature (see, e.g., \cite{Papinutto:2016xpq}\footnote{The convention used in this reference differs (compared to the current work) by a factor of 2 for the operators $Q_5$ and ${\cal Q}_5$.}).

\section{Calculation of Feynman  integrals for Green's functions in \sout{the} coordinate space}
\label{CalculationProcedure}
In this section, we briefly describe our methodology for calculating the two-point and three-point ``GIRS'' Green's functions defined in the previous section using dimensional regularization. 

There are two types of prototype scalar Feynman integrals that enter the calculation of the two-point functions $G^{\rm 2pt}_{\mathcal{O}_{\Gamma \Tilde{\Gamma}}; \mathcal{O}_{\Gamma' \Tilde{\Gamma'}}} (z)$ to the tree level and one loop, respectively:
\begin{equation}
I_1^D (\xi_1;\alpha_1) \equiv \ \int \frac{d^D p_1}{{(2 \pi)}^D} \frac{e^{i p_1 \cdot \xi_1}}{{(p_1^2)}^{\alpha_1}},
\end{equation}
\begin{equation}
I_2^D (\xi_1; \alpha_1,\alpha_2,\alpha_3,\alpha_4,\alpha_5) \equiv \ \int \frac{d^D p_1 \ d^D p_2 \ d^D p_3}{{(2 \pi)}^{3D}} \frac{e^{i p_3 \cdot \xi_1}}{{(p_1^2)}^{\alpha_1} \ {((-p_1 +p_3)^2)}^{\alpha_2} \ {({(-p_1 + p_2)}^2)}^{\alpha_3} \ {(p_2^2)}^{\alpha_4} \ {({(-p_2 + p_3)}^2)}^{\alpha_5}},
\end{equation}
where $D \equiv 4 -2 \epsilon$  is the number of spacetime dimensions and the vector $\xi_1$ satisfies: $\xi_1 \neq 0$. In our calculation, $I_2^D$ is only needed for integer values of the exponents $\alpha_i$.  Tensor integrals with an arbitrary number of momentum-loop components ${p_1}_\mu$, ${p_2}_\nu$, ${p_3}_\rho$ in the numerator can be reduced to scalars through derivatives w.r.t. components of $\xi_1$ in the above scalar integrals or integration by parts (see Eq. (45) in Ref.~\cite{Panagopoulos:2020qcn}). 

Integral $I_1^D$ is computed by introducing a Schwinger parameter: $1/{(p_1^2)}^{\alpha_1} = 1/\Gamma (\alpha_1) \int_0^\infty d\lambda \ \lambda^{\alpha_1 - 1} e^{-\lambda p_1^2}$, leading to the following resulting expression:
\begin{equation}
I_1^D (\xi_1;\alpha_1) = \frac{\Gamma(-\alpha_1 + D/2) \ {(\xi_1^2)}^{\alpha_1 - D/2}}{4^\alpha_1 \ \pi^{D/2} \ \Gamma (\alpha_1)}.
\end{equation}

Integral $I_2^D$ is calculated in two steps: first, the integration over $p_1$ and $p_2$ is performed, which is independent of the phase factor of the numerator. This inner two-loop integral is evaluated through the standard ``diamond''-type recursive formula of Ref.~\cite{Chetyrkin:1981qh}. The resulting expression depends on the scalar quantity $p_3^2$. Then, the remaining integral over $p_3$ takes the form of $I_1^D$.

The calculation of the three-point functions $G^{\rm 3pt}_{\mathcal{O}_{\Gamma'}; \mathcal{O}_{\Gamma\Tilde{\Gamma}}; \mathcal{O}_{\Gamma''}} (z,z')$ involve the following prototype scalar Feynman integrals, in addition to $I_1^D$: 
\begin{eqnarray}
I_3^D (\xi_1, \xi_2;\alpha_1,\alpha_2,\alpha_3) &\equiv & \ \int \frac{d^D p_1 \ d^D p_2}{{(2 \pi)}^{2D}} \frac{e^{i p_1 \cdot \xi_1} \ e^{i p_2 \cdot \xi_2}}{{(p_1^2)}^{\alpha_1} \ {({(-p_1 + p_2)}^2)}^{\alpha_2} \ {(p_2^2)}^{\alpha_3}}, \\
I_4^D (\xi_1, \xi_2;\alpha_1,\alpha_2,\alpha_3,\alpha_4,\alpha_5) &\equiv & \ \int \frac{d^D p_1 \ d^D p_2 \ d^D p_3}{{(2 \pi)}^{3D}} \frac{e^{i p_2 \cdot \xi_1} \ e^{i p_3 \cdot \xi_2}}{{(p_1^2)}^{\alpha_1} \ {(p_2^2)}^{\alpha_2} \ {({(-p_1 + p_2)}^2)}^{\alpha_3} \ {(p_3^2)}^{\alpha_4} \ {({(-p_1 + p_3)}^2)}^{\alpha_5}},
\label{I4d}
\end{eqnarray}
where the vectors $\xi_1,\ \xi_2$ satisfy: $\xi_1 \neq 0$, $\xi_2 \neq 0$, and $(\xi_1 + \xi_2) \neq 0$. Integral $I_4^D$ takes only integer values of $\alpha_i$ in the current calculation. As in the case of the two-point functions, tensor integrals can be reduced to scalars through derivatives w.r.t. components of $\xi_1$, $\xi_2$ of the scalar integrals, or integration by parts. 

The two-loop integral $I_3^D$ can be reduced to one-loop integral $J^D$ by using Schwinger parametrization:
\begin{equation}
  I_3^D (\xi_1, \xi_2;\alpha_1,\alpha_2,\alpha_3) = \frac{ \Gamma(D/2-s)}{4^{s+D/2} \ \pi^{3 D/2} \ \Gamma(s)} \ (\xi_1^2)^{s-\alpha_3} \ (\xi_2^2)^{s-\alpha_1} \ ((\xi_1 + \xi_2)^2)^{s-\alpha_2} \ J^D (\xi_1, \xi_2; \alpha_1, \alpha_2, \alpha_3),  
\end{equation}
where $s \equiv \alpha_1 + \alpha_2 + \alpha_3 -D/2$, and
\begin{equation}
    J^D (\xi_1, \xi_2; \alpha_1, \alpha_2, \alpha_3) \equiv \int d^D x \frac{1}{{((-x + \xi_1)^2)}^{\alpha_1} \ (x^2)^{\alpha_2} ((x + \xi_2)^2)^{\alpha_3}}.
\end{equation}
The ``triangle'' integral $J^D$ is well-studied in Refs.~\cite{Davydychev:1992xr,Usyukina:1994iw}. By using the recursive relations of Ref.~\cite{Davydychev:1992xr}, the integrals of type $J^D$ appearing in our calculation can be expressed in terms of the following master integrals, calculated in Ref.~\cite{Usyukina:1994iw} up to $\mathcal{O} (\epsilon)$: 
\begin{eqnarray}
J^{4-2\epsilon} (\xi_1,\xi_2; 1,1,1) &=& \frac{\pi^{2-\epsilon} \ \Gamma(1+\epsilon)}{(\xi_3^2)^{1+\epsilon}} \left\{ \Phi^{(1)} \left(\frac{\xi_1^2}{\xi_3^2}, \frac{\xi_2^2}{\xi_3^2}\right) + \epsilon \ \Psi^{(1)} \left(\frac{\xi_1^2}{\xi_3^2}, \frac{\xi_2^2}{\xi_3^2}\right) + \mathcal{O} (\epsilon^2) \right\}, \\
J^{4-2\epsilon} (\xi_1,\xi_2; 1,1+\epsilon,1) &=& \frac{\pi^{2-\epsilon} \ \Gamma(1+\epsilon)}{(\xi_3^2)^{1+2\epsilon}} \left\{ \Phi^{(1)} \left(\frac{\xi_1^2}{\xi_3^2}, \frac{\xi_2^2}{\xi_3^2}\right) \left(1-\frac{\epsilon}{2} \ln \left( \frac{\xi_1^2 \xi_2^2}{\xi_3^2}\right) \right) + \epsilon \ \Psi^{(1)} \left(\frac{\xi_1^2}{\xi_3^2}, \frac{\xi_2^2}{\xi_3^2}\right) + \mathcal{O} (\epsilon^2) \right\}, \\
J^{4-2\epsilon} (\xi_1,\xi_2; 1,\epsilon,1) &=& \frac{\pi^{2-\epsilon} \ \Gamma(1+\epsilon)}{(\xi_3^2)^{2\epsilon} \ 2 \ (1-3\epsilon)} \left\{ \frac{1}{\epsilon} - \epsilon \left[ \frac{\pi^2}{6} +  \ln\left(\frac{\xi_1^2}{\xi_3^2}\right) \ln\left(\frac{\xi_2^2}{\xi_3^2}\right) - \frac{2 \ \xi_1 \cdot \xi_2}{\xi_3^2} \ \Phi^{(1)} \left(\frac{\xi_1^2}{\xi_3^2}, \frac{\xi_2^2}{\xi_3^2}\right) \right] + \mathcal{O} (\epsilon^2) \right\}, \quad \ \,
\end{eqnarray}
where $\xi_3^2 \equiv (\xi_1 + \xi_2)^2$ and $\Phi^{(1)} (\xi_1^2/\xi_3^2, \xi_2^2/\xi_3^2)$, $\Psi^{(1)} (\xi_1^2/\xi_3^2), \xi_2^2/\xi_3^2)$ are polylogarithmic functions given in~\cite{Usyukina:1994iw}. Note that by summing all Feynman diagrams, $\Phi^{(1)}$ and $\Psi^{(1)}$ functions are canceled from the final expressions of the three-point Green's functions.

Integral $I_4^D$ is simplified by applying integration by parts w.r.t. $p_1$, thus, leading to the following recursive relation, which can eliminate inverse powers of $p_1^2$, or $p_2^2$, or $p_3^2$~\cite{Costa:2021iyv}:
 \begin{eqnarray}
 I_4^D (\xi_1, \xi_2; \alpha_1, \alpha_2, \alpha_3, \alpha_4, \alpha_5) &=& \frac{1}{-2 \alpha_1 - \alpha_3 -\alpha_5 + D} \cdot \nonumber \\
&& \hspace{-1cm} \Big[ \alpha_3 \Big( I_4^D(\xi_1, \xi_2; \alpha_1 - 1, \alpha_2, \alpha_3 + 1, \alpha_4, \alpha_5) - I_4^D(\xi_1, \xi_2; \alpha_1, \alpha_2 - 1, \alpha_3 + 1, \alpha_4, \alpha_5) \Big) + \nonumber \\
&& \hspace{-1cm} \ \ \alpha_5 \Big( I_4^D(\xi_1, \xi_2; \alpha_1 - 1, \alpha_2, \alpha_3, \alpha_4, \alpha_5 + 1) - I_4^D(\xi_1, \xi_2; \alpha_1, \alpha_2, \alpha_3, \alpha_4 - 1, \alpha_5 + 1) \Big) \Big]. \quad \quad
\label{I4value}
 \end{eqnarray}
In the case where $\alpha_1$, $\alpha_2$, $\alpha_4$ are positive integers, which is true in the computation at hand, an iterative implementation of Eq. \eqref{I4value} leads to terms with one propagator less. One momentum can then be integrated using a well-known one-loop formula (see Eqs. (A.1 -- A.2) in Ref.~\cite{Chetyrkin:1981qh}); the remaining integrals are of type $I_1^D$ or $I_3^D$.

\section{Results} 
\label{results}

In this section, we present perturbative results for the two-point and three-point Green's functions, along with the mixing matrices and conversion matrices between GIRS and $\MSbar$ schemes, utilizing DR in $D \equiv 4 - 2 \epsilon$ dimensions. Due to the very lengthy expressions of the renormalized Green’s functions, we include a supplemental Mathematica input file, named ``\texttt{GIRS\_Greens\_functions\_4-quark.m}'', containing the full expressions, while partial results are presented in the manuscript. Also, since there are various options of conditions that lead to unique solutions (within one-loop perturbation theory), we have chosen to present one set, while further options can be extracted from our results provided in the supplemental file.

\subsection{Bare Green's functions}

We present our results for the bare tree-level two-point Green's function of two four-quark operators with arbitrary Dirac matrices ($\Gamma, \Tilde{\Gamma}, \Gamma', \Tilde{\Gamma}'$) and arbitrary flavors ($f_i$, $f_i'$, $i=1,2,3,4$) carried by the quark fields. The result is given to all orders in $\epsilon$, and it depends, explicitly, on the $D$-vector $z \equiv y-x$, which connects the positions of the two operators:
\begin{eqnarray}   
&& \hspace{-2cm}\left\langle \left( \bar \psi_{f_1}(x) \Gamma \psi_{f_3} (x) \bar \psi_{f_2} (x) \Tilde{\Gamma} \psi_{f_4} (x) \right) \,   \left(\bar \psi_{f'_1}(y) \Gamma' \psi_{f'_3} (y) \bar \psi_{f'_2} (y) \Tilde{\Gamma}' \psi_{f'_4} (y)\right) \right\rangle^{\rm tree} = \frac{N_c \ \Gamma (2-\epsilon)^4}{16 \ \pi^{8-4\epsilon} \ (z^2)^{8-4\epsilon}} \times \nonumber \\
&& \Big\{\delta_{f_1 f_3'} \delta_{f_2 f_4'} \ [N_c \ \delta_{f_3 f_1'} \delta_{f_4 f_2'} \ {\rm tr} (\Gamma \slashed{z} \Gamma' \slashed{z}) \ {\rm tr} (\Tilde{\Gamma} \slashed{z} \Tilde{\Gamma}' \slashed{z}) - \delta_{f_3 f_2'} \delta_{f_4 f_1'} \ {\rm tr} (\Gamma \slashed{z} \Tilde{\Gamma}' \slashed{z} \Tilde{\Gamma} \slashed{z} \Gamma' \slashed{z})] \nonumber \\
&& + \delta_{f_1 f_4'} \delta_{f_2 f_3'} \ [N_c \ \delta_{f_3 f_2'} \delta_{f_4 f_1'} \ {\rm tr} (\Gamma \slashed{z} \Tilde{\Gamma}' \slashed{z}) \ {\rm tr} (\Tilde{\Gamma} \slashed{z} \Gamma' \slashed{z}) - \delta_{f_3 f_1'} \delta_{f_4 f_2'} \ {\rm tr} (\Gamma \slashed{z} \Gamma' \slashed{z} \Tilde{\Gamma} \slashed{z} \Tilde{\Gamma}' \slashed{z})]\Big\},
\label{bare2pttree}
\end{eqnarray}
where $N_c$ is the number of colors and $\Gamma(2-\epsilon)$ is Euler's gamma function.

The tree-level three-point Green's function of one four-quark and two quark bilinear operators for arbitrary Dirac matrices and flavors is given below to all orders in $\epsilon$ and in terms of the $D$-vectors $z \equiv x-y$ and $z' \equiv y-w$, which connect the four-quark operator with the left and right bilinear operators, respectively:
\begin{eqnarray}   
&&\hspace{-1cm} \langle \left( \bar \psi_{f'_1}(x) \Gamma' \psi_{f'_2} (x) \right) \, \left( \bar \psi_{f_1}(y) \Gamma \psi_{f_3} (y) \bar \psi_{f_2} (y) \Tilde{\Gamma} \psi_{f_4} (y) \right) \, \left( \bar \psi_{f''_1}(w) \Gamma'' \psi_{f''_2} (w) \right) \rangle^{\rm tree} = \frac{N_c \ \Gamma (2-\epsilon)^4}{16 \ \pi^{8-4\epsilon} \ (z^2)^{4-2\epsilon} \ ({z'}^2)^{4-2\epsilon}} \times \nonumber \\ 
&& \qquad \Big\{ \delta_{f_3 f'_1} \delta_{f_4 f''_1} [N_c \ \delta_{f_1 f'_2} \delta_{f_2 f''_2} \ {\rm tr} (\Gamma' \slashed{z} \Gamma \slashed{z}) \ {\rm tr} (\Tilde{\Gamma} \slashed{z'} \Gamma'' \slashed{z'}) - \delta_{f_2 f_2'} \delta_{f_1 f''_2} \ {\rm tr} (\Gamma' \slashed{z} \Tilde{\Gamma} \slashed{z'} \Gamma'' \slashed{z'} \Gamma \slashed{z})] \nonumber \\
&& \qquad + \delta_{f_4 f'_1} \delta_{f_3 f''_1} [N_c \ \delta_{f_2 f'_2} \delta_{f_1 f''_2} \ {\rm tr} (\Gamma' \slashed{z} \Tilde{\Gamma} \slashed{z}) \ {\rm tr} (\Gamma \slashed{z'} \Gamma'' \slashed{z'}) - \delta_{f_1 f_2'} \delta_{f_2 f''_2} \ {\rm tr} (\Gamma' \slashed{z} \Gamma \slashed{z'} \Gamma'' \slashed{z'} \Tilde{\Gamma} \slashed{z})] \Big\}.
\label{bare3pttree}
\end{eqnarray}
In order to extract the renormalization matrices to one-loop order, we require only the above tree-level expressions up to $\mathcal{O}(\epsilon^1)$.

The corresponding one-loop expressions are more complex and more lengthy, and thus, we do not (directly) provide the explicit expressions in the manuscript. In particular, the one-loop three-point functions are difficult to express in a closed form without expanding over $\epsilon$. For the determination of the one-loop renormalization matrices, we need only up to $\mathcal{O} (\epsilon^0)$ contributions of the bare one-loop Green's functions. Divergent $\mathcal{O} (1/\epsilon)$ terms can be read from the one-loop coefficients of the $\MSbar$ renormalization matrix (see Table~\ref{tab:ZMSbar}). Finite $\mathcal{O} (\epsilon^0)$ terms can be read from the $\MSbar$-renormalized Green’s functions (see Eqs.
(\ref{2ptCons} -- \ref{3ptViol}) and Tables~\ref{tab:GFs2ptCons} -- \ref{tab:GFs3ptViol}). 

\subsection{Mixing matrices in the $\MSbar$ scheme}

An outcome of our calculation is the one-loop coefficients of the mixing matrices ${(Z^{S=\pm 1})}^\MSbar$ and ${(\mathcal{Z}^{S=\pm 1})}^\MSbar$ in the $\MSbar$ scheme. By isolating the pole terms (negative powers of $\epsilon$ in the Laurent series expansion) in the bare two-point and three-point Green's functions, we extract the mixing coefficients for the Parity Conserving operators by solving the following system of conditions:
\begin{eqnarray}
\sum_{k,l = 1}^5 (Z_{ik}^{S=\pm 1})^\MSbar (Z_{jl}^{S=\pm 1})^\MSbar G^{\rm 2pt}_{Q_k^{S=\pm 1}; Q_l^{S=\pm 1}} (z) \Big{|}_{\epsilon^{-n}} &=& 0,   \quad n \in  \mathbb{Z}^{+} \, , \\
 (Z_{{\cal O}_{\Gamma}}^\MSbar)^2 \ \sum_{k=1}^5 \,  (Z_{ik}^{S=\pm 1})^\MSbar \, G^{\rm 3pt}_{\mathcal{O}_\Gamma; Q_k^{S=\pm 1}; \mathcal{O}_\Gamma} (z,z')\Big{|}_{\epsilon^{-n}}&=&0,   \quad n \in  \mathbb{Z}^{+} \, ,
\end{eqnarray}
where $i, j$ run from 1 to 5 and $i \leq j$. Similar conditions are considered for the Parity Violating operators, where the two bilinear operators in the three-point function are chosen to differ by $\gamma_5$. As we mentioned in section~\ref{conditions}, the bilinear operators must be chosen in such a way as to give nonzero Green's functions. Even though we can construct multiple systems of conditions for different $\Gamma$ matrices, all must give the same unique solution. We have confirmed that, indeed all Green's functions calculated in this work give a consistent solution, provided below:
\begin{equation}
{(Z_{ij}^{S=\pm 1})}^{\MSbar}  = \delta_{ij} + \frac{g_{\MSbar}^2}{16 \pi^2 \epsilon} z_{ij}^{\pm} + \mathcal{O} (g_{\MSbar}^4), \qquad {(\mathcal{Z}_{ij}^{S=\pm 1})}^{\MSbar}  = \delta_{ij} + \frac{g_{\MSbar}^2}{16 \pi^2 \epsilon} \tilde{z}_{ij}^{\pm} + \mathcal{O} (g_{\MSbar}^4),
\label{Zmatrix}
\end{equation}
where the nonzero coefficients $z_{ij}^{\pm}, \tilde{z}_{ij}^{\pm}$ are given in Table \ref{tab:ZMSbar} ($C_F = (N_c^2 - 1)/(2 N_c)$).
\begin{table}[thb!]
  \centering
  \begin{tabular}{c|c|c}
  \hline
\quad $i$ \quad & \quad $j$ \quad & \quad $z_{ij}^{\pm} = \tilde{z}_{ij}^{\pm}$ \quad \\ [1ex] 
\hline
\hline
$1$ & $1$ & $-3 (1 \mp N_c)/N_c$ \\ [1ex]
\hline
$2$ & $2$ & $3/N_c$ \\ [1ex]
$2$ & $3$ & $\pm 6$ \\ [1ex]
$3$ & $2$ & $0$ \\ [1ex]
$3$ & $3$ & $-6 C_F$ \\ [1ex]
\hline
$4$ & $4$ & $-3 (2 C_F \mp 1)$ \\ [1ex]
$4$ & $5$ & $-(2 \mp N_c)/(2 N_c)$ \\ [1ex]
$5$ & $4$ & $-6 (2 \pm N_c)/N_c$ \\ [1ex]
$5$ & $5$ & $2 C_F \pm 3$ \\ [1ex]
\hline
  \end{tabular}
  \caption{Numerical values of the coefficients $z_{ij}^{\pm}$, $\tilde{z}_{ij}^{\pm}$ appearing in the nonvanishing blocks of Eq.~\eqref{Zmatrix}.}
  \label{tab:ZMSbar}
\end{table}

We observe that the $\MSbar$ mixing matrices of Parity Conserving and Parity Violating operators coincide for both $S=+1$ and $S=-1$, and they take the block diagonal form of $\mathcal{Z}^{S =\pm 1}$ in Eq.~\eqref{MixingMatrix}. Our results agree with previous calculations in Refs~\cite{Ciuchini:1997bw,Papinutto:2016xpq}. 

We note that in our calculation, we have employed the t’Hooft-Veltman prescription~\cite{tHooft:1972tcz} for defining $\gamma_5$ in $D$ dimensions, which does not violate Ward identities involving pseudoscalar and axial-vector operators. Hence, the following commutation/anticommutation relations of $\gamma_5$ are employed:
\begin{equation}
    \{ \gamma_5, \gamma_\mu \}=0, \ \mu=1,2,3,4, \qquad [\gamma_5, \gamma_\mu]=0, \ \mu>4.
\end{equation}
We also note that Lorentz indices appearing in the definition of the four-quark operators and quark bilinear operators are taken to lie in 4 instead of $D$ dimensions in order to handle potential mixing with evanescent operators in dimensional regularization; for studies of such operators, see Refs.~\cite{Herrlich:1994kh,Ciuchini:1997bw,Lehner:2011fz,Ciuchini:1995cd,Lin:2024mws}.

\subsection{$\MSbar$-renormalized Green's functions} 

By removing the pole parts ($1/\epsilon$) in the bare Green’s function one defines the $\MSbar$-renormalized Green’s functions. As an example, we provide one two-point and one three-point Green's function renormalized in $\MSbar$; they depend on the scales $z$ and/or $z'$ corresponding to the separations between the operators that present in each Green's function, as well as on the $\MSbar$ renormalization scale $\bar{\mu}$ appearing in the renormalization of the coupling constant in $D$ dimensions: $g_R = \mu^{-\epsilon} Z_g^{-1} g_B$ ($g_B$ ($g_R$) is the bare (renormalized) coupling constant, $\mu = \bar \mu \sqrt{e^{\gamma_E}/ 4\pi}$). The remaining Green's functions can be found in the supplemental file.

\begin{eqnarray}
  {[G^{\rm 2pt}_{Q_1^{S=\pm 1}; Q_1^{S=\pm 1}} (z)]}^\MSbar &=& \frac{4 N_c}{\pi^8 {(z^2)}^6} \left(\delta_{f_1 f'_4} \,\delta_{f_2 f'_3} \pm \delta_{f_1 f'_3} \,\delta_{f_2 f'_4}\right) \left(\delta_{f_3 f'_2} \,\delta_{f_4 f'_1} \pm \delta_{f_3 f'_1} \,\delta_{f_4 f'_2}\right) \times \nonumber \\
&& \left\{\pm 1 + N_c + 2 \ \frac{g_{\MSbar}^2 \ C_F}{16 \pi^2} \ \Big[\pm 6 + 7 N_c \mp 6 \left(\ln (\bar{\mu}^2 z^2) + 2 \gamma_E - 2 \ln(2) \right) \Big] + \mathcal{O} (g_{\MSbar}^4) \right\},
\end{eqnarray}
\begin{eqnarray}
{[G^{\rm 3pt}_{V_\mu;Q_1^{S=\pm 1}; V_\mu} (z,z')]}^\MSbar &=& \frac{N_c}{\pi^8 (z^2)^3 (z'^2)^3} \left(\delta_{f'_1 f_4} \,\delta_{f''_1 f_3} \pm \delta_{f'_1 f_3} \,\delta_{f''_1 f_4}\right) \left(\delta_{f'_2 f_2} \,\delta_{f''_2 f_1} \pm \delta_{f'_2 f_1} \,\delta_{f''_2 f_2}\right) \times \nonumber \\
&& \Bigg\{\frac{N_c \pm 1}{2} \ \Big[1 - 2 \frac{z_\mu}{z^2} - 2 \frac{z'_\mu}{z'^2} + 4 \frac{(z \cdot z') z_\mu z'_\mu}{z^2 z'^2} \Big] \pm \ \frac{g_{\MSbar}^2 \ C_F}{16 \pi^2} \ \Big[1 -2 \frac{{(z_\mu + z'_\mu)}^2}{{(z+z')}^2} \Big] \nonumber \\
&& \pm \ \frac{g_{\MSbar}^2 \ C_F}{16 \pi^2} \ \Big[1 - 2 \frac{z_\mu}{z^2} - 2 \frac{z'_\mu}{z'^2} + 4 \frac{(z \cdot z') z_\mu z'_\mu}{z^2 z'^2} \Big] \times \nonumber \\
&& \qquad \qquad \quad \left[2 - 3 \left(\ln\left(\frac{\bar{\mu}^2 z^2 z'^2}{{(z+z')}^2}\right) + 2 \gamma_E - 2 \ln(2) \mp N_c\right)\right] + \mathcal{O} (g_{\MSbar}^4) \Bigg\},
\end{eqnarray}
where the flavor indices follow the conventions of Eqs. (\ref{bare2pttree}, \ref{bare3pttree}).

We also provide the $\MSbar$-renormalized two-point and three-point Green's functions after integration over timeslices (see Eqs.~(\ref{2pt_int} -- \ref{3pt_int}))\footnote{Explicit formulae for integration over timeslices can be found in our Ref.~\cite{Costa:2021iyv}.}, which are relevant for the extraction of the conversion matrices. These are written in a compact form for all four-quark and quark bilinear operators, as follows:
\begin{eqnarray}
[\Tilde{G}^{\rm 2pt}_{Q_i^{S=\pm 1}; Q_j^{S=\pm 1}} (t)]^{\MSbar} &=& \frac{N_c}{\pi^6 |t|^9} \left(\delta_{f_1 f'_4} \,\delta_{f_2 f'_3} \pm \delta_{f_1 f'_3} \,\delta_{f_2 f'_4}\right) \left(\delta_{f_3 f'_2} \,\delta_{f_4 f'_1} \pm \delta_{f_3 f'_1} \,\delta_{f_4 f'_2}\right) \Bigg\{\left(\bmb{a_{ij;0}^{\pm}} + \bmb{a_{ij;1}^{\pm}} N_c \right) \nonumber \\
&& + \frac{g_{\MSbar}^2 \ C_F}{16 \pi^2} \ \Big[\left(\bmp{b_{ij;0}^{\pm}} + \bmp{b_{ij;1}^{\pm}} N_c \right) + \left(\ln (\bar{\mu}^2 t^2) + 2 \gamma_E \right) \left(\bmr{c_{ij;0}^{\pm}} + \bmr{c_{ij;1}^{\pm}} N_c \right)\Big] + \mathcal{O} (g_{\MSbar}^4) \Bigg\}, \label{2ptCons} \\
{[\Tilde{G}^{\rm 2pt}_{\mathcal{Q}_i^{S=\pm 1}; \mathcal{Q}_j^{S=\pm 1}} (t)]}^{\MSbar} &=& \frac{N_c}{\pi^6 |t|^9} \left(\delta_{f_1 f'_4} \,\delta_{f_2 f'_3} \pm \delta_{f_1 f'_3} \,\delta_{f_2 f'_4}\right) \left(\delta_{f_3 f'_2} \,\delta_{f_4 f'_1} \pm {(-1)}^{\delta_{i2} + \delta_{i3}} \delta_{f_3 f'_1} \,\delta_{f_4 f'_2}\right) \Bigg\{\left(\bmb{\tilde{a}_{ij;0}^{\pm}} + \bmb{\tilde{a}_{ij;1}^{\pm}} N_c \right) \nonumber \\
&& + \frac{g_{\MSbar}^2 \ C_F}{16 \pi^2} \ \left[\left(\bmp{\tilde{b}_{ij;0}^{\pm}} + \bmp{\tilde{b}_{ij;1}^{\pm}} N_c \right) + \left(\ln (\bar{\mu}^2 t^2) + 2 \gamma_E \right) \left(\bmr{\tilde{c}_{ij;0}^{\pm}} + \bmr{\tilde{c}_{ij;1}^{\pm}} N_c \right)\right] + \mathcal{O} (g_{\MSbar}^4) \Bigg\},\label{2ptViol} \\
{[\Tilde{G}^{\rm 3pt}_{\mathcal{O}_\Gamma;Q_i^{S=\pm 1}; \mathcal{O}_{\Gamma}} (t,t)]}^{\MSbar} &=& \frac{N_c}{\pi^4 t^6} \left(\delta_{f'_1 f_4} \,\delta_{f''_1 f_3} \pm \delta_{f'_1 f_3} \,\delta_{f''_1 f_4}\right) \left(\delta_{f'_2 f_2} \,\delta_{f''_2 f_1} \pm \delta_{f'_2 f_1} \,\delta_{f''_2 f_2}\right) \Bigg\{\left(\bmb{d_{i\Gamma;0}^{\pm}} + \bmb{d_{i\Gamma;1}^{\pm}} N_c \right) \nonumber \\
&& + \frac{g_{\MSbar}^2 \ C_F}{16 \pi^2} \ \Big[\left(\bmp{e_{i\Gamma;0}^{\pm}} + \bmp{e_{i\Gamma;1}^{\pm}} N_c \right) + \left(\ln (\bar{\mu}^2 t^2) + 2 \gamma_E \right) \left(\bmr{f_{i\Gamma;0}^{\pm}} + \bmr{f_{i\Gamma;1}^{\pm}} N_c \right)\Big] + \mathcal{O} (g_{\MSbar}^4) \Bigg\}, \label{3ptCons} \\
{[\Tilde{G}^{\rm 3pt}_{\mathcal{O}_\Gamma;\mathcal{Q}_i^{S=\pm 1}; \mathcal{O}_{\Gamma \gamma_5}} (t,t)]}^{\MSbar} &=& \frac{N_c}{\pi^4 t^6} \left(\delta_{f'_1 f_4} \,\delta_{f''_1 f_3} \pm \delta_{f'_1 f_3} \,\delta_{f''_1 f_4}\right) \left(\delta_{f'_2 f_2} \,\delta_{f''_2 f_1} \pm {(-1)}^{\delta_{i2} + \delta_{i3}} \delta_{f'_2 f_1} \,\delta_{f''_2 f_2}\right) \Bigg\{\left(\bmb{\tilde{d}_{i\Gamma;0}^{\pm}} + \bmb{\tilde{d}_{i\Gamma;1}^{\pm}} N_c \right) \nonumber \\
&& + \frac{g_{\MSbar}^2 \ C_F}{16 \pi^2} \ \left[\left(\bmp{\tilde{e}_{i\Gamma;0}^{\pm}} + \bmp{\tilde{e}_{i\Gamma;1}^{\pm}} N_c \right) + \left(\ln (\bar{\mu}^2 t^2) + 2 \gamma_E \right) \left(\bmr{\tilde{f}_{i\Gamma;0}^{\pm}} + \bmr{\tilde{f}_{i\Gamma;1}^{\pm}} N_c \right)\right] + \mathcal{O} (g_{\MSbar}^4) \Bigg\}, \label{3ptViol}
\end{eqnarray}
where the coefficients $a_{ij;k}^{\pm}$, $b_{ij;k}^{\pm}$, $c_{ij;k}^{\pm}$, $\tilde{a}_{ij;k}^{\pm}$, $\tilde{b}_{ij;k}^{\pm}$, $\tilde{c}_{ij;k}^{\pm}$, $d_{i\Gamma;k}^{\pm}$, $e_{i\Gamma;k}^{\pm}$, $f_{i\Gamma;k}^{\pm}$, $\tilde{d}_{i\Gamma;k}^{\pm}$, $\tilde{e}_{i\Gamma;k}^{\pm}$, $\tilde{f}_{i\Gamma;k}^{\pm}$ are given in Tables~\ref{tab:GFs2ptCons} -- \ref{tab:GFs3ptViol}. Note that the three-point functions, which include temporal components of vector ($V_4$) and/or axial-vector ($A_4$) operators vanish (under integration over timeslices) and thus, they omitted from Tables~\ref{tab:GFs3ptCons} and \ref{tab:GFs3ptViol}. \\

\begin{table}[thb!]
  \centering
  \begin{tabular}{c|c|c|c|c|c|c|c}
  \hline
\quad $i$ \quad & \quad $j$ \quad & \quad $\bmb{a_{ij;0}^{\pm}}$ \quad & \quad $\bmb{a_{ij;1}^{\pm}}$ \quad & \quad $\bmp{b_{ij;0}^{\pm}}$ \quad & \quad $\bmp{b_{ij;1}^{\pm}}$ \quad & \quad $\bmr{c_{ij;0}^{\pm}}$ \quad & \quad $\bmr{c_{ij;1}^{\pm}}$ \\ [1ex] 
\hline
\hline
$1$ & $1$ & $\pm 7/32$ & $7/32$ & $\pm 869/160$ & $49/16$ & $\mp 21/8$ & $0$ \\ [1ex] 
$1$ & $2$ & $0$ & $0$ & $\mp 7/4$ & $-7/4$ & $0$ & $0$ \\ [1ex] 
$1$ & $3$ & $0$ & $0$ & $\pm 7/8$ & $0$ & $0$ & $0$ \\ [1ex]
$1$ & $4$ & $0$ & $0$ & $\pm 7/8$ & $0$ & $0$ & $0$ \\ [1ex]
$1$ & $5$ & $0$ & $0$ & $\pm 21/4$ & $0$ & $0$ & $0$ \\ [1ex]
$2$ & $2$ & $0$ & $7/32$ & $\pm 7/4$ & $49/16$ & $0$ & $0$ \\ [1ex]
$2$ & $3$ & $\mp 7/64$ & $0$ & $\mp 391/320$ & $0$ & $\mp 21/16$ & $0$ \\ [1ex]
$2$ & $4$ & $0$ & $0$ & $\pm 7/4$ & $0$ & $0$ & $0$ \\ [1ex]
$2$ & $5$ & $0$ & $0$ & $0$ & $0$ & $0$ & $0$ \\ [1ex]
$3$ & $3$ & $0$ & $7/128$ & $\pm 7/16$ & $251/640$ & $0$ & $21/32$ \\ [1ex]
$3$ & $4$ & $0$ & $0$ & $\pm 7/16$ & $-7/8$ & $0$ & $0$ \\ [1ex]
$3$ & $5$ & $0$ & $0$ & $\mp 21/8$ & $0$ & $0$ & $0$ \\ [1ex]
$4$ & $4$ & $\mp 7/256$ & $7/128$ & $\pm 87/1280$ & $251/640$ & $\mp 63/64$ & $21/32$ \\ [1ex]
$4$ & $5$ & $\pm 21/128$ & $0$ & $\pm 1651/640$ & $0$ & $\pm 21/32$ & $0$ \\ [1ex]
$5$ & $5$ & $\pm 21/64$ & $21/32$ & $\pm 3563/320$ & $1709/160$ & $\mp 147/16$ & $-21/8$ \\ [1ex]
\hline
  \end{tabular}
  \caption{Numerical values of the coefficients $a_{ij;0}^{\pm}$, $a_{ij;1}^{\pm}$, $b_{ij;0}^{\pm}$, $b_{ij;1}^{\pm}$, $c_{ij;0}^{\pm}$, $c_{ij;1}^{\pm}$ appearing in Eq.~(\eqref{2ptCons}).}
  \label{tab:GFs2ptCons}
\end{table}

\newpage

\begin{table}[thb!]
  \centering
  \begin{tabular}{c|c|c|c|c|c|c|c}
  \hline
\quad $i$ \quad & \quad $j$ \quad & \quad $\bmb{\tilde{a}_{ij;0}^{\pm}}$ \quad & \quad $\bmb{\tilde{a}_{ij;1}^{\pm}}$ \quad & \quad $\bmp{\tilde{b}_{ij;0}^{\pm}}$ \quad & \quad $\bmp{\tilde{b}_{ij;1}^{\pm}}$ \quad & \quad $\bmr{\tilde{c}_{ij;0}^{\pm}}$ \quad & \quad $\bmr{\tilde{c}_{ij;1}^{\pm}}$ \\ [1ex] 
\hline
\hline
$1$ & $1$ & $\pm 7/32$ & $7/32$ & $\pm 869/160$ & $49/16$ & $\mp 21/8$ & $0$ \\ [1ex] 
$1$ & $2$ & $0$ & $0$ & $0$ & $0$ & $0$ & $0$ \\ [1ex] 
$1$ & $3$ & $0$ & $0$ & $0$ & $0$ & $0$ & $0$ \\ [1ex]
$1$ & $4$ & $0$ & $0$ & $0$ & $0$ & $0$ & $0$ \\ [1ex]
$1$ & $5$ & $0$ & $0$ & $0$ & $0$ & $0$ & $0$ \\ [1ex]
$2$ & $2$ & $0$ & $-7/32$ & $0$ & $-49/16$ & $0$ & $0$ \\ [1ex]
$2$ & $3$ & $\pm 7/64$ & $0$ & $\pm 391/320$ & $0$ & $\pm 21/16$ & $0$ \\ [1ex]
$2$ & $4$ & $0$ & $0$ & $0$ & $0$ & $0$ & $0$ \\ [1ex]
$2$ & $5$ & $0$ & $0$ & $0$ & $0$ & $0$ & $0$ \\ [1ex]
$3$ & $3$ & $0$ & $-7/128$ & $0$ & $-251/640$ & $0$ & $-21/32$ \\ [1ex]
$3$ & $4$ & $0$ & $0$ & $0$ & $0$ & $0$ & $0$ \\ [1ex]
$3$ & $5$ & $0$ & $0$ & $0$ & $0$ & $0$ & $0$ \\ [1ex]
$4$ & $4$ & $\mp 7/256$ & $7/128$ & $\pm 87/1280$ & $251/640$ & $\mp 63/64$ & $21/32$ \\ [1ex]
$4$ & $5$ & $\pm 21/128$ & $0$ & $\pm 1651/640$ & $0$ & $\pm 21/32$ & $0$ \\ [1ex]
$5$ & $5$ & $\pm 21/64$ & $21/32$ & $\pm 3563/320$ & $1709/160$ & $\mp 147/16$ & $-21/8$ \\ [1ex]
\hline
  \end{tabular}
  \caption{Numerical values of the coefficients $\tilde{a}_{ij;0}^{\pm}$, $\tilde{a}_{ij;1}^{\pm}$, $\tilde{b}_{ij;0}^{\pm}$, $\tilde{b}_{ij;1}^{\pm}$, $\tilde{c}_{ij;0}^{\pm}$, $\tilde{c}_{ij;1}^{\pm}$ appearing in Eq.~(\eqref{2ptViol}).}
  \label{tab:GFs2ptViol}
\end{table}

\newpage

\begin{table}[ht!]
  \centering
  \begin{tabular}{c|c|c|c|c|c|c|c}
  \hline
\quad $i$ \quad & \quad $\Gamma$ \quad & \quad $\bmb{d_{i\Gamma;0}^{\pm}}$ \quad & \quad $\bmb{d_{i\Gamma;1}^{\pm}}$ \quad & \quad $\bmp{e_{i\Gamma;0}^{\pm}}$ \quad & \quad $\bmp{e_{i\Gamma;1}^{\pm}}$ \quad & \quad $\bmr{f_{i\Gamma;0}^{\pm}}$ \quad & \quad $\bmr{f_{i\Gamma;1}^{\pm}}$ \\ [1ex] 
\hline
\hline
$1$ & $S$ & $0$ & $0$ & $\pm 1/2$ & $0$ & $0$ & $0$ \\ [1ex]
$2$ & $S$ & $\mp 1/16$ & $0$ & $0$ & $0$ & $\mp 3/4$ & $0$ \\ [1ex] 
$3$ & $S$ & $0$ & $1/32$ & $\pm 1/4$ & $-1/16$ & $0$ & $3/8$ \\ [1ex]
$4$ & $S$ & $\mp 1/64$ & $1/32$ & $\mp 3/8 \ (1/8 - \ln(2))$ & $-1/16$ & $\mp 3/8$ & $3/8$ \\ [1ex]
$5$ & $S$ & $\pm 3/32$ & $0$ & $\mp 3/4 \ (1/8 - \ln(2))$ & $0$ & $\pm 3/4$ & $0$ \\ [1ex]
$1$ & $P$ & $0$ & $0$ & $0$ & $0$ & $0$ & $0$ \\ [1ex]
$2$ & $P$ & $\pm 1/16$ & $0$ & $\pm 2$ & $0$ & $\pm 3/4$ & $0$ \\ [1ex]
$3$ & $P$ & $0$ & $-1/32$ & $0$ & $-15/16$ & $0$ & $-3/8$ \\ [1ex]
$4$ & $P$ & $\mp 1/64$ & $1/32$ & $\mp 3/8 \ (35/24 - \ln(2))$ & $15/16$ & $\mp 3/8$ & $3/8$ \\ [1ex]
$5$ & $P$ & $\pm 3/32$ & $0$ & $\pm 3/4 \ (93/24 + \ln(2))$ & $0$ & $\pm 3/4$ & $0$ \\ [1ex]
$1$ & $V_j$ & $\pm 1/72$ & $1/72$ & $\pm 1/6 \ (41/48 + \ln(2))$ & $1/12$ & $\mp 1/12$ & $0$ \\ [1ex]
$2$ & $V_j$ & $0$ & $1/72$ & $0$ & $1/12$ & $0$ & $0$ \\ [1ex]
$3$ & $V_j$ & $\mp 1/144$ & $0$ & $\mp 1/12 \ (23/48 - \ln(2))$ & $0$ & $\mp 1/24$ & $0$ \\ [1ex]
$4$ & $V_j$ & $0$ & $0$ & $\pm 1/12$ & $0$ & $0$ & $0$ \\ [1ex]
$5$ & $V_j$ & $0$ & $0$ & $\pm 1/6$ & $0$ & $0$ & $0$ \\ [1ex]
$1$ & $A_j$ & $\pm 1/72$ & $1/72$ & $\pm 1/6 \ (35/16 + \ln(2))$ & $11/36$ & $\mp 1/12$ & $0$ \\ [1ex]
$2$ & $A_j$ & $0$ & $-1/72$ & $\mp 1/9$ & $-11/36$ & $0$ & $0$ \\ [1ex]
$3$ & $A_j$ & $\pm 1/144$ & $0$ & $\pm 1/12 \ (29/16 - \ln(2))$ & $0$ & $\pm 1/24$ & $0$ \\ [1ex]
$4$ & $A_j$ & $0$ & $0$ & $\mp 1/36$ & $0$ & $0$ & $0$ \\ [1ex]
$5$ & $A_j$ & $0$ & $0$ & $\pm 1/6$ & $0$ & $0$ & $0$ \\ [1ex]
$1$ & $T_{jk}$ & $0$ & $0$ & $\pm 1/36$ & $0$ & $0$ & $0$ \\ [1ex]
$2$ & $T_{jk}$ & $0$ & $0$ & $\pm 11/192$ & $0$ & $0$ & $0$ \\ [1ex]
$3$ & $T_{jk}$ & $0$ & $0$ & $\mp 1/72$ & $0$ & $0$ & $0$ \\ [1ex]
$4$ & $T_{jk}$ & $\pm 1/576$ & $0$ & $\pm 1/72 \ (15/8 - \ln(2))$ & $0$ & $0$ & $0$ \\ [1ex]
$5$ & $T_{jk}$ & $\pm 1/288$ & $1/144$ & $\pm 1/12 \ (89/72 + \ln(2))$ & $25/216$ & $\mp 1/18$ & $-1/36$ \\ [1ex]
$1$ & $T_{j4}$ & $0$ & $0$ & $\pm 1/36$ & $0$ & $0$ & $0$ \\ [1ex]
$2$ & $T_{j4}$ & $0$ & $0$ & $\mp 11/192$ & $0$ & $0$ & $0$ \\ [1ex]
$3$ & $T_{j4}$ & $0$ & $0$ & $\mp 1/72$ & $0$ & $0$ & $0$ \\ [1ex]
$4$ & $T_{j4}$ & $\pm 1/576$ & $0$ & $\pm 1/72 \ (15/8 - \ln(2))$ & $0$ & $0$ & $0$ \\ [1ex]
$5$ & $T_{j4}$ & $\pm 1/288$ & $1/144$ & $\pm 1/12 \ (89/72 + \ln(2))$ & $25/216$ & $\mp 1/18$ & $-1/36$ \\ [1ex]
\hline
  \end{tabular}
  \caption{Numerical values of the coefficients $d_{i\Gamma;l}^{\pm}$, $e_{i\Gamma;l}^{\pm}$, $f_{i\Gamma;l}^{\pm}$, appearing in Eq.~(\eqref{3ptCons}).}
  \label{tab:GFs3ptCons}
\end{table}

\newpage

\begin{table}[ht!]
  \centering
  \begin{tabular}{c|c|c|c|c|c|c|c}
  \hline
\quad $i$ \quad & \quad $\Gamma$ \quad & \quad $\bmb{\tilde{d}_{i\Gamma;0}^{\pm}}$ \quad & \quad $\bmb{\tilde{d}_{i\Gamma;1}^{\pm}}$ \quad & \quad $\bmp{\tilde{e}_{i\Gamma;0}^{\pm}}$ \quad & \quad $\bmp{\tilde{e}_{i\Gamma;1}^{\pm}}$ \quad & \quad $\bmr{\tilde{f}_{i\Gamma;0}^{\pm}}$ \quad & \quad $\bmr{\tilde{f}_{i\Gamma;1}^{\pm}}$ \\ [1ex] 
\hline
\hline
$1$ & $S$ & $0$ & $0$ & $0$ & $0$ & $0$ & $0$ \\ [1ex]
$2$ & $S$ & $\pm 1/16$ & $0$ & $\pm 1$ & $0$ & $\pm 3/4$ & $0$ \\ [1ex] 
$3$ & $S$ & $0$ & $-1/32$ & $0$ & $-7/16$ & $0$ & $-3/8$ \\ [1ex]
$4$ & $S$ & $\pm 1/64$ & $-1/32$ & $\pm 3/8 (19/24 - \ln(2))$ & $-7/16$ & $\pm 3/8$ & $-3/8$ \\ [1ex]
$5$ & $S$ & $\mp 3/32$ & $0$ & $\mp 3/4 (15/8 + \ln(2))$ & $0$ & $\mp 3/4$ & $0$ \\ [1ex]
$1$ & $V_j$ & $\pm 1/72$ & $1/72$ & $\pm 1/6 (73/48 + \ln(2))$ & $7/36$ & $\mp 1/12$ & $0$ \\ [1ex]
$2$ & $V_j$ & $0$ & $-1/72$ & $0$ & $-7/36$ & $0$ & $0$ \\ [1ex]
$3$ & $V_j$ & $\pm 1/144$ & $0$ & $\pm 1/12 (55/48 - \ln(2))$ & $0$ & $\pm 1/24$ & $0$ \\ [1ex]
$4$ & $V_j$ & $0$ & $0$ & $0$ & $0$ & $0$ & $0$ \\ [1ex]
$5$ & $V_j$ & $0$ & $0$ & $0$ & $0$ & $0$ & $0$ \\ [1ex]
$1$ & $T_{jk}$ & $0$ & $0$ & $0$ & $0$ & $0$ & $0$ \\ [1ex]
$2$ & $T_{jk}$ & $0$ & $0$ & $\mp 11/192$ & $0$ & $0$ & $0$ \\ [1ex]
$3$ & $T_{jk}$ & $0$ & $0$ & $0$ & $0$ & $0$ & $0$ \\ [1ex]
$4$ & $T_{jk}$ & $\mp 1/576$ & $0$ & $\mp 1/72 \ (15/8 - \ln(2))$ & $0$ & $0$ & $0$ \\ [1ex]
$5$ & $T_{jk}$ & $\mp 1/288$ & $-1/144$ & $\mp 1/12 \ (89/72 + \ln(2))$ & $-25/216$ & $\pm 1/18$ & $1/36$ \\ [1ex]
$1$ & $T_{j4}$ & $0$ & $0$ & $0$ & $0$ & $0$ & $0$ \\ [1ex]
$2$ & $T_{j4}$ & $0$ & $0$ & $\pm 11/192$ & $0$ & $0$ & $0$ \\ [1ex]
$3$ & $T_{j4}$ & $0$ & $0$ & $0$ & $0$ & $0$ & $0$ \\ [1ex]
$4$ & $T_{j4}$ & $\mp 1/576$ & $0$ & $\mp 1/72 \ (15/8 - \ln(2))$ & $0$ & $0$ & $0$ \\ [1ex]
$5$ & $T_{j4}$ & $\mp 1/288$ & $-1/144$ & $\mp 1/12 \ (89/72 + \ln(2))$ & $-25/216$ & $\pm 1/18$ & $1/36$ \\ [1ex]
\hline
  \end{tabular}
  \caption{Numerical values of the coefficients $\tilde{d}_{i\Gamma;l}^{\pm}$, $\tilde{e}_{i\Gamma;l}^{\pm}$, $\tilde{f}_{i\Gamma;l}^{\pm}$ appearing in Eq.~(\eqref{3ptViol}).}
  \label{tab:GFs3ptViol}
\end{table}

\begin{figure}[ht!]
\includegraphics[scale=0.6]{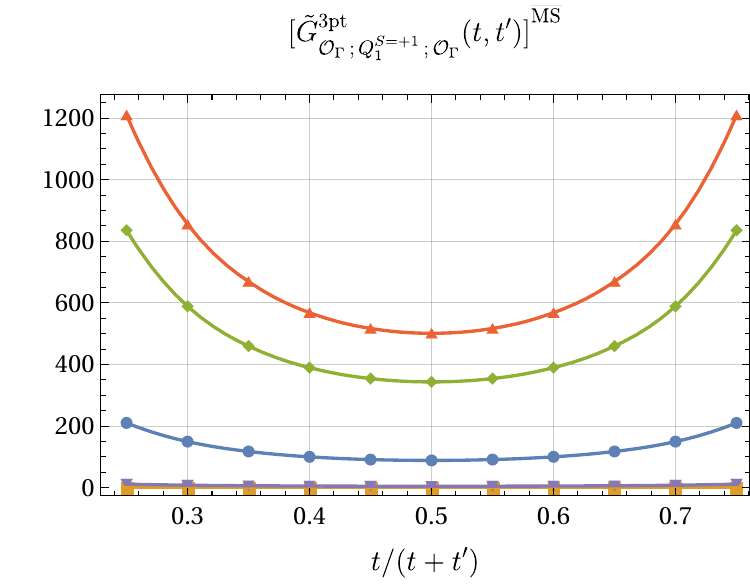} \hspace{0.5cm}
\includegraphics[scale=0.61]{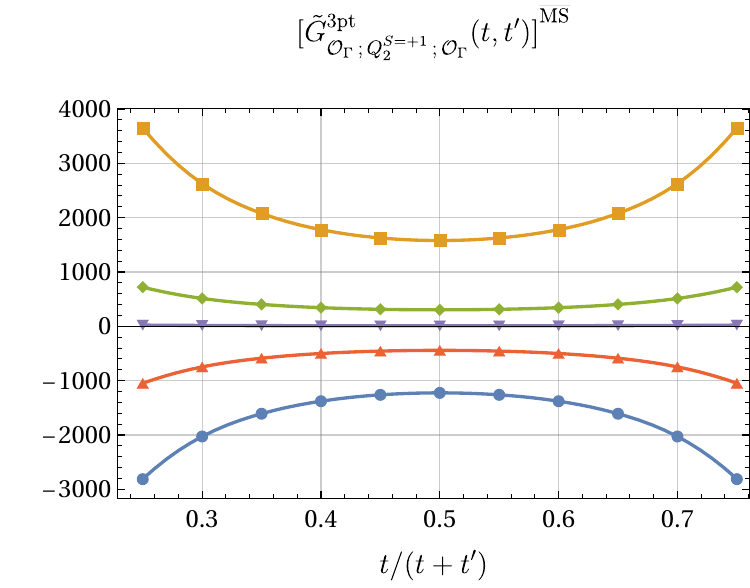} \\
\vspace{0.5cm}
\includegraphics[scale=0.61]{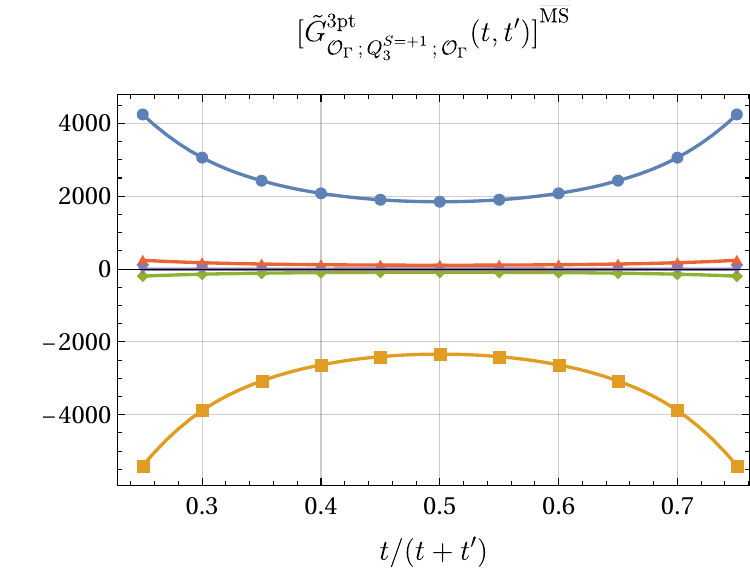} \hspace{0.65cm}
\includegraphics[scale=0.6]{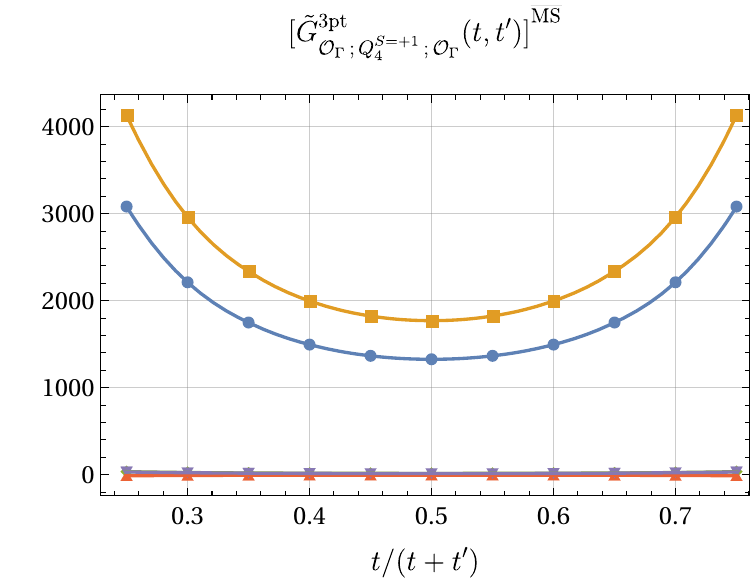} \\
\vspace{0.4cm} 
\includegraphics[scale=0.61]{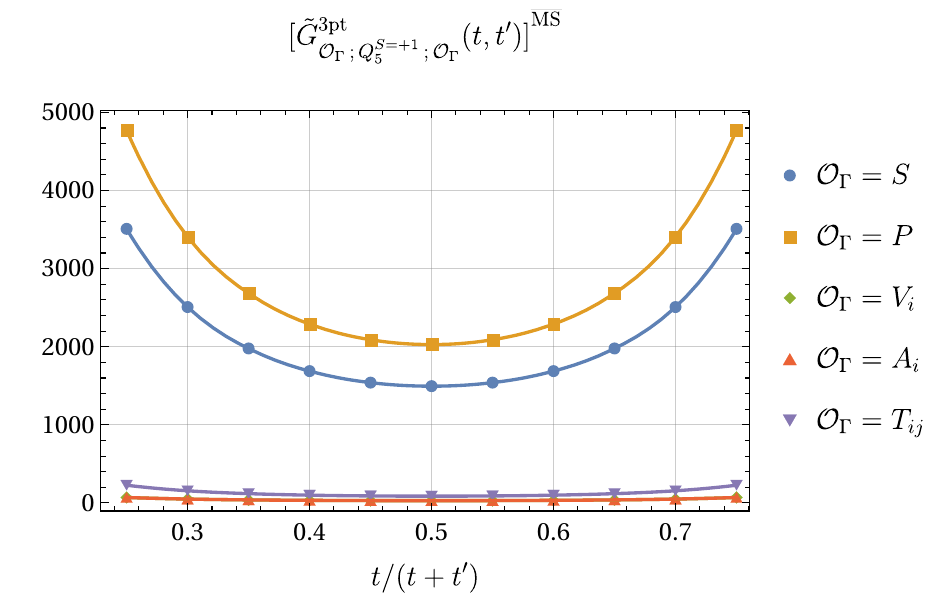} \hspace{-2.3cm} 
\caption{Plots of three-point functions ${[\Tilde {G}^{\rm 3 pt} _ {\mathcal{O}_\Gamma;Q_ i^{S = + 1}; \mathcal{O}_\Gamma} (t,t')]}^{\MSbar}$, $i \in [1,5]$, as a function of $t/(t+t')$, for fixed $t+t'$\,. A common factor of $N_c/(\pi^8 \ (t + t')^6) \times \left(\delta_{f'_1 f_4} \,\delta_{f''_1 f_3} + \delta_{f'_1 f_3} \,\delta_{f''_1 f_4}\right) \left(\delta_{f'_2 f_2} \,\delta_{f''_2 f_1} + \delta_{f'_2 f_1} \,\delta_{f''_2 f_2}\right)$ is excluded from all graphs. Here, we set $N_c = 3$, $g_{\MSbar}^2 = 6/\beta$, $\beta = 1.788$, $\bar{\mu} = 2 \ {\rm GeV}$, $(t+t') = T/2$, $T=64 a$, $a=0.07957 \ {\rm fm}$.}
\label{fig:plots_3pt}
\end{figure}

For simplicity, we have presented algebraic results for the three-point functions at $t=t'$. In Fig.~\ref{fig:plots_3pt}, we examine the dependence of the three-point functions on more general relative values of $t$ and $t'$. As an example, we provide plots for the $\MSbar$-renormalized three-point functions of the Parity Conserving operators for $S=+1$ as a function of $t/(t+t')$, keeping $t+t'$ constant. All other three-point functions (of Parity Conserving operators with $S=-1$, or Parity Violating operators with $S=\pm 1$) have similar behavior. We have employed certain values of the
free parameters used in lattice simulations: $N_c = 3$, $g_{\MSbar}^2 = 6/\beta$, $\beta = 1.788$, $\bar{\mu} = 2 \ {\rm GeV}$, $(t+t') = T/2$ ($T$ is the temporal lattice size), $T=64 a$ ($a$ is the lattice spacing), $a=0.07957 \ {\rm fm}$.

We observe in Fig.~\ref{fig:plots_3pt} that the three-point functions are symmetric over $t/(t+t')=0.5$ (or equivalently, $t=t'$), as expected. Also, they have a divergent behavior when $t/(t+t')$ tends to 0 or 1 because they approach contact points. Even though the $\MSbar$-renormalized Green's functions have a strong dependence on $t$ and $t'$, the renormalization functions in the $\MSbar$ scheme must be $t, t'$-independent; this is a powerful crosscheck that can be examined in the nonperturbative investigations on the lattice, given that appropriate conversion functions are employed. 

Some further observations are: Green's functions with $Q_{i \neq 1}$ take larger absolute values when scalar (S) or pseudoscalar (P) operators are considered, while for $Q_1$, the three-point functions with vector ($V_i$) or axial-vector ($A_i$) operators (where $i$ is a spatial direction) give the highest values. Green's functions with tensor ($T_{ij}$) operators have much smaller values compared to all other three-point functions. Our findings on the magnitude and size of the one-loop three-point functions for each $\Gamma$ can give an input to the corresponding nonperturbative calculations: We expect that larger contributions in the perturbative calculations will give a better signal in lattice simulations. However, some of the large contributions observed in the perturbative calculations are a consequence of mixing; thus, it is not obvious that a reliable set of renormalization conditions can be extracted by using as criterion the size of the Green's functions.

\newpage

\subsection{Conversion matrices}
\label{conv_matrices}

The one-loop conversion matrices between different variants of GIRS and the $\MSbar$ scheme are extracted from our results by rewriting the GIRS conditions (Eqs.~(\ref{cond2pt} -- \ref{cond3pt})) in terms of the conversion matrices, as defined in Eq. (\eqref{Cdef}):
\bea
{[\Tilde{G}^{\rm 2pt}_{Q_i^{S=\pm 1}; Q_j^{S=\pm 1}} (t)]}^\MSbar &=& \sum_{k,l = 1}^5 (C_{ik}^{S \pm 1})^{\MSbar, {\rm GIRS}} \ (C_{jl}^{S \pm 1})^{\MSbar, {\rm GIRS}} \ [\Tilde{G}^{\rm 2pt}_{Q_k^{S=\pm 1};Q_l^{S=\pm 1}} (t)]^{\rm tree}, \\
{[\Tilde{G}^{\rm 3pt}_{\mathcal{O}_\Gamma; Q_i^{S=\pm 1}; \mathcal{O}_\Gamma} (t,t)]}^\MSbar &=& (C_{\mathcal{O}_\Gamma}^{\MSbar,{\rm GIRS}})^2 \ \sum_{k=1}^5 \ (C_{ik}^{S \pm 1})^{\MSbar, {\rm GIRS}} \ [\Tilde{G}^{\rm 3pt}_{\mathcal{O}_\Gamma;Q_k^{S=\pm 1}; \mathcal{O}_\Gamma} (t,t)]^{\rm tree}, \\
{[\Tilde{G}^{\rm 2pt}_{\mathcal{Q}_i^{S=\pm 1}; \mathcal{Q}_j^{S=\pm 1}} (t)]}^\MSbar &=& \sum_{k,l = 1}^5 (\Tilde{C}_{ik}^{S \pm 1})^{\MSbar, {\rm GIRS}} \ (\Tilde{C}_{jl}^{S \pm 1})^{\MSbar, {\rm GIRS}} \ [\Tilde{G}^{\rm 2pt}_{\mathcal{Q}_k^{S=\pm 1};\mathcal{Q}_l^{S=\pm 1}} (t)]^{\rm tree}, \\
{[\Tilde{G}^{\rm 3pt}_{\mathcal{O}_\Gamma; \mathcal{Q}_i^{S=\pm 1}; \mathcal{O}_{\Gamma \gamma_5}} (t,t)]}^\MSbar  &=& (C_{\mathcal{O}_\Gamma}^{\MSbar,{\rm GIRS}}) \ (C_{\mathcal{O}_{\Gamma \gamma_5}}^{\MSbar,{\rm GIRS}}) \ \sum_{k=1}^5 \ (\Tilde{C}_{ik}^{S \pm 1})^{\MSbar, {\rm GIRS}} \  [\Tilde{G}^{\rm 3pt}_{\mathcal{O}_\Gamma;\mathcal{Q}_k^{S=\pm 1}; \mathcal{O}_{\Gamma \gamma_5}} (t,t)]^{\rm tree},
\eea
where $C_{\mathcal{O}_\Gamma}^{\MSbar,{\rm GIRS}}$ is the conversion factor of the quark bilinear operator $\mathcal{O}_\Gamma$ calculated to one loop in Ref.~\cite{Costa:2021iyv}:
\bea
C_S^{\MSbar,{\rm GIRS}} &=& 1 + \frac{g_\MSbar^2 \ C_F}{16 \pi^2} \ \left(-\frac{1}{2} + 3 \ln (\bar{\mu}^2 t^2) + 6 \gamma_E\right) + \mathcal{O} (g_\MSbar^4), \\
C_P^{\MSbar,{\rm GIRS}} &=& 1 + \frac{g_\MSbar^2 \ C_F}{16 \pi^2} \ \left(\frac{15}{2} + 3 \ln (\bar{\mu}^2 t^2) + 6 \gamma_E\right) + \mathcal{O} (g_\MSbar^4), \\
C_V^{\MSbar,{\rm GIRS}} &=& 1 + \frac{g_\MSbar^2 \ C_F}{16 \pi^2} \ \frac{3}{2} + \mathcal{O} (g_\MSbar^4), \\
C_A^{\MSbar,{\rm GIRS}} &=& 1 + \frac{g_\MSbar^2 \ C_F}{16 \pi^2} \ \frac{11}{2} + \mathcal{O} (g_\MSbar^4), \\
C_T^{\MSbar,{\rm GIRS}} &=& 1 + \frac{g_\MSbar^2 \ C_F}{16 \pi^2} \ \left(\frac{25}{6} - \ln (\bar{\mu}^2 t^2) - 2 \gamma_E\right) + \mathcal{O} (g_\MSbar^4).
\eea
Note that the conversion matrix $(\Tilde{C}^{S \pm 1})^{\MSbar, {\rm GIRS}}$ has the block diagonal form of $\mathcal{Z}^{S=\pm1}$ (see Eq.~(\eqref{MixingMatrix})). As we discussed in section \ref{conditions}, there are a lot of different choices of three-point Green's functions that can be included in the renormalization conditions, giving a different version of GIRS. In particular, for the Parity Conserving operators ($Q_i^{S=\pm 1}$), where 15 conditions are obtained from the two-point functions, there are $30!/(10! \ 20!) = 30,045,015$ choices for obtaining the remaining 10 conditions from the three-point functions (see Table \ref{tab:GFs3ptCons}). However, some choices include linear dependent or incompatible conditions leading to infinite or no solutions, respectively. By examining all cases in one-loop perturbation theory, we conclude that there are 205,088 choices of conditions, which give a unique solution.

Even though all solvable systems of conditions are acceptable, it is natural to set a criterion in order to select options which have better behavior compared to others. Such a criterion can be the size of the mixing contributions. To this end, we evaluate the sum of squares of the off-diagonal coefficients in the conversion matrices for all the accepted cases, and we choose the cases with the smallest values. We found that, in general, the sums of squares among different choices are comparable. We also observed that the mixing is less pronounced for the operators with $S=-1$, as compared to $S=+1$. 

From the options that give the smallest sum of squares of the off-diagonal coefficients (smallest mixing contributions), we choose one to present below. We avoid including tensor operators in the selected set of conditions, which are typically more noisy in simulations. Also, we prefer to have more scalar or pseudoscalar operators which are computationally cheaper compared to other bilinear operators. The selected set of conditions includes the following 10 renormalized three-point functions:
\begin{eqnarray*}
&&  \Tilde{G}^{\rm 3pt}_{S;Q_1^{S=\pm 1}; S} (t,t), \qquad \Tilde{G}^{\rm 3pt}_{P;Q_1^{S=\pm 1}; P} (t,t), \qquad \Tilde{G}^{\rm 3pt}_{V_i;Q_1^{S=\pm 1}; V_i} (t,t), \qquad \Tilde{G}^{\rm 3pt}_{S;Q_2^{S=\pm 1}; S} (t,t), \qquad \Tilde{G}^{\rm 3pt}_{P;Q_2^{S=\pm 1}; P} (t,t), \\
&&  \Tilde{G}^{\rm 3pt}_{S;Q_3^{S=\pm 1}; S} (t,t), \qquad \Tilde{G}^{\rm 3pt}_{S;Q_5^{S=\pm 1}; S} (t,t), \qquad \Tilde{G}^{\rm 3pt}_{P;Q_5^{S=\pm 1}; P} (t,t), \qquad \Tilde{G}^{\rm 3pt}_{V_i;Q_5^{S=\pm 1}; V_i} (t,t), \qquad \Tilde{G}^{\rm 3pt}_{A_i;Q_5^{S=\pm 1}; A_i} (t,t),
\end{eqnarray*}
and the solution reads:
\begin{equation}
    (C_{ij}^{S \pm 1})^{\MSbar, {\rm GIRS}} = \delta_{ij} + \frac{g_\MSbar^2}{16 \pi^2} \sum_{k=-1}^{+1} \left[ \bmb{g_{ij;k}^\pm} + \left(\ln (\bar{\mu}^2 t^2) + 2 \gamma_E \right) \ \bmr{h_{ij;k}^\pm} \right] N_c^k + \mathcal{O} (g_\MSbar^4),
    \label{Ccons}
\end{equation}
where the coefficients $g_{ij;k}^{\pm}$, $h_{ij;k}^{\pm}$, are given in Table~\ref{tab:CmatrixCons}.    

\begin{table}[ht!]
  \centering
  \begin{tabular}{c|c|c|c|c|c|c|c}
  \hline
\quad $i$ \quad & \quad $j$ \quad & \quad $\bmb{g_{ij;-1}^{\pm}}$ \quad & \quad $\bmb{g_{ij;0}^{\pm}}$ \quad & \quad $\bmb{g_{ij;+1}^{\pm}}$ \quad & \quad $\bmr{h_{ij;-1}^{\pm}}$ \quad & \quad $\bmr{h_{ij;0}^{\pm}}$ \quad & \quad $\bmr{h_{ij;+1}^{\pm}}$ \\ [1ex] 
\hline
\hline
$1$ & $1$ & $-869/140$ & $\pm 379/140$ & $7/2$ & $3$ & $\mp 3$ & $0$ \\
$1$ & $2$ & $2$ & $\mp (723/280 - 6 \ln (2))$ & $-2$ & $0$ & $0$ & $0$ \\
$1$ & $3$ & $-723/140 + 12 \ln (2)$ & $0$ & $0$ & $0$ & $0$ & $0$ \\
$1$ & $4$ & $-4$ & $\pm 4$ & $0$ & $0$ & $0$ & $0$ \\
$1$ & $5$ & $-2$ & $\pm 2$ & $0$ & $0$ & $0$ & $0$ \\
$2$ & $1$ & $397/280 + 6 \ln (2)$ & $\pm (163/280 -6 \ln (2))$ & $-2$ & $0$ & $0$ & $0$ \\
$2$ & $2$ & $-9/2$ & $\pm 2$ & $7/2$ & $-3$ & $0$ & $0$ \\
$2$ & $3$ & $4$ & $\mp 2$ & $0$ & $0$ & $\mp 6$ & $0$ \\
$2$ & $4$ & $4$ & $\pm 8$ & $0$ & $0$ & $0$ & $0$ \\
$2$ & $5$ & $-2$ & $0$ & $0$ & $0$ & $0$ & $0$ \\
$3$ & $1$ & $-1$ & $\pm 1$ & $0$ & $0$ & $0$ & $0$ \\
$3$ & $2$ & $1$ & $\pm 99/280$ & $0$ & $0$ & $0$ & $0$ \\
$3$ & $3$ & $-38/35$ & $\pm 2$ & $251/140$ & $-3$ & $0$ & $3$ \\
$3$ & $4$ & $4$ & $\pm 239/280$ & $-321/140$ & $0$ & $0$ & $0$ \\
$3$ & $5$ & $0$ & $\mp 239/560$ & $0$ & $0$ & $0$ & $0$ \\
$4$ & $1$ & $-1$ & $\pm 1$ & $0$ & $0$ & $0$ & $0$ \\
$4$ & $2$ & $1$ & $\mp 239/280$ & $0$ & $0$ & $0$ & $0$ \\
$4$ & $3$ & $4$ & $\pm 2$ & $-799/140$ & $0$ & $0$ & $0$ \\
$4$ & $4$ & $-307/112 + 3 \ln (2)$ & $\pm 169/140$ & $251/140$ & $-3$ & $\mp 3$ & $3$ \\
$4$ & $5$ & $-269/480 + 1/2 \ \ln (2)$ & $\pm (869/1680 -\ln (2))$ & $0$ & $1$ & $\mp 1/2$ & $0$ \\
$5$ & $1$ & $-6$ & $\pm 6$ & $0$ & $0$ & $0$ & $0$ \\
$5$ & $2$ & $-6$ & $0$ & $0$ & $0$ & $0$ & $0$ \\
$5$ & $3$ & $0$ & $\mp 12$ & $0$ & $0$ & $0$ & $0$ \\
$5$ & $4$ & $-269/40 + 6 \ln (2)$ & $\mp (29/140 - 12 \ln (2))$ & $0$ & $12$ & $\pm 6$ & $0$ \\
$5$ & $5$ & $-1229/240-3 \ln (2)$ & $\pm 309/140$ & $1709/420$ & $1$ & $\mp 3$ & $-1$ \\
      \hline
  \end{tabular}
  \caption{Numerical values of the coefficients $g_{ij;k}^{\pm}$, $h_{ij;k}^{\pm}$ appearing in Eq.~( \eqref{Ccons}).}
  \label{tab:CmatrixCons}
\end{table}

In the case of Parity Violating operators, the number of possible sets of conditions is much smaller since the $5\times5$ mixing matrices are decomposed into three blocks of $1\times1$, and two $2\times2$ sub-matrices, as explained in section \ref{formula}. For the $1\times1$ block, we consider the condition with the corresponding two-point function, and thus, there is no need to involve any three-point functions. For the $2\times2$ blocks, there are 8 choices (for each block) for obtaining 1 condition from the three-point functions (see Table \ref{tab:GFs3ptViol}), in addition to the 3 conditions obtained from the two-point functions. However, only two (three) options for the block that involves $\{\mathcal{Q}_2, \mathcal{Q}_3 \}$ ($\{\mathcal{Q}_4, \mathcal{Q}_5 \}$) give a unique solution. By applying the same criterion, as in the Parity Conserving operators, for restricting the number of possible sets of conditions, we conclude that the block of $\{\mathcal{Q}_2, \mathcal{Q}_3 \}$ has smaller mixing contributions compared to the block of $\{\mathcal{Q}_4, \mathcal{Q}_5 \}$ for the Parity Violating operators with $S=+1$, and vice versa for the operators with $S=-1$. 

The option that gives the smallest sum of squares of the off-diagonal coefficients include the following renormalized three-point functions:
\begin{equation*}
\Tilde{G}^{\rm 3pt}_{S;\mathcal{Q}_2^{S=\pm 1}; P} (t,t), \qquad \Tilde{G}^{\rm 3pt}_{S;\mathcal{Q}_5^{S=\pm 1}; P} (t,t),
\end{equation*}
and the solution reads:
\begin{equation}
    (\Tilde{C}_{ij}^{S \pm 1})^{\MSbar, {\rm GIRS}} = \delta_{ij} + \frac{g_\MSbar^2}{16 \pi^2} \sum_{k=-1}^{+1} \left[ \bmb{\tilde{g}_{ij;k}^\pm} + \left(\ln (\bar{\mu}^2 t^2) + 2 \gamma_E \right) \ \bmr{\tilde{h}_{ij;k}^\pm} \right] N_c^k + \mathcal{O} (g_\MSbar^4),
    \label{Cviol}
\end{equation}
where the coefficients $\tilde{g}_{ij;k}^{\pm}$, $\tilde{h}_{ij;k}^{\pm}$, are given in Table~\ref{tab:CmatrixViol}.    

\begin{table}[ht!]
  \centering
  \begin{tabular}{c|c|c|c|c|c|c|c}
  \hline
\quad $i$ \quad & \quad $j$ \quad & \quad $\bmb{\tilde{g}_{ij;-1}^{\pm}}$ \quad & \quad $\bmb{\tilde{g}_{ij;0}^{\pm}}$ \quad & \quad $\bmb{\tilde{g}_{ij;+1}^{\pm}}$ \quad & \quad $\bmr{\tilde{h}_{ij;-1}^{\pm}}$ \quad & \quad $\bmr{\tilde{h}_{ij;0}^{\pm}}$ \quad & \quad $\bmr{\tilde{h}_{ij;+1}^{\pm}}$ \\ [1ex] 
\hline
\hline
$1$ & $1$ & $-869/140$ & $\pm 379/140$ & $7/2$ & $3$ & $\mp 3$ & $0$ \\
$1$ & $2$ & $0$ & $0$ & $0$ & $0$ & $0$ & $0$ \\
$1$ & $3$ & $0$ & $0$ & $0$ & $0$ & $0$ & $0$ \\
$1$ & $4$ & $0$ & $0$ & $0$ & $0$ & $0$ & $0$ \\
$1$ & $5$ & $0$ & $0$ & $0$ & $0$ & $0$ & $0$ \\
$2$ & $1$ & $0$ & $0$ & $0$ & $0$ & $0$ & $0$ \\
$2$ & $2$ & $-9/2$ & $0$ & $7/2$ & $-3$ & $0$ & $0$ \\
$2$ & $3$ & $0$ & $\mp 2$ & $0$ & $0$ & $\mp 6$ & $0$ \\
$2$ & $4$ & $0$ & $0$ & $0$ & $0$ & $0$ & $0$ \\
$2$ & $5$ & $0$ & $0$ & $0$ & $0$ & $0$ & $0$ \\
$3$ & $1$ & $0$ & $0$ & $0$ & $0$ & $0$ & $0$ \\
$3$ & $2$ & $0$ & $\pm 99/280$ & $0$ & $0$ & $0$ & $0$ \\
$3$ & $3$ & $-38/35$ & $0$ & $251/140$ & $-3$ & $0$ & $3$ \\
$3$ & $4$ & $0$ & $0$ & $0$ & $0$ & $0$ & $0$ \\
$3$ & $5$ & $0$ & $0$ & $0$ & $0$ & $0$ & $0$ \\
$4$ & $1$ & $0$ & $0$ & $0$ & $0$ & $0$ & $0$ \\
$4$ & $2$ & $0$ & $0$ & $0$ & $0$ & $0$ & $0$ \\
$4$ & $3$ & $0$ & $0$ & $0$ & $0$ & $0$ & $0$ \\
$4$ & $4$ & $-307/112 + 3 \ln (2)$ & $\pm 169/140$ & $251/140$ & $-3$ & $\mp 3$ & $3$ \\
$4$ & $5$ & $-269/480 + 1/2 \ \ln (2)$ & $\pm (869/1680 -\ln (2))$ & $0$ & $1$ & $\mp 1/2$ & $0$ \\
$5$ & $1$ & $0$ & $0$ & $0$ & $0$ & $0$ & $0$ \\
$5$ & $2$ & $0$ & $0$ & $0$ & $0$ & $0$ & $0$ \\
$5$ & $3$ & $0$ & $0$ & $0$ & $0$ & $0$ & $0$ \\
$5$ & $4$ & $-269/40 + 6 \ln (2)$ & $\mp (29/140 - 12 \ln (2))$ & $0$ & $12$ & $\pm 6$ & $0$ \\
$5$ & $5$ & $-1229/240-3 \ln (2)$ & $\pm 309/140$ & $1709/420$ & $1$ & $\mp 3$ & $-1$ \\
\hline
  \end{tabular}
  \caption{Numerical values of the coefficients $\tilde{g}_{ij;k}^{\pm}$, $\tilde{h}_{ij;k}^{\pm}$ appearing in Eq.~(\eqref{Cviol}).}
  \label{tab:CmatrixViol}
\end{table}

Other accepted options include the renormalized three-point functions of:
\begin{equation*}
\Tilde{G}^{\rm 3pt}_{V_i;\mathcal{Q}_3^{S=\pm 1}; A_i} (t,t) \quad {\rm and} \quad \Tilde{G}^{\rm 3pt}_{T_{ij};\mathcal{Q}_4^{S=\pm 1}; T'_{ij}} (t,t) \quad \left({\rm or} \quad \Tilde{G}^{\rm 3pt}_{T_{i4};\mathcal{Q}_4^{S=\pm 1}; T'_{i4}} (t,t) \right).
\end{equation*}

\newpage

\subsection{Anomalous dimensions}

The NLO ($g^4$) anomalous dimensions of the four-quark operators in the GIRS scheme can be extracted from the combination of our NLO ($g^2$) results for the conversion matrix between GIRS and $\MSbar$, with the NLO ($g^4$) anomalous dimensions in the $\MSbar$ scheme, as dictated in Eqs. (\ref{gamma_cons} -- \ref{gamma_viol}). For the selected GIRS version (given in the previous subsection), the NLO anomalous dimensions read: 
\begin{eqnarray}
    (\gamma_1^{\pm,{\rm GIRS}})_{ij} &=& \frac{1}{(16 \pi^2)^2} \sum_{k=-2}^{+2} \sum_{l=0}^1 \left[ \bmb{p_{ij;kl}^\pm} + \left(\ln (c^2) + 2 \gamma_E \right) \ \bmr{q_{ij;kl}^\pm} \right] N_c^k N_f^l, \label{anomdim1} \\
    (\tilde{\gamma}_1^{\pm,{\rm GIRS}})_{ij} &=& \frac{1}{(16 \pi^2)^2} \sum_{k=-2}^{+2} \sum_{l=0}^1 \left[ \bmb{\tilde{p}_{ij;kl}^\pm} + \left(\ln (c^2) + 2 \gamma_E \right) \ \bmr{\tilde{q}_{ij;kl}^\pm} \right] N_c^k N_f^l,
    \label{anomdim2}
\end{eqnarray}
where we set $\bar{\mu} \, t \equiv c$ ($=$ constant), and the coefficients $p_{ij;kl}^{\pm}$, $q_{ij;kl}^{\pm}$, $\tilde{p}_{ij;kl}^{\pm}$, $\tilde{q}_{ij;kl}^{\pm}$ are given in Tables~\ref{tab:anomdimCons1} --  \ref{tab:anomdimq}.

\newpage

\begin{table}[ht!]
  \centering
  \begin{tabular}{c|c|c|c|c|c|c}
  \hline
\quad $i$ \quad & \quad $j$ \quad & \quad $\bmb{p_{ij;-2 0}^{\pm}}$ \quad & \quad $\bmb{p_{ij;-1 0}^{\pm}}$ \quad & \quad $\bmb{p_{ij;0 0}^{\pm}}$ \quad & \quad $\bmb{p_{ij;+1 0}^{\pm}}$ \quad & \quad $\bmb{p_{ij;+2 0}^{\pm}}$ \\ [1ex] 
\hline
\hline
$1$ & $1$ & $0$ & $0$ & $9559/210$ & $\mp 4169/210$ & $-77/3$ \\
$1$ & $2$ & $-24$ & $\pm (3009/70 - 72 \ln (2))$ & $-2587/420 + 36 \ln (2)$ & $\pm (971/140 - 44 \ln(2))$ & $44/3$ \\
$1$ & $3$ & $2169/35 - 144 \ln (2)$ & $\mp (3849/70 - 72 \ln(2))$ & $2651/70 - 88 \ln(2)$ & $\pm 24$ & $0$ \\
$1$ & $4$ & $0$ & $\mp 24$ & $88/3$ & $\mp16/3$ & $0$ \\
$1$ & $5$ & $0$ & $\pm 4$ & $44/3$ & $\mp 56/3$ & $0$ \\
$2$ & $1$ & $1191/70 + 72 \ln (2)$ & $\mp (1893/140 + 108 \ln (2))$ & $-5437/210 - 8 \ln(2)$ & $\pm (3247/420 + 44 \ln(2))$ & $44/3$ \\
$2$ & $2$ & $15/2$ & $\pm 12$ & $6308/105$ & $\mp 44/3$ & $-77/3$ \\
$2$ & $3$ & $0$ & $\pm 1224/35$ & $-16/3$ & $\pm 5129/105$ & $0$ \\
$2$ & $4$ & $-48$ & $0$ & $-9049/210$ & $\mp 40009/105$ & $0$ \\
$2$ & $5$ & $-8$ & $\pm 24$ & $2329/420$ & $0$ & $0$ \\
$3$ & $1$ & $-12$ & $\pm 18$ & $22/3$ & $\mp 40/3$ & $0$ \\
$3$ & $2$ & $0$ & $\pm 9$ & $-40/3$ & $\pm 365/28$ & $0$ \\
$3$ & $3$ & $15/2$ & $\mp 12$ & $8773/105$ & $\mp 44/3$ & $-4933/105$ \\
$3$ & $4$ & $0$ & $\mp 2397/70$ & $-8311/210$ & $\pm 3149/420$ & $1177/70$ \\
$3$ & $5$ & $8$ & $\mp 799/140$ & $-403/140$ & $\pm 7423/840$ & $0$ \\
$4$ & $1$ & $0$ & $\mp 6$ & $58/3$ & $\mp 40/3$ & $0$ \\
$4$ & $2$ & $12$ & $0$ & $-7751/420$ & $\pm 239/21$ & $0$ \\
$4$ & $3$ & $0$ & $\pm 36$ & $-4009/210$ & $\mp 10271/210$ & $8789/210$ \\
$4$ & $4$ & $-107/2$ & $\mp (1767/140 + 36 \ln(2))$ & $12001/168 - 22 \ln(2)$ & $\pm 1877/70$ & $-4933/105$ \\
$4$ & $5$ & $-811/105 + 16 \ln (2)$ & $\pm (2131/120 - 14 \ln (2))$ & $-207/112 - 23/3 \ln(2)$ & $\mp (1599/280 - 46/3 \ln(2))$ & $0$ \\
$5$ & $1$ & $0$ & $\pm 12$ & $20$ & $\mp 32$ & $0$ \\
$5$ & $2$ & $24$ & $\mp 963/35$ & $2957/70$ & $0$ & $0$ \\
$5$ & $3$ & $-96$ & $\pm 72$ & $1434/35$ & $\pm 2117/35$ & $0$ \\
$5$ & $4$ & $-5716/35 - 192 \ln (2)$ & $\mp (1069/10 + 168 \ln (2))$ & $-23179/420 + 4 \ln(2)$ & $\mp (7033/210 - 8 \ln(2))$ & $0$ \\
$5$ & $5$ & $21/2$ & $\pm (6247/140 + 36 \ln(2))$ & $3839/360 + 22 \ln(2)$ & $\pm 337/70$ & $-3397/315$ \\
      \hline
  \end{tabular}
  \caption{Numerical values of the coefficients $p_{ij;k0}^{\pm}$ appearing in Eq.~\eqref{anomdim1}.}
  \label{tab:anomdimCons1}
\end{table}

\newpage

\begin{table}[ht!]
  \centering
\begin{tabular}{c|c|c|c|c|c|c}
  \hline
\quad $i$ \quad & \quad $j$ \quad & \quad $\bmb{p_{ij;-2 1}^{\pm}}$ \quad & \quad $\bmb{p_{ij;-1 1}^{\pm}}$ \quad & \quad $\bmb{p_{ij;0 1}^{\pm}}$ \quad & \quad $\bmb{p_{ij;+1 1}^{\pm}}$ \quad & \quad $\bmb{p_{ij;+2 1}^{\pm}}$ \\ [1ex] 
\hline
\hline
$1$ & $1$ & $0$ & $-869/105$ & $\pm 379/105$ & $14/3$ & $0$ \\
$1$ & $2$ & $0$ & $8/3$ & $\mp (241/70 - 8 \ln (2))$ & $-8/3$ & $0$ \\
$1$ & $3$ & $0$ & $-241/35 + 16 \ln(2)$ & $0$ & $0$ & $0$ \\
$1$ & $4$ & $0$ & $-16/3$ & $\pm 16/3$ & $0$ & $0$ \\
$1$ & $5$ & $0$ & $-8/3$ & $\pm 8/3$ & $0$ & $0$ \\
$2$ & $1$ & $0$ & $397/210 + 8 \ln (2)$ & $\pm (163/210 - 8 \ln(2))$ & $-8/3$ & $0$ \\
$2$ & $2$ & $0$ & $-40/3$ & $\pm 8/3$ & $14/3$ & $0$ \\
$2$ & $3$ & $0$ & $16/3$ & $\mp 52/3$ & $0$ & $0$ \\
$2$ & $4$ & $0$ & $16/3$ & $\pm 32/3$ & $0$ & $0$ \\
$2$ & $5$ & $0$ & $-8/3$ & $0$ & $0$ & $0$ \\
$3$ & $1$ & $0$ & $-4/3$ & $\pm 4/3$ & $0$ & $0$ \\
$3$ & $2$ & $0$ & $4/3$ & $\mp 107/70$ & $0$ & $0$ \\
$3$ & $3$ & $0$ & $-922/105$ & $\pm 8/3$ & $601/105$ & $0$ \\
$3$ & $4$ & $0$ & $16/3$ & $\pm 239/210$ & $-107/35$ & $0$ \\
$3$ & $5$ & $0$ & $0$ & $\mp 239/420$ & $0$ & $0$ \\
$4$ & $1$ & $0$ & $-4/3$ & $\pm 4/3$ & $0$ & $0$ \\
$4$ & $2$ & $0$ & $4/3$ & $\mp 239/210$ & $0$ & $0$ \\
$4$ & $3$ & $0$ & $16/3$ & $\pm 8/3$ & $-799/105$ & $0$ \\
$4$ & $4$ & $0$ & $-587/84 + 4 \ln(2)$ & $\pm 33/35$ & $601/105$ & $0$ \\
$4$ & $5$ & $0$ & $-21/40 + 2/3 \ln(2)$ & $\pm (81/140 - 4/3 \ln(2))$ & $0$ & $0$ \\
$5$ & $1$ & $0$ & $-8$ & $\pm 8$ & $0$ & $0$ \\
$5$ & $2$ & $0$ & $-8$ & $0$ & $0$ & $0$ \\
$5$ & $3$ & $0$ & $0$ & $\mp 16$ & $0$ & $0$ \\
$5$ & $4$ & $0$ & $611/30 + 8 \ln (2)$ & $\pm (2351/105 + 16 \ln(2))$ & $0$ & $0$ \\
$5$ & $5$ & $0$ & $-1189/180 - 4 \ln(2)$ & $\mp 107/35$ & $799/315$ & $0$ \\
      \hline
  \end{tabular}
  \caption{Numerical values of the coefficients $p_{ij;k1}^{\pm}$ appearing in Eq.~\eqref{anomdim1}.}
  \label{tab:anomdimCons2}
\end{table}

\newpage

\begin{table}[ht!]
  \centering
  \begin{tabular}{c|c|c|c|c|c|c}
  \hline
\quad $i$ \quad & \quad $j$ \quad & \quad $\bmb{\tilde{p}_{ij;-2 0}^{\pm}}$ \quad & \quad $\bmb{\tilde{p}_{ij;-1 0}^{\pm}}$ \quad & \quad $\bmb{\tilde{p}_{ij;0 0}^{\pm}}$ \quad & \quad $\bmb{\tilde{p}_{ij;+1 0}^{\pm}}$ \quad & \quad $\bmb{\tilde{p}_{ij;+2 0}^{\pm}}$ \\ [1ex] 
\hline
\hline
$1$ & $1$ & $0$ & $0$ & $9559/210$ & $\mp 4169/210$ & $-77/3$ \\
$1$ & $2$ & $0$ & $0$ & $0$ & $0$ & $0$ \\
$1$ & $3$ & $0$ & $0$ & $0$ & $0$ & $0$ \\
$1$ & $4$ & $0$ & $0$ & $0$ & $0$ & $0$ \\
$1$ & $5$ & $0$ & $0$ & $0$ & $0$ & $0$ \\
$2$ & $1$ & $0$ & $0$ & $0$ & $0$ & $0$ \\
$2$ & $2$ & $15/2$ & $0$ & $6308/105$ & $0$ & $-77/3$ \\
$2$ & $3$ & $0$ & $\pm 1224/35$ & $0$ & $\pm 5129/105$ & $0$ \\
$2$ & $4$ & $0$ & $0$ & $0$ & $0$ & $0$ \\
$2$ & $5$ & $0$ & $0$ & $0$ & $0$ & $0$ \\
$3$ & $1$ & $0$ & $0$ & $0$ & $0$ & $0$ \\
$3$ & $2$ & $0$ & $\pm 9$ & $0$ & $\pm 365/28$ & $0$ \\
$3$ & $3$ & $15/2$ & $0$ & $8773/105$ & $0$ & $-4933/105$ \\
$3$ & $4$ & $0$ & $0$ & $0$ & $0$ & $0$ \\
$3$ & $5$ & $0$ & $0$ & $0$ & $0$ & $0$ \\
$4$ & $1$ & $0$ & $0$ & $0$ & $0$ & $0$ \\
$4$ & $2$ & $0$ & $0$ & $0$ & $0$ & $0$ \\
$4$ & $3$ & $0$ & $0$ & $0$ & $0$ & $0$ \\
$4$ & $4$ & $-107/2$ & $\mp (1767/140 + 36 \ln(2))$ & $12001/168 - 22 \ln(2)$ & $\pm 1877/70$ & $-4933/105$ \\
$4$ & $5$ & $-811/105 + 16 \ln(2)$ & $\pm (2131/120 - 14 \ln(2))$ & $-207/112-23/3 \ln(2)$ & $\mp (1599/280 - 46/3 \ln(2))$ & $0$ \\
$5$ & $1$ & $0$ & $0$ & $0$ & $0$ & $0$ \\
$5$ & $2$ & $0$ & $0$ & $0$ & $0$ & $0$ \\
$5$ & $3$ & $0$ & $0$ & $0$ & $0$ & $0$ \\
$5$ & $4$ & $-5716/35 - 192 \ln(2)$ & $\mp (1069/10 + 168 \ln(2))$ & $-23179/420 + 4 \ln(2)$ & $\mp (7033/210 - 8 \ln(2))$ & $0$ \\
$5$ & $5$ & $21/2$ & $\pm (6247/140 + 36 \ln(2))$ & $3839/360 + 22 \ln(2)$ & $\pm 337/70$ & $-3397/315$ \\
      \hline
  \end{tabular}
  \caption{Numerical values of the coefficients $\tilde{p}_{ij;k0}^{\pm}$ appearing in Eq.~\eqref{anomdim2}.}
  \label{tab:anomdimViol1}
\end{table}

\newpage

\begin{table}[ht!]
  \centering
\begin{tabular}{c|c|c|c|c|c|c}
  \hline
\quad $i$ \quad & \quad $j$ \quad & \quad $\bmb{\tilde{p}_{ij;-2 1}^{\pm}}$ \quad & \quad $\bmb{\tilde{p}_{ij;-1 1}^{\pm}}$ \quad & \quad $\bmb{\tilde{p}_{ij;0 1}^{\pm}}$ \quad & \quad $\bmb{\tilde{p}_{ij;+1 1}^{\pm}}$ \quad & \quad $\bmb{\tilde{p}_{ij;+2 1}^{\pm}}$ \\ [1ex] 
\hline
\hline
$1$ & $1$ & $0$ & $-869/105$ & $\pm 379/105$ & $14/3$ & $0$ \\
$1$ & $2$ & $0$ & $0$ & $0$ & $0$ & $0$ \\
$1$ & $3$ & $0$ & $0$ & $0$ & $0$ & $0$ \\
$1$ & $4$ & $0$ & $0$ & $0$ & $0$ & $0$ \\
$1$ & $5$ & $0$ & $0$ & $0$ & $0$ & $0$ \\
$2$ & $1$ & $0$ & $0$ & $0$ & $0$ & $0$ \\
$2$ & $2$ & $0$ & $-40/3$ & $0$ & $14/3$ & $0$ \\
$2$ & $3$ & $0$ & $0$ & $\mp 52/3$ & $0$ & $0$ \\
$2$ & $4$ & $0$ & $0$ & $0$ & $0$ & $0$ \\
$2$ & $5$ & $0$ & $0$ & $0$ & $0$ & $0$ \\
$3$ & $1$ & $0$ & $0$ & $0$ & $0$ & $0$ \\
$3$ & $2$ & $0$ & $0$ & $\mp 107/70$ & $0$ & $0$ \\
$3$ & $3$ & $0$ & $-922/105$ & $0$ & $601/105$ & $0$ \\
$3$ & $4$ & $0$ & $0$ & $0$ & $0$ & $0$ \\
$3$ & $5$ & $0$ & $0$ & $0$ & $0$ & $0$ \\
$4$ & $1$ & $0$ & $0$ & $0$ & $0$ & $0$ \\
$4$ & $2$ & $0$ & $0$ & $0$ & $0$ & $0$ \\
$4$ & $3$ & $0$ & $0$ & $0$ & $0$ & $0$ \\
$4$ & $4$ & $0$ & $-587/84 + 4 \ln(2)$ & $\pm 33/35$ & $601/105$ & $0$ \\
$4$ & $5$ & $0$ & $-21/40 + 2/3 \ln(2)$ & $\pm (81/140-4/3 \ln(2))$ & $0$ & $0$ \\
$5$ & $1$ & $0$ & $0$ & $0$ & $0$ & $0$ \\
$5$ & $2$ & $0$ & $0$ & $0$ & $0$ & $0$ \\
$5$ & $3$ & $0$ & $0$ & $0$ & $0$ & $0$ \\
$5$ & $4$ & $0$ & $611/30 + 8 \ln(2))$ & $\pm (2351/105 + 16 \ln(2))$ & $0$ & $0$ \\
$5$ & $5$ & $0$ & $-1189/180 - 4 \ln(2)$ & $\mp 107/35$ & $799/315$ & $0$ \\
      \hline
  \end{tabular}
  \caption{Numerical values of the coefficients $\tilde{p}_{ij;k1}^{\pm}$ appearing in Eq.~\eqref{anomdim2}.}
  \label{tab:anomdimViol2}
\end{table}

\newpage

\begin{table}[ht!]
  \centering
  \begin{tabular}{c|c|c|c|c}
  \hline
\quad $i$ \quad & \quad $j$ \quad & \quad $\bmr{q_{ij;-1 1}^{\pm} = \tilde{q}_{ij;-1 1}^{\pm}}$ \quad & $\bmr{q_{ij;0 1}^{\pm} = \tilde{q}_{ij;0 1}^{\pm}}$ \quad \qquad & $\bmr{q_{ij;+1 1}^{\pm} = \tilde{q}_{ij;+1 1}^{\pm}}$ \\ [1ex] 
& & \quad $\bmr{= (-11/2) \ q_{ij;0 0}^{\pm}}$ \quad & \quad $\bmr{= (-11/2) \ q_{ij;+1 0}^{\pm}}$ \quad & \quad $\bmr{= (-11/2) \ q_{ij;+2 0}^{\pm}}$ \\ [1ex]
& & \quad $\bmr{= (-11/2) \ \tilde{q}_{ij;0 0}^{\pm}}$ \quad & \quad $\bmr{= (-11/2) \ \tilde{q}_{ij;+1 0}^{\pm}}$ \quad & \quad $\bmr{= (-11/2) \ \tilde{q}_{ij;+2 0}^{\pm}}$ \\ [1ex]
\hline
\hline
$1$ & $1$ & $4$ & $\mp 4$ & $0$ \\
$1$ & $2$ & $0$ & $0$ & $0$ \\
$1$ & $3$ & $0$ & $0$ & $0$ \\
$1$ & $4$ & $0$ & $0$ & $0$ \\
$1$ & $5$ & $0$ & $0$ & $0$ \\
$2$ & $1$ & $0$ & $0$ & $0$ \\
$2$ & $2$ & $-4$ & $0$ & $0$ \\
$2$ & $3$ & $0$ & $\mp 8$ & $0$ \\
$2$ & $4$ & $0$ & $0$ & $0$ \\
$2$ & $5$ & $0$ & $0$ & $0$ \\
$3$ & $1$ & $0$ & $0$ & $0$ \\
$3$ & $2$ & $0$ & $0$ & $0$ \\
$3$ & $3$ & $-4$ & $0$ & $4$ \\
$3$ & $4$ & $0$ & $0$ & $0$ \\
$3$ & $5$ & $0$ & $0$ & $0$ \\
$4$ & $1$ & $0$ & $0$ & $0$ \\
$4$ & $2$ & $0$ & $0$ & $0$ \\
$4$ & $3$ & $0$ & $0$ & $0$ \\
$4$ & $4$ & $-4$ & $\mp 4$ & $4$ \\
$4$ & $5$ & $4/3$ & $\mp 2/3$ & $0$ \\
$5$ & $1$ & $0$ & $0$ & $0$ \\
$5$ & $2$ & $0$ & $0$ & $0$ \\
$5$ & $3$ & $0$ & $0$ & $0$ \\
$5$ & $4$ & $16$ & $\pm 8$ & $0$ \\
$5$ & $5$ & $4/3$ & $\mp 4$ & $-4/3$ \\
      \hline
  \end{tabular}
  \caption{Numerical values of the coefficients $q_{ij;kl}^{\pm} = \tilde{q}_{ij;kl}^{\pm}$ appearing in Eqs. (\ref{anomdim1} -- \ref{anomdim2}). The coefficients $q^\pm_{ij; -2\,0}$, ${\tilde q}^\pm_{ij;-2\,0}$, $q^\pm_{ij; -1\,0}$, ${\tilde q}^\pm_{ij;-1\,0}$, $q^\pm_{ij; -2\,1}$, ${\tilde q}^\pm_{ij;-2\,1}$, $q^\pm_{ij; +2\,1}$, ${\tilde q}^\pm_{ij;+2\,1}$ are all zero.}
  \label{tab:anomdimq}
\end{table}

\section{Conclusions -- Future plans} 
\label{conclusions}

In this work, we present a comprehensive study of the renormalization of four-quark operators involved in $\Delta F = 2$ processes by using a gauge-invariant renormalization scheme (GIRS). The analysis is based on a one-loop perturbative calculation of two-point Green's functions involving products of two four-quark operators, as well as three-point Green’s functions with one four-quark and two bilinear operators. The computations are performed in dimensional regularization. Operator mixing between four-quark operators, which share the same symmetry properties, was addressed through a set of renormalization conditions involving the Green's functions under study. We found a variety of acceptable renormalization prescriptions within GIRS, which are applicable in both perturbative and nonperturbative data. We present a specific choice of them in the manuscript, which can lead to smaller mixing contributions; all other choices can be directly inferred from the results for the Green's functions, provided in a supplemental file. This calculation enables us to derive the one-loop conversion matrices connecting GIRS results to their $\MSbar$ counterparts. Our results can be employed in nonperturbative investigations on the lattice.
For this purpose, integrations over time slices have been performed in the Green's functions of the proposed GIRS scheme, and the effect on the corresponding conversion factors has been calculated; this procedure is expected to reduce statistical noise in the nonperturbative evaluation of the relevant Green's functions, summed over time slices.
Thus, this study not only advances the theoretical framework for the renormalization procedure of four-quark operators but also offers a practical tool for improving the reliability of physical quantities calculated in lattice QCD. 

A natural extension of this work will involve the investigation of four-quark operators with $\Delta F = 1$ and $\Delta F = 0$. This investigation also encompasses the mixing with lower-dimensional operators, including local and extended quark bilinear operators, the chromomagnetic operator, and the energy-momentum tensor.

\newpage

\begin{acknowledgments}
The project (EXCELLENCE/0421/0025) is implemented under the programme of social cohesion ``THALIA 2021-2027'' co-funded by the European Union through the Cyprus Research and Innovation Foundation (RIF). M.~Constantinou acknowledges financial support from the U.S. Department of Energy, Office of Nuclear Physics, Early Career Award under Grant No.\ DE-SC0020405. The results are generated within the FEDILA software (project: CONCEPT/0823/0052), which is also implemented under the same ``THALIA 2021-2027'' programme, co-funded by the European Union through RIF. G.S. and M.~Constantinou acknowledge funding under the project 3D-nucleon, contract number EXCELLENCE/0421/0043, co-financed by the European Regional Development Fund and the Republic of Cyprus through the Cyprus RIF.
\end{acknowledgments}

\label{Bibliography}
\bibliography{Bibliography}

\begin{thebibliography}{60}%
\makeatletter
\providecommand \@ifxundefined [1]{%
 \@ifx{#1\undefined}
}%
\providecommand \@ifnum [1]{%
 \ifnum #1\expandafter \@firstoftwo
 \else \expandafter \@secondoftwo
 \fi
}%
\providecommand \@ifx [1]{%
 \ifx #1\expandafter \@firstoftwo
 \else \expandafter \@secondoftwo
 \fi
}%
\providecommand \natexlab [1]{#1}%
\providecommand \enquote  [1]{``#1''}%
\providecommand \bibnamefont  [1]{#1}%
\providecommand \bibfnamefont [1]{#1}%
\providecommand \citenamefont [1]{#1}%
\providecommand \href@noop [0]{\@secondoftwo}%
\providecommand \href [0]{\begingroup \@sanitize@url \@href}%
\providecommand \@href[1]{\@@startlink{#1}\@@href}%
\providecommand \@@href[1]{\endgroup#1\@@endlink}%
\providecommand \@sanitize@url [0]{\catcode `\\12\catcode `\$12\catcode
  `\&12\catcode `\#12\catcode `\^12\catcode `\_12\catcode `\%12\relax}%
\providecommand \@@startlink[1]{}%
\providecommand \@@endlink[0]{}%
\providecommand \url  [0]{\begingroup\@sanitize@url \@url }%
\providecommand \@url [1]{\endgroup\@href {#1}{\urlprefix }}%
\providecommand \urlprefix  [0]{URL }%
\providecommand \Eprint [0]{\href }%
\providecommand \doibase [0]{https://doi.org/}%
\providecommand \selectlanguage [0]{\@gobble}%
\providecommand \bibinfo  [0]{\@secondoftwo}%
\providecommand \bibfield  [0]{\@secondoftwo}%
\providecommand \translation [1]{[#1]}%
\providecommand \BibitemOpen [0]{}%
\providecommand \bibitemStop [0]{}%
\providecommand \bibitemNoStop [0]{.\EOS\space}%
\providecommand \EOS [0]{\spacefactor3000\relax}%
\providecommand \BibitemShut  [1]{\csname bibitem#1\endcsname}%
\let\auto@bib@innerbib\@empty
\bibitem [{\citenamefont {Buchalla}\ \emph {et~al.}(1996)\citenamefont
  {Buchalla}, \citenamefont {Buras},\ and\ \citenamefont
  {Lautenbacher}}]{Buchalla:1995vs}%
  \BibitemOpen
  \bibfield  {author} {\bibinfo {author} {\bibfnamefont {G.}~\bibnamefont
  {Buchalla}}, \bibinfo {author} {\bibfnamefont {A.~J.}\ \bibnamefont
  {Buras}},\ and\ \bibinfo {author} {\bibfnamefont {M.~E.}\ \bibnamefont
  {Lautenbacher}},\ }\bibfield  {title} {\bibinfo {title} {{Weak decays beyond
  leading logarithms}},\ }\href {https://doi.org/10.1103/RevModPhys.68.1125}
  {\bibfield  {journal} {\bibinfo  {journal} {Rev. Mod. Phys.}\ }\textbf
  {\bibinfo {volume} {68}},\ \bibinfo {pages} {1125} (\bibinfo {year}
  {1996})},\ \Eprint {https://arxiv.org/abs/hep-ph/9512380}
  {arXiv:hep-ph/9512380} \BibitemShut {NoStop}%
\bibitem [{\citenamefont {Lehner}\ \emph {et~al.}(2019)\citenamefont {Lehner}
  \emph {et~al.}}]{USQCD:2019hyg}%
  \BibitemOpen
  \bibfield  {author} {\bibinfo {author} {\bibfnamefont {C.}~\bibnamefont
  {Lehner}} \emph {et~al.} (\bibinfo {collaboration} {USQCD}),\ }\bibfield
  {title} {\bibinfo {title} {{Opportunities for Lattice QCD in Quark and Lepton
  Flavor Physics}},\ }\href {https://doi.org/10.1140/epja/i2019-12891-2}
  {\bibfield  {journal} {\bibinfo  {journal} {Eur. Phys. J. A}\ }\textbf
  {\bibinfo {volume} {55}},\ \bibinfo {pages} {195} (\bibinfo {year} {2019})},\
  \Eprint {https://arxiv.org/abs/1904.09479} {arXiv:1904.09479 [hep-lat]}
  \BibitemShut {NoStop}%
\bibitem [{\citenamefont {Aoki}\ \emph {et~al.}(2022)\citenamefont {Aoki} \emph
  {et~al.}}]{FlavourLatticeAveragingGroupFLAG:2021npn}%
  \BibitemOpen
  \bibfield  {author} {\bibinfo {author} {\bibfnamefont {Y.}~\bibnamefont
  {Aoki}} \emph {et~al.} (\bibinfo {collaboration} {Flavour Lattice Averaging
  Group (FLAG)}),\ }\bibfield  {title} {\bibinfo {title} {{FLAG Review 2021}},\
  }\href {https://doi.org/10.1140/epjc/s10052-022-10536-1} {\bibfield
  {journal} {\bibinfo  {journal} {Eur. Phys. J. C}\ }\textbf {\bibinfo {volume}
  {82}},\ \bibinfo {pages} {869} (\bibinfo {year} {2022})},\ \Eprint
  {https://arxiv.org/abs/2111.09849} {arXiv:2111.09849 [hep-lat]} \BibitemShut
  {NoStop}%
\bibitem [{\citenamefont {Aaij}\ \emph
  {et~al.}(2023{\natexlab{a}})\citenamefont {Aaij} \emph
  {et~al.}}]{LHCb:2022aki}%
  \BibitemOpen
  \bibfield  {author} {\bibinfo {author} {\bibfnamefont {R.}~\bibnamefont
  {Aaij}} \emph {et~al.} (\bibinfo {collaboration} {LHCb}),\ }\bibfield
  {title} {\bibinfo {title} {{Observation of a Resonant Structure near the
  Ds+Ds- Threshold in the B+\textrightarrow{}Ds+Ds-K+ Decay}},\ }\href
  {https://doi.org/10.1103/PhysRevLett.131.071901} {\bibfield  {journal}
  {\bibinfo  {journal} {Phys. Rev. Lett.}\ }\textbf {\bibinfo {volume} {131}},\
  \bibinfo {pages} {071901} (\bibinfo {year} {2023}{\natexlab{a}})},\ \Eprint
  {https://arxiv.org/abs/2210.15153} {arXiv:2210.15153 [hep-ex]} \BibitemShut
  {NoStop}%
\bibitem [{\citenamefont {Aaij}\ \emph
  {et~al.}(2023{\natexlab{b}})\citenamefont {Aaij} \emph
  {et~al.}}]{LHCb:2022dvn}%
  \BibitemOpen
  \bibfield  {author} {\bibinfo {author} {\bibfnamefont {R.}~\bibnamefont
  {Aaij}} \emph {et~al.} (\bibinfo {collaboration} {LHCb}),\ }\bibfield
  {title} {\bibinfo {title} {{First observation of the
  B+\textrightarrow{}Ds+Ds-K+ decay}},\ }\href
  {https://doi.org/10.1103/PhysRevD.108.034012} {\bibfield  {journal} {\bibinfo
   {journal} {Phys. Rev. D}\ }\textbf {\bibinfo {volume} {108}},\ \bibinfo
  {pages} {034012} (\bibinfo {year} {2023}{\natexlab{b}})},\ \Eprint
  {https://arxiv.org/abs/2211.05034} {arXiv:2211.05034 [hep-ex]} \BibitemShut
  {NoStop}%
\bibitem [{\citenamefont {Aaij}\ \emph
  {et~al.}(2023{\natexlab{c}})\citenamefont {Aaij} \emph
  {et~al.}}]{LHCb:2022sfr}%
  \BibitemOpen
  \bibfield  {author} {\bibinfo {author} {\bibfnamefont {R.}~\bibnamefont
  {Aaij}} \emph {et~al.} (\bibinfo {collaboration} {LHCb}),\ }\bibfield
  {title} {\bibinfo {title} {{First Observation of a Doubly Charged Tetraquark
  and Its Neutral Partner}},\ }\href
  {https://doi.org/10.1103/PhysRevLett.131.041902} {\bibfield  {journal}
  {\bibinfo  {journal} {Phys. Rev. Lett.}\ }\textbf {\bibinfo {volume} {131}},\
  \bibinfo {pages} {041902} (\bibinfo {year} {2023}{\natexlab{c}})},\ \Eprint
  {https://arxiv.org/abs/2212.02716} {arXiv:2212.02716 [hep-ex]} \BibitemShut
  {NoStop}%
\bibitem [{\citenamefont {Aaij}\ \emph
  {et~al.}(2023{\natexlab{d}})\citenamefont {Aaij} \emph
  {et~al.}}]{LHCb:2022lzp}%
  \BibitemOpen
  \bibfield  {author} {\bibinfo {author} {\bibfnamefont {R.}~\bibnamefont
  {Aaij}} \emph {et~al.} (\bibinfo {collaboration} {LHCb}),\ }\bibfield
  {title} {\bibinfo {title} {{Amplitude analysis of
  B0\textrightarrow{}D\textasciimacron{}0Ds+\ensuremath{\pi}- and
  B+\textrightarrow{}D-Ds+\ensuremath{\pi}+ decays}},\ }\href
  {https://doi.org/10.1103/PhysRevD.108.012017} {\bibfield  {journal} {\bibinfo
   {journal} {Phys. Rev. D}\ }\textbf {\bibinfo {volume} {108}},\ \bibinfo
  {pages} {012017} (\bibinfo {year} {2023}{\natexlab{d}})},\ \Eprint
  {https://arxiv.org/abs/2212.02717} {arXiv:2212.02717 [hep-ex]} \BibitemShut
  {NoStop}%
\bibitem [{\citenamefont {Ishikawa}\ \emph {et~al.}(2011)\citenamefont
  {Ishikawa}, \citenamefont {Aoki}, \citenamefont {Flynn}, \citenamefont
  {Izubuchi},\ and\ \citenamefont {Loktik}}]{Ishikawa:2011dd}%
  \BibitemOpen
  \bibfield  {author} {\bibinfo {author} {\bibfnamefont {T.}~\bibnamefont
  {Ishikawa}}, \bibinfo {author} {\bibfnamefont {Y.}~\bibnamefont {Aoki}},
  \bibinfo {author} {\bibfnamefont {J.~M.}\ \bibnamefont {Flynn}}, \bibinfo
  {author} {\bibfnamefont {T.}~\bibnamefont {Izubuchi}},\ and\ \bibinfo
  {author} {\bibfnamefont {O.}~\bibnamefont {Loktik}},\ }\bibfield  {title}
  {\bibinfo {title} {{One-loop operator matching in the static heavy and
  domain-wall light quark system with O(a) improvement}},\ }\href
  {https://doi.org/10.1007/JHEP05(2011)040} {\bibfield  {journal} {\bibinfo
  {journal} {JHEP}\ }\textbf {\bibinfo {volume} {05}},\ \bibinfo {pages}
  {040}},\ \Eprint {https://arxiv.org/abs/1101.1072} {arXiv:1101.1072
  [hep-lat]} \BibitemShut {NoStop}%
\bibitem [{\citenamefont {Di~Pierro}\ and\ \citenamefont
  {Sachrajda}(1998)}]{DiPierro:1998ty}%
  \BibitemOpen
  \bibfield  {author} {\bibinfo {author} {\bibfnamefont {M.}~\bibnamefont
  {Di~Pierro}}\ and\ \bibinfo {author} {\bibfnamefont {C.~T.}\ \bibnamefont
  {Sachrajda}} (\bibinfo {collaboration} {UKQCD}),\ }\bibfield  {title}
  {\bibinfo {title} {{A Lattice study of spectator effects in inclusive decays
  of B mesons}},\ }\href {https://doi.org/10.1016/S0550-3213(98)00580-X}
  {\bibfield  {journal} {\bibinfo  {journal} {Nucl. Phys. B}\ }\textbf
  {\bibinfo {volume} {534}},\ \bibinfo {pages} {373} (\bibinfo {year}
  {1998})},\ \Eprint {https://arxiv.org/abs/hep-lat/9805028}
  {arXiv:hep-lat/9805028} \BibitemShut {NoStop}%
\bibitem [{\citenamefont {Di~Pierro}\ \emph {et~al.}(1999)\citenamefont
  {Di~Pierro}, \citenamefont {Sachrajda},\ and\ \citenamefont
  {Michael}}]{DiPierro:1999tb}%
  \BibitemOpen
  \bibfield  {author} {\bibinfo {author} {\bibfnamefont {M.}~\bibnamefont
  {Di~Pierro}}, \bibinfo {author} {\bibfnamefont {C.~T.}\ \bibnamefont
  {Sachrajda}},\ and\ \bibinfo {author} {\bibfnamefont {C.}~\bibnamefont
  {Michael}} (\bibinfo {collaboration} {UKQCD}),\ }\bibfield  {title} {\bibinfo
  {title} {{An Exploratory lattice study of spectator effects in inclusive
  decays of the $\Lambda_b$ baryon}},\ }\href
  {https://doi.org/10.1016/S0370-2693(99)01166-1} {\bibfield  {journal}
  {\bibinfo  {journal} {Phys. Lett. B}\ }\textbf {\bibinfo {volume} {468}},\
  \bibinfo {pages} {143} (\bibinfo {year} {1999})},\ \bibinfo {note} {[Erratum:
  Phys.Lett.B 525, 360--360 (2002)]},\ \Eprint
  {https://arxiv.org/abs/hep-lat/9906031} {arXiv:hep-lat/9906031} \BibitemShut
  {NoStop}%
\bibitem [{\citenamefont {Gimenez}\ and\ \citenamefont
  {Reyes}(1999)}]{Gimenez:1998mw}%
  \BibitemOpen
  \bibfield  {author} {\bibinfo {author} {\bibfnamefont {V.}~\bibnamefont
  {Gimenez}}\ and\ \bibinfo {author} {\bibfnamefont {J.}~\bibnamefont
  {Reyes}},\ }\bibfield  {title} {\bibinfo {title} {{Calculation of the
  continuum lattice HQET matching for the complete basis of four fermion
  operators: Reanalysis of the B0 - anti-B0 mixing}},\ }\href
  {https://doi.org/10.1016/S0550-3213(98)00867-0} {\bibfield  {journal}
  {\bibinfo  {journal} {Nucl. Phys. B}\ }\textbf {\bibinfo {volume} {545}},\
  \bibinfo {pages} {576} (\bibinfo {year} {1999})},\ \Eprint
  {https://arxiv.org/abs/hep-lat/9806023} {arXiv:hep-lat/9806023} \BibitemShut
  {NoStop}%
\bibitem [{\citenamefont {Ishizuka}\ \emph {et~al.}(2018)\citenamefont
  {Ishizuka}, \citenamefont {Ishikawa}, \citenamefont {Ukawa},\ and\
  \citenamefont {Yoshi\'e}}]{Ishizuka:2018qbn}%
  \BibitemOpen
  \bibfield  {author} {\bibinfo {author} {\bibfnamefont {N.}~\bibnamefont
  {Ishizuka}}, \bibinfo {author} {\bibfnamefont {K.~I.}\ \bibnamefont
  {Ishikawa}}, \bibinfo {author} {\bibfnamefont {A.}~\bibnamefont {Ukawa}},\
  and\ \bibinfo {author} {\bibfnamefont {T.}~\bibnamefont {Yoshi\'e}},\
  }\bibfield  {title} {\bibinfo {title} {{Calculation of $K \to \pi\pi$ decay
  amplitudes with improved Wilson fermion action in non-zero momentum frame in
  lattice QCD}},\ }\href {https://doi.org/10.1103/PhysRevD.98.114512}
  {\bibfield  {journal} {\bibinfo  {journal} {Phys. Rev. D}\ }\textbf {\bibinfo
  {volume} {98}},\ \bibinfo {pages} {114512} (\bibinfo {year} {2018})},\
  \Eprint {https://arxiv.org/abs/1809.03893} {arXiv:1809.03893 [hep-lat]}
  \BibitemShut {NoStop}%
\bibitem [{\citenamefont {Carrasco}\ \emph {et~al.}(2015)\citenamefont
  {Carrasco}, \citenamefont {Dimopoulos}, \citenamefont {Frezzotti},
  \citenamefont {Lubicz}, \citenamefont {Rossi}, \citenamefont {Simula},\ and\
  \citenamefont {Tarantino}}]{Carrasco:2015pra}%
  \BibitemOpen
  \bibfield  {author} {\bibinfo {author} {\bibfnamefont {N.}~\bibnamefont
  {Carrasco}}, \bibinfo {author} {\bibfnamefont {P.}~\bibnamefont
  {Dimopoulos}}, \bibinfo {author} {\bibfnamefont {R.}~\bibnamefont
  {Frezzotti}}, \bibinfo {author} {\bibfnamefont {V.}~\bibnamefont {Lubicz}},
  \bibinfo {author} {\bibfnamefont {G.~C.}\ \bibnamefont {Rossi}}, \bibinfo
  {author} {\bibfnamefont {S.}~\bibnamefont {Simula}},\ and\ \bibinfo {author}
  {\bibfnamefont {C.}~\bibnamefont {Tarantino}} (\bibinfo {collaboration}
  {ETM}),\ }\bibfield  {title} {\bibinfo {title} {{\ensuremath{\Delta}S=2 and
  \ensuremath{\Delta}C=2 bag parameters in the standard model and beyond from
  N$_f$=2+1+1 twisted-mass lattice QCD}},\ }\href
  {https://doi.org/10.1103/PhysRevD.92.034516} {\bibfield  {journal} {\bibinfo
  {journal} {Phys. Rev. D}\ }\textbf {\bibinfo {volume} {92}},\ \bibinfo
  {pages} {034516} (\bibinfo {year} {2015})},\ \Eprint
  {https://arxiv.org/abs/1505.06639} {arXiv:1505.06639 [hep-lat]} \BibitemShut
  {NoStop}%
\bibitem [{\citenamefont {Gupta}\ \emph {et~al.}(1993)\citenamefont {Gupta},
  \citenamefont {Daniel}, \citenamefont {Kilcup}, \citenamefont {Patel},\ and\
  \citenamefont {Sharpe}}]{Gupta:1992bd}%
  \BibitemOpen
  \bibfield  {author} {\bibinfo {author} {\bibfnamefont {R.}~\bibnamefont
  {Gupta}}, \bibinfo {author} {\bibfnamefont {D.}~\bibnamefont {Daniel}},
  \bibinfo {author} {\bibfnamefont {G.~W.}\ \bibnamefont {Kilcup}}, \bibinfo
  {author} {\bibfnamefont {A.}~\bibnamefont {Patel}},\ and\ \bibinfo {author}
  {\bibfnamefont {S.~R.}\ \bibnamefont {Sharpe}},\ }\bibfield  {title}
  {\bibinfo {title} {{The Kaon B parameter with Wilson fermions}},\ }\href
  {https://doi.org/10.1103/PhysRevD.47.5113} {\bibfield  {journal} {\bibinfo
  {journal} {Phys. Rev. D}\ }\textbf {\bibinfo {volume} {47}},\ \bibinfo
  {pages} {5113} (\bibinfo {year} {1993})},\ \Eprint
  {https://arxiv.org/abs/hep-lat/9210018} {arXiv:hep-lat/9210018} \BibitemShut
  {NoStop}%
\bibitem [{\citenamefont {Becirevic}\ \emph {et~al.}(2002)\citenamefont
  {Becirevic}, \citenamefont {Boucaud}, \citenamefont {Gimenez}, \citenamefont
  {Lubicz}, \citenamefont {Martinelli},\ and\ \citenamefont
  {Papinutto}}]{Becirevic:2001re}%
  \BibitemOpen
  \bibfield  {author} {\bibinfo {author} {\bibfnamefont {D.}~\bibnamefont
  {Becirevic}}, \bibinfo {author} {\bibfnamefont {P.}~\bibnamefont {Boucaud}},
  \bibinfo {author} {\bibfnamefont {V.}~\bibnamefont {Gimenez}}, \bibinfo
  {author} {\bibfnamefont {V.}~\bibnamefont {Lubicz}}, \bibinfo {author}
  {\bibfnamefont {G.}~\bibnamefont {Martinelli}},\ and\ \bibinfo {author}
  {\bibfnamefont {M.}~\bibnamefont {Papinutto}},\ }\bibfield  {title} {\bibinfo
  {title} {{Matrix elements of $\Delta$S = 2 operators with Wilson fermions}},\
  }\href {https://doi.org/10.1016/S0920-5632(01)01703-0} {\bibfield  {journal}
  {\bibinfo  {journal} {Nucl. Phys. B Proc. Suppl.}\ }\textbf {\bibinfo
  {volume} {106}},\ \bibinfo {pages} {326} (\bibinfo {year} {2002})},\ \Eprint
  {https://arxiv.org/abs/hep-lat/0110006} {arXiv:hep-lat/0110006} \BibitemShut
  {NoStop}%
\bibitem [{\citenamefont {Aoki}\ \emph {et~al.}(2011)\citenamefont {Aoki} \emph
  {et~al.}}]{Aoki:2010pe}%
  \BibitemOpen
  \bibfield  {author} {\bibinfo {author} {\bibfnamefont {Y.}~\bibnamefont
  {Aoki}} \emph {et~al.},\ }\bibfield  {title} {\bibinfo {title} {{Continuum
  Limit of $B_K$ from 2+1 Flavor Domain Wall QCD}},\ }\href
  {https://doi.org/10.1103/PhysRevD.84.014503} {\bibfield  {journal} {\bibinfo
  {journal} {Phys. Rev. D}\ }\textbf {\bibinfo {volume} {84}},\ \bibinfo
  {pages} {014503} (\bibinfo {year} {2011})},\ \Eprint
  {https://arxiv.org/abs/1012.4178} {arXiv:1012.4178 [hep-lat]} \BibitemShut
  {NoStop}%
\bibitem [{\citenamefont {Constantinou}\ \emph
  {et~al.}(2011{\natexlab{a}})\citenamefont {Constantinou} \emph
  {et~al.}}]{ETM:2010ubf}%
  \BibitemOpen
  \bibfield  {author} {\bibinfo {author} {\bibfnamefont {M.}~\bibnamefont
  {Constantinou}} \emph {et~al.} (\bibinfo {collaboration} {ETM}),\ }\bibfield
  {title} {\bibinfo {title} {{$B_K$-parameter from $N_f$ = 2 twisted mass
  lattice QCD}},\ }\href {https://doi.org/10.1103/PhysRevD.83.014505}
  {\bibfield  {journal} {\bibinfo  {journal} {Phys. Rev. D}\ }\textbf {\bibinfo
  {volume} {83}},\ \bibinfo {pages} {014505} (\bibinfo {year}
  {2011}{\natexlab{a}})},\ \Eprint {https://arxiv.org/abs/1009.5606}
  {arXiv:1009.5606 [hep-lat]} \BibitemShut {NoStop}%
\bibitem [{\citenamefont {Aebischer}\ \emph {et~al.}(2020)\citenamefont
  {Aebischer}, \citenamefont {Bobeth}, \citenamefont {Buras},\ and\
  \citenamefont {Kumar}}]{Aebischer:2020dsw}%
  \BibitemOpen
  \bibfield  {author} {\bibinfo {author} {\bibfnamefont {J.}~\bibnamefont
  {Aebischer}}, \bibinfo {author} {\bibfnamefont {C.}~\bibnamefont {Bobeth}},
  \bibinfo {author} {\bibfnamefont {A.~J.}\ \bibnamefont {Buras}},\ and\
  \bibinfo {author} {\bibfnamefont {J.}~\bibnamefont {Kumar}},\ }\bibfield
  {title} {\bibinfo {title} {{SMEFT ATLAS of $\Delta$F = 2 transitions}},\
  }\href {https://doi.org/10.1007/JHEP12(2020)187} {\bibfield  {journal}
  {\bibinfo  {journal} {JHEP}\ }\textbf {\bibinfo {volume} {12}},\ \bibinfo
  {pages} {187}},\ \Eprint {https://arxiv.org/abs/2009.07276} {arXiv:2009.07276
  [hep-ph]} \BibitemShut {NoStop}%
\bibitem [{\citenamefont {Suzuki}\ \emph {et~al.}(2020)\citenamefont {Suzuki},
  \citenamefont {Taniguchi}, \citenamefont {Suzuki},\ and\ \citenamefont
  {Kanaya}}]{Suzuki:2020zue}%
  \BibitemOpen
  \bibfield  {author} {\bibinfo {author} {\bibfnamefont {A.}~\bibnamefont
  {Suzuki}}, \bibinfo {author} {\bibfnamefont {Y.}~\bibnamefont {Taniguchi}},
  \bibinfo {author} {\bibfnamefont {H.}~\bibnamefont {Suzuki}},\ and\ \bibinfo
  {author} {\bibfnamefont {K.}~\bibnamefont {Kanaya}},\ }\bibfield  {title}
  {\bibinfo {title} {{Four quark operators for kaon bag parameter with gradient
  flow}},\ }\href {https://doi.org/10.1103/PhysRevD.102.034508} {\bibfield
  {journal} {\bibinfo  {journal} {Phys. Rev. D}\ }\textbf {\bibinfo {volume}
  {102}},\ \bibinfo {pages} {034508} (\bibinfo {year} {2020})},\ \Eprint
  {https://arxiv.org/abs/2006.06999} {arXiv:2006.06999 [hep-lat]} \BibitemShut
  {NoStop}%
\bibitem [{\citenamefont {Neubert}\ and\ \citenamefont
  {Sachrajda}(1997)}]{Neubert:1996we}%
  \BibitemOpen
  \bibfield  {author} {\bibinfo {author} {\bibfnamefont {M.}~\bibnamefont
  {Neubert}}\ and\ \bibinfo {author} {\bibfnamefont {C.~T.}\ \bibnamefont
  {Sachrajda}},\ }\bibfield  {title} {\bibinfo {title} {{Spectator effects in
  inclusive decays of beauty hadrons}},\ }\href
  {https://doi.org/10.1016/S0550-3213(96)00559-7} {\bibfield  {journal}
  {\bibinfo  {journal} {Nucl. Phys. B}\ }\textbf {\bibinfo {volume} {483}},\
  \bibinfo {pages} {339} (\bibinfo {year} {1997})},\ \Eprint
  {https://arxiv.org/abs/hep-ph/9603202} {arXiv:hep-ph/9603202} \BibitemShut
  {NoStop}%
\bibitem [{\citenamefont {Lenz}(2015)}]{Lenz:2014jha}%
  \BibitemOpen
  \bibfield  {author} {\bibinfo {author} {\bibfnamefont {A.}~\bibnamefont
  {Lenz}},\ }\bibfield  {title} {\bibinfo {title} {{Lifetimes and heavy quark
  expansion}},\ }\href {https://doi.org/10.1142/S0217751X15430058} {\bibfield
  {journal} {\bibinfo  {journal} {Int. J. Mod. Phys. A}\ }\textbf {\bibinfo
  {volume} {30}},\ \bibinfo {pages} {1543005} (\bibinfo {year} {2015})},\
  \Eprint {https://arxiv.org/abs/1405.3601} {arXiv:1405.3601 [hep-ph]}
  \BibitemShut {NoStop}%
\bibitem [{\citenamefont {Gabbiani}\ \emph {et~al.}(1996)\citenamefont
  {Gabbiani}, \citenamefont {Gabrielli}, \citenamefont {Masiero},\ and\
  \citenamefont {Silvestrini}}]{Gabbiani:1996hi}%
  \BibitemOpen
  \bibfield  {author} {\bibinfo {author} {\bibfnamefont {F.}~\bibnamefont
  {Gabbiani}}, \bibinfo {author} {\bibfnamefont {E.}~\bibnamefont {Gabrielli}},
  \bibinfo {author} {\bibfnamefont {A.}~\bibnamefont {Masiero}},\ and\ \bibinfo
  {author} {\bibfnamefont {L.}~\bibnamefont {Silvestrini}},\ }\bibfield
  {title} {\bibinfo {title} {{A Complete analysis of FCNC and CP constraints in
  general SUSY extensions of the standard model}},\ }\href
  {https://doi.org/10.1016/0550-3213(96)00390-2} {\bibfield  {journal}
  {\bibinfo  {journal} {Nucl. Phys. B}\ }\textbf {\bibinfo {volume} {477}},\
  \bibinfo {pages} {321} (\bibinfo {year} {1996})},\ \Eprint
  {https://arxiv.org/abs/hep-ph/9604387} {arXiv:hep-ph/9604387} \BibitemShut
  {NoStop}%
\bibitem [{\citenamefont {Ciuchini}\ \emph {et~al.}(1998)\citenamefont
  {Ciuchini}, \citenamefont {Franco}, \citenamefont {Lubicz}, \citenamefont
  {Martinelli}, \citenamefont {Scimemi},\ and\ \citenamefont
  {Silvestrini}}]{Ciuchini:1997bw}%
  \BibitemOpen
  \bibfield  {author} {\bibinfo {author} {\bibfnamefont {M.}~\bibnamefont
  {Ciuchini}}, \bibinfo {author} {\bibfnamefont {E.}~\bibnamefont {Franco}},
  \bibinfo {author} {\bibfnamefont {V.}~\bibnamefont {Lubicz}}, \bibinfo
  {author} {\bibfnamefont {G.}~\bibnamefont {Martinelli}}, \bibinfo {author}
  {\bibfnamefont {I.}~\bibnamefont {Scimemi}},\ and\ \bibinfo {author}
  {\bibfnamefont {L.}~\bibnamefont {Silvestrini}},\ }\bibfield  {title}
  {\bibinfo {title} {{Next-to-leading order QCD corrections to Delta F = 2
  effective Hamiltonians}},\ }\href
  {https://doi.org/10.1016/S0550-3213(98)00161-8} {\bibfield  {journal}
  {\bibinfo  {journal} {Nucl. Phys. B}\ }\textbf {\bibinfo {volume} {523}},\
  \bibinfo {pages} {501} (\bibinfo {year} {1998})},\ \Eprint
  {https://arxiv.org/abs/hep-ph/9711402} {arXiv:hep-ph/9711402} \BibitemShut
  {NoStop}%
\bibitem [{\citenamefont {Capitani}\ and\ \citenamefont
  {Giusti}(2000)}]{Capitani:2000da}%
  \BibitemOpen
  \bibfield  {author} {\bibinfo {author} {\bibfnamefont {S.}~\bibnamefont
  {Capitani}}\ and\ \bibinfo {author} {\bibfnamefont {L.}~\bibnamefont
  {Giusti}},\ }\bibfield  {title} {\bibinfo {title} {{Perturbative
  renormalization of weak Hamiltonian four fermion operators with overlap
  fermions}},\ }\href {https://doi.org/10.1103/PhysRevD.62.114506} {\bibfield
  {journal} {\bibinfo  {journal} {Phys. Rev. D}\ }\textbf {\bibinfo {volume}
  {62}},\ \bibinfo {pages} {114506} (\bibinfo {year} {2000})},\ \Eprint
  {https://arxiv.org/abs/hep-lat/0007011} {arXiv:hep-lat/0007011} \BibitemShut
  {NoStop}%
\bibitem [{\citenamefont {Nakamura}\ and\ \citenamefont
  {Kuramashi}(2006)}]{Nakamura:2006zx}%
  \BibitemOpen
  \bibfield  {author} {\bibinfo {author} {\bibfnamefont {Y.}~\bibnamefont
  {Nakamura}}\ and\ \bibinfo {author} {\bibfnamefont {Y.}~\bibnamefont
  {Kuramashi}},\ }\bibfield  {title} {\bibinfo {title} {{Perturbative
  renormalization factors for generic $\Delta$s = 2 four-quark operators in
  domain-wall QCD with improved gauge action}},\ }\href
  {https://doi.org/10.1103/PhysRevD.73.094502} {\bibfield  {journal} {\bibinfo
  {journal} {Phys. Rev. D}\ }\textbf {\bibinfo {volume} {73}},\ \bibinfo
  {pages} {094502} (\bibinfo {year} {2006})},\ \Eprint
  {https://arxiv.org/abs/hep-lat/0603012} {arXiv:hep-lat/0603012} \BibitemShut
  {NoStop}%
\bibitem [{\citenamefont {Taniguchi}(2012)}]{Taniguchi:2012xm}%
  \BibitemOpen
  \bibfield  {author} {\bibinfo {author} {\bibfnamefont {Y.}~\bibnamefont
  {Taniguchi}},\ }\bibfield  {title} {\bibinfo {title} {{Perturbative
  renormalization factors of four-quark operators for improved Wilson fermion
  action and Iwasaki gauge action}},\ }\href
  {https://doi.org/10.1007/JHEP04(2012)143} {\bibfield  {journal} {\bibinfo
  {journal} {JHEP}\ }\textbf {\bibinfo {volume} {04}},\ \bibinfo {pages}
  {143}},\ \Eprint {https://arxiv.org/abs/1203.1401} {arXiv:1203.1401
  [hep-lat]} \BibitemShut {NoStop}%
\bibitem [{\citenamefont {Donini}\ \emph {et~al.}(1999)\citenamefont {Donini},
  \citenamefont {Gimenez}, \citenamefont {Martinelli}, \citenamefont {Talevi},\
  and\ \citenamefont {Vladikas}}]{Donini:1999sf}%
  \BibitemOpen
  \bibfield  {author} {\bibinfo {author} {\bibfnamefont {A.}~\bibnamefont
  {Donini}}, \bibinfo {author} {\bibfnamefont {V.}~\bibnamefont {Gimenez}},
  \bibinfo {author} {\bibfnamefont {G.}~\bibnamefont {Martinelli}}, \bibinfo
  {author} {\bibfnamefont {M.}~\bibnamefont {Talevi}},\ and\ \bibinfo {author}
  {\bibfnamefont {A.}~\bibnamefont {Vladikas}},\ }\bibfield  {title} {\bibinfo
  {title} {{Nonperturbative renormalization of lattice four fermion operators
  without power subtractions}},\ }\href {https://doi.org/10.1007/s100529900097}
  {\bibfield  {journal} {\bibinfo  {journal} {Eur. Phys. J. C}\ }\textbf
  {\bibinfo {volume} {10}},\ \bibinfo {pages} {121} (\bibinfo {year} {1999})},\
  \Eprint {https://arxiv.org/abs/hep-lat/9902030} {arXiv:hep-lat/9902030}
  \BibitemShut {NoStop}%
\bibitem [{\citenamefont {Becirevic}\ \emph {et~al.}(2003)\citenamefont
  {Becirevic}, \citenamefont {Gimenez}, \citenamefont {Lubicz}, \citenamefont
  {Martinelli}, \citenamefont {Papinutto},\ and\ \citenamefont
  {Reyes}}]{Becirevic:2002qr}%
  \BibitemOpen
  \bibfield  {author} {\bibinfo {author} {\bibfnamefont {D.}~\bibnamefont
  {Becirevic}}, \bibinfo {author} {\bibfnamefont {V.}~\bibnamefont {Gimenez}},
  \bibinfo {author} {\bibfnamefont {V.}~\bibnamefont {Lubicz}}, \bibinfo
  {author} {\bibfnamefont {G.}~\bibnamefont {Martinelli}}, \bibinfo {author}
  {\bibfnamefont {M.}~\bibnamefont {Papinutto}},\ and\ \bibinfo {author}
  {\bibfnamefont {J.}~\bibnamefont {Reyes}} (\bibinfo {collaboration}
  {SPQcdR}),\ }\bibfield  {title} {\bibinfo {title} {{Nonperturbative
  renormalization of four fermion operators and B0 - anti-B0 mixing with Wilson
  fermions}},\ }\href {https://doi.org/10.1016/S0920-5632(03)01641-4}
  {\bibfield  {journal} {\bibinfo  {journal} {Nucl. Phys. B Proc. Suppl.}\
  }\textbf {\bibinfo {volume} {119}},\ \bibinfo {pages} {619} (\bibinfo {year}
  {2003})},\ \Eprint {https://arxiv.org/abs/hep-lat/0209131}
  {arXiv:hep-lat/0209131} \BibitemShut {NoStop}%
\bibitem [{\citenamefont {Constantinou}\ \emph
  {et~al.}(2011{\natexlab{b}})\citenamefont {Constantinou}, \citenamefont
  {Dimopoulos}, \citenamefont {Frezzotti}, \citenamefont {Lubicz},
  \citenamefont {Panagopoulos}, \citenamefont {Skouroupathis},\ and\
  \citenamefont {Stylianou}}]{Constantinou:2010zs}%
  \BibitemOpen
  \bibfield  {author} {\bibinfo {author} {\bibfnamefont {M.}~\bibnamefont
  {Constantinou}}, \bibinfo {author} {\bibfnamefont {P.}~\bibnamefont
  {Dimopoulos}}, \bibinfo {author} {\bibfnamefont {R.}~\bibnamefont
  {Frezzotti}}, \bibinfo {author} {\bibfnamefont {V.}~\bibnamefont {Lubicz}},
  \bibinfo {author} {\bibfnamefont {H.}~\bibnamefont {Panagopoulos}}, \bibinfo
  {author} {\bibfnamefont {A.}~\bibnamefont {Skouroupathis}},\ and\ \bibinfo
  {author} {\bibfnamefont {F.}~\bibnamefont {Stylianou}},\ }\bibfield  {title}
  {\bibinfo {title} {{Perturbative renormalization factors and
  $\mathcal{O}(a^2)$ corrections for lattice 4-fermion operators with improved
  fermion/gluon actions}},\ }\href {https://doi.org/10.1103/PhysRevD.83.074503}
  {\bibfield  {journal} {\bibinfo  {journal} {Phys. Rev. D}\ }\textbf {\bibinfo
  {volume} {83}},\ \bibinfo {pages} {074503} (\bibinfo {year}
  {2011}{\natexlab{b}})},\ \Eprint {https://arxiv.org/abs/1011.6059}
  {arXiv:1011.6059 [hep-lat]} \BibitemShut {NoStop}%
\bibitem [{\citenamefont {Lehner}\ and\ \citenamefont
  {Sturm}(2011)}]{Lehner:2011fz}%
  \BibitemOpen
  \bibfield  {author} {\bibinfo {author} {\bibfnamefont {C.}~\bibnamefont
  {Lehner}}\ and\ \bibinfo {author} {\bibfnamefont {C.}~\bibnamefont {Sturm}},\
  }\bibfield  {title} {\bibinfo {title} {{Matching factors for Delta S=1
  four-quark operators in RI/SMOM schemes}},\ }\href
  {https://doi.org/10.1103/PhysRevD.84.014001} {\bibfield  {journal} {\bibinfo
  {journal} {Phys. Rev. D}\ }\textbf {\bibinfo {volume} {84}},\ \bibinfo
  {pages} {014001} (\bibinfo {year} {2011})},\ \Eprint
  {https://arxiv.org/abs/1104.4948} {arXiv:1104.4948 [hep-ph]} \BibitemShut
  {NoStop}%
\bibitem [{\citenamefont {Boyle}\ \emph {et~al.}(2017)\citenamefont {Boyle},
  \citenamefont {Garron}, \citenamefont {Hudspith}, \citenamefont {Lehner},\
  and\ \citenamefont {Lytle}}]{Boyle:2017skn}%
  \BibitemOpen
  \bibfield  {author} {\bibinfo {author} {\bibfnamefont {P.~A.}\ \bibnamefont
  {Boyle}}, \bibinfo {author} {\bibfnamefont {N.}~\bibnamefont {Garron}},
  \bibinfo {author} {\bibfnamefont {R.~J.}\ \bibnamefont {Hudspith}}, \bibinfo
  {author} {\bibfnamefont {C.}~\bibnamefont {Lehner}},\ and\ \bibinfo {author}
  {\bibfnamefont {A.~T.}\ \bibnamefont {Lytle}} (\bibinfo {collaboration} {RBC,
  UKQCD}),\ }\bibfield  {title} {\bibinfo {title} {{Neutral kaon mixing beyond
  the Standard Model with n$_{f}$ = 2 + 1 chiral fermions. Part 2: non
  perturbative renormalisation of the $\Delta F=2$ four-quark operators}},\
  }\href {https://doi.org/10.1007/JHEP10(2017)054} {\bibfield  {journal}
  {\bibinfo  {journal} {JHEP}\ }\textbf {\bibinfo {volume} {10}},\ \bibinfo
  {pages} {054}},\ \Eprint {https://arxiv.org/abs/1708.03552} {arXiv:1708.03552
  [hep-lat]} \BibitemShut {NoStop}%
\bibitem [{\citenamefont {Garron}(2018)}]{Garron:2018tst}%
  \BibitemOpen
  \bibfield  {author} {\bibinfo {author} {\bibfnamefont {N.}~\bibnamefont
  {Garron}},\ }\bibfield  {title} {\bibinfo {title} {{Fierz transformations and
  renormalization schemes for fourquark operators}},\ }\href
  {https://doi.org/10.1051/epjconf/201817510005} {\bibfield  {journal}
  {\bibinfo  {journal} {EPJ Web Conf.}\ }\textbf {\bibinfo {volume} {175}},\
  \bibinfo {pages} {10005} (\bibinfo {year} {2018})}\BibitemShut {NoStop}%
\bibitem [{\citenamefont {Boyle}\ \emph {et~al.}(2024)\citenamefont {Boyle},
  \citenamefont {Erben}, \citenamefont {Flynn}, \citenamefont {Garron},
  \citenamefont {Kettle}, \citenamefont {Mukherjee},\ and\ \citenamefont
  {Tsang}}]{Boyle:2024gge}%
  \BibitemOpen
  \bibfield  {author} {\bibinfo {author} {\bibfnamefont {P.~A.}\ \bibnamefont
  {Boyle}}, \bibinfo {author} {\bibfnamefont {F.}~\bibnamefont {Erben}},
  \bibinfo {author} {\bibfnamefont {J.~M.}\ \bibnamefont {Flynn}}, \bibinfo
  {author} {\bibfnamefont {N.}~\bibnamefont {Garron}}, \bibinfo {author}
  {\bibfnamefont {J.}~\bibnamefont {Kettle}}, \bibinfo {author} {\bibfnamefont
  {R.}~\bibnamefont {Mukherjee}},\ and\ \bibinfo {author} {\bibfnamefont
  {J.~T.}\ \bibnamefont {Tsang}},\ }\bibfield  {title} {\bibinfo {title} {{Kaon
  mixing beyond the standard model with physical masses}},\ }\href@noop {} {\
  (\bibinfo {year} {2024})},\ \Eprint {https://arxiv.org/abs/2404.02297}
  {arXiv:2404.02297 [hep-lat]} \BibitemShut {NoStop}%
\bibitem [{\citenamefont {Guagnelli}\ \emph {et~al.}(2003)\citenamefont
  {Guagnelli}, \citenamefont {Heitger}, \citenamefont {Pena}, \citenamefont
  {Sint},\ and\ \citenamefont {Vladikas}}]{Guagnelli:2002rw}%
  \BibitemOpen
  \bibfield  {author} {\bibinfo {author} {\bibfnamefont {M.}~\bibnamefont
  {Guagnelli}}, \bibinfo {author} {\bibfnamefont {J.}~\bibnamefont {Heitger}},
  \bibinfo {author} {\bibfnamefont {C.}~\bibnamefont {Pena}}, \bibinfo {author}
  {\bibfnamefont {S.}~\bibnamefont {Sint}},\ and\ \bibinfo {author}
  {\bibfnamefont {A.}~\bibnamefont {Vladikas}} (\bibinfo {collaboration}
  {ALPHA}),\ }\bibfield  {title} {\bibinfo {title} {{Nonperturbative scale
  evolution of four fermion operators}},\ }\href
  {https://doi.org/10.1016/S0920-5632(03)01578-0} {\bibfield  {journal}
  {\bibinfo  {journal} {Nucl. Phys. B Proc. Suppl.}\ }\textbf {\bibinfo
  {volume} {119}},\ \bibinfo {pages} {436} (\bibinfo {year} {2003})},\ \Eprint
  {https://arxiv.org/abs/hep-lat/0209046} {arXiv:hep-lat/0209046} \BibitemShut
  {NoStop}%
\bibitem [{\citenamefont {Guagnelli}\ \emph {et~al.}(2006)\citenamefont
  {Guagnelli}, \citenamefont {Heitger}, \citenamefont {Pena}, \citenamefont
  {Sint},\ and\ \citenamefont {Vladikas}}]{Guagnelli:2005zc}%
  \BibitemOpen
  \bibfield  {author} {\bibinfo {author} {\bibfnamefont {M.}~\bibnamefont
  {Guagnelli}}, \bibinfo {author} {\bibfnamefont {J.}~\bibnamefont {Heitger}},
  \bibinfo {author} {\bibfnamefont {C.}~\bibnamefont {Pena}}, \bibinfo {author}
  {\bibfnamefont {S.}~\bibnamefont {Sint}},\ and\ \bibinfo {author}
  {\bibfnamefont {A.}~\bibnamefont {Vladikas}} (\bibinfo {collaboration}
  {ALPHA}),\ }\bibfield  {title} {\bibinfo {title} {{Non-perturbative
  renormalization of left-left four-fermion operators in quenched lattice
  QCD}},\ }\href {https://doi.org/10.1088/1126-6708/2006/03/088} {\bibfield
  {journal} {\bibinfo  {journal} {JHEP}\ }\textbf {\bibinfo {volume} {03}},\
  \bibinfo {pages} {088}},\ \Eprint {https://arxiv.org/abs/hep-lat/0505002}
  {arXiv:hep-lat/0505002} \BibitemShut {NoStop}%
\bibitem [{\citenamefont {Palombi}\ \emph
  {et~al.}(2006{\natexlab{a}})\citenamefont {Palombi}, \citenamefont {Pena},\
  and\ \citenamefont {Sint}}]{Palombi:2005zd}%
  \BibitemOpen
  \bibfield  {author} {\bibinfo {author} {\bibfnamefont {F.}~\bibnamefont
  {Palombi}}, \bibinfo {author} {\bibfnamefont {C.}~\bibnamefont {Pena}},\ and\
  \bibinfo {author} {\bibfnamefont {S.}~\bibnamefont {Sint}},\ }\bibfield
  {title} {\bibinfo {title} {{A Perturbative study of two four-quark operators
  in finite volume renormalization schemes}},\ }\href
  {https://doi.org/10.1088/1126-6708/2006/03/089} {\bibfield  {journal}
  {\bibinfo  {journal} {JHEP}\ }\textbf {\bibinfo {volume} {03}},\ \bibinfo
  {pages} {089}},\ \Eprint {https://arxiv.org/abs/hep-lat/0505003}
  {arXiv:hep-lat/0505003} \BibitemShut {NoStop}%
\bibitem [{\citenamefont {Dimopoulos}\ \emph {et~al.}(2006)\citenamefont
  {Dimopoulos}, \citenamefont {Giusti}, \citenamefont {Hernandez},
  \citenamefont {Palombi}, \citenamefont {Pena}, \citenamefont {Vladikas},
  \citenamefont {Wennekers},\ and\ \citenamefont {Wittig}}]{Dimopoulos:2006ma}%
  \BibitemOpen
  \bibfield  {author} {\bibinfo {author} {\bibfnamefont {P.}~\bibnamefont
  {Dimopoulos}}, \bibinfo {author} {\bibfnamefont {L.}~\bibnamefont {Giusti}},
  \bibinfo {author} {\bibfnamefont {P.}~\bibnamefont {Hernandez}}, \bibinfo
  {author} {\bibfnamefont {F.}~\bibnamefont {Palombi}}, \bibinfo {author}
  {\bibfnamefont {C.}~\bibnamefont {Pena}}, \bibinfo {author} {\bibfnamefont
  {A.}~\bibnamefont {Vladikas}}, \bibinfo {author} {\bibfnamefont
  {J.}~\bibnamefont {Wennekers}},\ and\ \bibinfo {author} {\bibfnamefont
  {H.}~\bibnamefont {Wittig}},\ }\bibfield  {title} {\bibinfo {title}
  {{Non-perturbative renormalisation of left-left four-fermion operators with
  Neuberger fermions}},\ }\href
  {https://doi.org/10.1016/j.physletb.2006.08.009} {\bibfield  {journal}
  {\bibinfo  {journal} {Phys. Lett. B}\ }\textbf {\bibinfo {volume} {641}},\
  \bibinfo {pages} {118} (\bibinfo {year} {2006})},\ \Eprint
  {https://arxiv.org/abs/hep-lat/0607028} {arXiv:hep-lat/0607028} \BibitemShut
  {NoStop}%
\bibitem [{\citenamefont {Palombi}\ \emph
  {et~al.}(2006{\natexlab{b}})\citenamefont {Palombi}, \citenamefont
  {Papinutto}, \citenamefont {Pena},\ and\ \citenamefont
  {Wittig}}]{Palombi:2006pu}%
  \BibitemOpen
  \bibfield  {author} {\bibinfo {author} {\bibfnamefont {F.}~\bibnamefont
  {Palombi}}, \bibinfo {author} {\bibfnamefont {M.}~\bibnamefont {Papinutto}},
  \bibinfo {author} {\bibfnamefont {C.}~\bibnamefont {Pena}},\ and\ \bibinfo
  {author} {\bibfnamefont {H.}~\bibnamefont {Wittig}},\ }\bibfield  {title}
  {\bibinfo {title} {{A Strategy for implementing non-perturbative
  renormalisation of heavy-light four-quark operators in the static
  approximation}},\ }\href {https://doi.org/10.1088/1126-6708/2006/08/017}
  {\bibfield  {journal} {\bibinfo  {journal} {JHEP}\ }\textbf {\bibinfo
  {volume} {08}},\ \bibinfo {pages} {017}},\ \Eprint
  {https://arxiv.org/abs/hep-lat/0604014} {arXiv:hep-lat/0604014} \BibitemShut
  {NoStop}%
\bibitem [{\citenamefont {Palombi}\ \emph {et~al.}(2007)\citenamefont
  {Palombi}, \citenamefont {Papinutto}, \citenamefont {Pena},\ and\
  \citenamefont {Wittig}}]{Palombi:2007dr}%
  \BibitemOpen
  \bibfield  {author} {\bibinfo {author} {\bibfnamefont {F.}~\bibnamefont
  {Palombi}}, \bibinfo {author} {\bibfnamefont {M.}~\bibnamefont {Papinutto}},
  \bibinfo {author} {\bibfnamefont {C.}~\bibnamefont {Pena}},\ and\ \bibinfo
  {author} {\bibfnamefont {H.}~\bibnamefont {Wittig}},\ }\bibfield  {title}
  {\bibinfo {title} {{Non-perturbative renormalization of static-light
  four-fermion operators in quenched lattice QCD}},\ }\href
  {https://doi.org/10.1088/1126-6708/2007/09/062} {\bibfield  {journal}
  {\bibinfo  {journal} {JHEP}\ }\textbf {\bibinfo {volume} {09}},\ \bibinfo
  {pages} {062}},\ \Eprint {https://arxiv.org/abs/0706.4153} {arXiv:0706.4153
  [hep-lat]} \BibitemShut {NoStop}%
\bibitem [{\citenamefont {Dimopoulos}\ \emph {et~al.}(2008)\citenamefont
  {Dimopoulos}, \citenamefont {Herdoiza}, \citenamefont {Palombi},
  \citenamefont {Papinutto}, \citenamefont {Pena}, \citenamefont {Vladikas},\
  and\ \citenamefont {Wittig}}]{Dimopoulos:2007ht}%
  \BibitemOpen
  \bibfield  {author} {\bibinfo {author} {\bibfnamefont {P.}~\bibnamefont
  {Dimopoulos}}, \bibinfo {author} {\bibfnamefont {G.}~\bibnamefont
  {Herdoiza}}, \bibinfo {author} {\bibfnamefont {F.}~\bibnamefont {Palombi}},
  \bibinfo {author} {\bibfnamefont {M.}~\bibnamefont {Papinutto}}, \bibinfo
  {author} {\bibfnamefont {C.}~\bibnamefont {Pena}}, \bibinfo {author}
  {\bibfnamefont {A.}~\bibnamefont {Vladikas}},\ and\ \bibinfo {author}
  {\bibfnamefont {H.}~\bibnamefont {Wittig}} (\bibinfo {collaboration}
  {ALPHA}),\ }\bibfield  {title} {\bibinfo {title} {{Non-perturbative
  renormalisation of $\Delta$F=2 four-fermion operators in two-flavour QCD}},\
  }\href {https://doi.org/10.1088/1126-6708/2008/05/065} {\bibfield  {journal}
  {\bibinfo  {journal} {JHEP}\ }\textbf {\bibinfo {volume} {05}},\ \bibinfo
  {pages} {065}},\ \Eprint {https://arxiv.org/abs/0712.2429} {arXiv:0712.2429
  [hep-lat]} \BibitemShut {NoStop}%
\bibitem [{\citenamefont {Papinutto}\ \emph {et~al.}(2017)\citenamefont
  {Papinutto}, \citenamefont {Pena},\ and\ \citenamefont
  {Preti}}]{Papinutto:2016xpq}%
  \BibitemOpen
  \bibfield  {author} {\bibinfo {author} {\bibfnamefont {M.}~\bibnamefont
  {Papinutto}}, \bibinfo {author} {\bibfnamefont {C.}~\bibnamefont {Pena}},\
  and\ \bibinfo {author} {\bibfnamefont {D.}~\bibnamefont {Preti}},\ }\bibfield
   {title} {\bibinfo {title} {{On the perturbative renormalization of
  four-quark operators for new physics}},\ }\href
  {https://doi.org/10.1140/epjc/s10052-017-4930-6} {\bibfield  {journal}
  {\bibinfo  {journal} {Eur. Phys. J. C}\ }\textbf {\bibinfo {volume} {77}},\
  \bibinfo {pages} {376} (\bibinfo {year} {2017})},\ \bibinfo {note} {[Erratum:
  Eur.Phys.J.C 78, 21 (2018)]},\ \Eprint {https://arxiv.org/abs/1612.06461}
  {arXiv:1612.06461 [hep-lat]} \BibitemShut {NoStop}%
\bibitem [{\citenamefont {Dalla~Brida}\ \emph {et~al.}(2016)\citenamefont
  {Dalla~Brida}, \citenamefont {Papinutto},\ and\ \citenamefont
  {Vilaseca}}]{DallaBrida:2016zxs}%
  \BibitemOpen
  \bibfield  {author} {\bibinfo {author} {\bibfnamefont {M.}~\bibnamefont
  {Dalla~Brida}}, \bibinfo {author} {\bibfnamefont {M.}~\bibnamefont
  {Papinutto}},\ and\ \bibinfo {author} {\bibfnamefont {P.}~\bibnamefont
  {Vilaseca}},\ }\bibfield  {title} {\bibinfo {title} {{Perturbative
  renormalization of $\Delta F = 2$ four-fermion operators with the chirally
  rotated Schr\"odinger functional}},\ }\href@noop {} {\  (\bibinfo {year}
  {2016})},\ \Eprint {https://arxiv.org/abs/1605.09053} {arXiv:1605.09053
  [hep-lat]} \BibitemShut {NoStop}%
\bibitem [{\citenamefont {Dimopoulos}\ \emph {et~al.}(2018)\citenamefont
  {Dimopoulos}, \citenamefont {Herdo\'\i{}za}, \citenamefont {Papinutto},
  \citenamefont {Pena}, \citenamefont {Preti},\ and\ \citenamefont
  {Vladikas}}]{Dimopoulos:2018zef}%
  \BibitemOpen
  \bibfield  {author} {\bibinfo {author} {\bibfnamefont {P.}~\bibnamefont
  {Dimopoulos}}, \bibinfo {author} {\bibfnamefont {G.}~\bibnamefont
  {Herdo\'\i{}za}}, \bibinfo {author} {\bibfnamefont {M.}~\bibnamefont
  {Papinutto}}, \bibinfo {author} {\bibfnamefont {C.}~\bibnamefont {Pena}},
  \bibinfo {author} {\bibfnamefont {D.}~\bibnamefont {Preti}},\ and\ \bibinfo
  {author} {\bibfnamefont {A.}~\bibnamefont {Vladikas}} (\bibinfo
  {collaboration} {ALPHA}),\ }\bibfield  {title} {\bibinfo {title}
  {{Non-Perturbative Renormalisation and Running of BSM Four-Quark Operators in
  $N_f = 2$ QCD}},\ }\href {https://doi.org/10.1140/epjc/s10052-018-6002-y}
  {\bibfield  {journal} {\bibinfo  {journal} {Eur. Phys. J. C}\ }\textbf
  {\bibinfo {volume} {78}},\ \bibinfo {pages} {579} (\bibinfo {year} {2018})},\
  \Eprint {https://arxiv.org/abs/1801.09455} {arXiv:1801.09455 [hep-lat]}
  \BibitemShut {NoStop}%
\bibitem [{\citenamefont {Marinelli}\ \emph {et~al.}(2023)\citenamefont
  {Marinelli} \emph {et~al.}}]{Marinelli:2023}%
  \BibitemOpen
  \bibfield  {author} {\bibinfo {author} {\bibfnamefont {R.}~\bibnamefont
  {Marinelli}} \emph {et~al.},\ }\bibfield  {title} {\bibinfo {title}
  {{Operator mixing and non-perturbative running of $\Delta$F=2 four-fermion
  operators}},\ }\href@noop {} {\bibfield  {journal} {\bibinfo  {journal}
  {PoS}\ }\textbf {\bibinfo {volume} {LATTICE2023}},\ \bibinfo {pages} {270}
  (\bibinfo {year} {2023})}\BibitemShut {NoStop}%
\bibitem [{\citenamefont {Costa}\ \emph
  {et~al.}(2021{\natexlab{a}})\citenamefont {Costa}, \citenamefont
  {Karpasitis}, \citenamefont {Pafitis}, \citenamefont {Panagopoulos},
  \citenamefont {Panagopoulos}, \citenamefont {Skouroupathis},\ and\
  \citenamefont {Spanoudes}}]{Costa:2021iyv}%
  \BibitemOpen
  \bibfield  {author} {\bibinfo {author} {\bibfnamefont {M.}~\bibnamefont
  {Costa}}, \bibinfo {author} {\bibfnamefont {I.}~\bibnamefont {Karpasitis}},
  \bibinfo {author} {\bibfnamefont {T.}~\bibnamefont {Pafitis}}, \bibinfo
  {author} {\bibfnamefont {G.}~\bibnamefont {Panagopoulos}}, \bibinfo {author}
  {\bibfnamefont {H.}~\bibnamefont {Panagopoulos}}, \bibinfo {author}
  {\bibfnamefont {A.}~\bibnamefont {Skouroupathis}},\ and\ \bibinfo {author}
  {\bibfnamefont {G.}~\bibnamefont {Spanoudes}},\ }\bibfield  {title} {\bibinfo
  {title} {{Gauge-invariant renormalization scheme in QCD: Application to
  fermion bilinears and the energy-momentum tensor}},\ }\href
  {https://doi.org/10.1103/PhysRevD.103.094509} {\bibfield  {journal} {\bibinfo
   {journal} {Phys. Rev. D}\ }\textbf {\bibinfo {volume} {103}},\ \bibinfo
  {pages} {094509} (\bibinfo {year} {2021}{\natexlab{a}})},\ \Eprint
  {https://arxiv.org/abs/2102.00858} {arXiv:2102.00858 [hep-lat]} \BibitemShut
  {NoStop}%
\bibitem [{\citenamefont {Lin}\ \emph {et~al.}(2024)\citenamefont {Lin},
  \citenamefont {Detmold},\ and\ \citenamefont {Meinel}}]{Lin:2024mws}%
  \BibitemOpen
  \bibfield  {author} {\bibinfo {author} {\bibfnamefont {J.}~\bibnamefont
  {Lin}}, \bibinfo {author} {\bibfnamefont {W.}~\bibnamefont {Detmold}},\ and\
  \bibinfo {author} {\bibfnamefont {S.}~\bibnamefont {Meinel}},\ }\bibfield
  {title} {\bibinfo {title} {{Position-space renormalization schemes for
  four-quark operators in HQET}},\ }\href@noop {} {\  (\bibinfo {year}
  {2024})},\ \Eprint {https://arxiv.org/abs/2404.16191} {arXiv:2404.16191
  [hep-lat]} \BibitemShut {NoStop}%
\bibitem [{\citenamefont {Frezzotti}\ and\ \citenamefont
  {Rossi}(2004)}]{Frezzotti:2004wz}%
  \BibitemOpen
  \bibfield  {author} {\bibinfo {author} {\bibfnamefont {R.}~\bibnamefont
  {Frezzotti}}\ and\ \bibinfo {author} {\bibfnamefont {G.~C.}\ \bibnamefont
  {Rossi}},\ }\bibfield  {title} {\bibinfo {title} {{Chirally improving Wilson
  fermions. II. Four-quark operators}},\ }\href
  {https://doi.org/10.1088/1126-6708/2004/10/070} {\bibfield  {journal}
  {\bibinfo  {journal} {JHEP}\ }\textbf {\bibinfo {volume} {10}},\ \bibinfo
  {pages} {070}},\ \Eprint {https://arxiv.org/abs/hep-lat/0407002}
  {arXiv:hep-lat/0407002} \BibitemShut {NoStop}%
\bibitem [{\citenamefont {Gimenez}\ \emph {et~al.}(2004)\citenamefont
  {Gimenez}, \citenamefont {Giusti}, \citenamefont {Guerriero}, \citenamefont
  {Lubicz}, \citenamefont {Martinelli}, \citenamefont {Petrarca}, \citenamefont
  {Reyes}, \citenamefont {Taglienti},\ and\ \citenamefont
  {Trevigne}}]{Gimenez:2004me}%
  \BibitemOpen
  \bibfield  {author} {\bibinfo {author} {\bibfnamefont {V.}~\bibnamefont
  {Gimenez}}, \bibinfo {author} {\bibfnamefont {L.}~\bibnamefont {Giusti}},
  \bibinfo {author} {\bibfnamefont {S.}~\bibnamefont {Guerriero}}, \bibinfo
  {author} {\bibfnamefont {V.}~\bibnamefont {Lubicz}}, \bibinfo {author}
  {\bibfnamefont {G.}~\bibnamefont {Martinelli}}, \bibinfo {author}
  {\bibfnamefont {S.}~\bibnamefont {Petrarca}}, \bibinfo {author}
  {\bibfnamefont {J.}~\bibnamefont {Reyes}}, \bibinfo {author} {\bibfnamefont
  {B.}~\bibnamefont {Taglienti}},\ and\ \bibinfo {author} {\bibfnamefont
  {E.}~\bibnamefont {Trevigne}},\ }\bibfield  {title} {\bibinfo {title}
  {{Non-perturbative renormalization of lattice operators in coordinate
  space}},\ }\href {https://doi.org/10.1016/j.physletb.2004.07.053} {\bibfield
  {journal} {\bibinfo  {journal} {Phys. Lett. B}\ }\textbf {\bibinfo {volume}
  {598}},\ \bibinfo {pages} {227} (\bibinfo {year} {2004})},\ \Eprint
  {https://arxiv.org/abs/hep-lat/0406019} {arXiv:hep-lat/0406019} \BibitemShut
  {NoStop}%
\bibitem [{\citenamefont {Chetyrkin}\ and\ \citenamefont
  {Maier}(2011)}]{Chetyrkin:2010dx}%
  \BibitemOpen
  \bibfield  {author} {\bibinfo {author} {\bibfnamefont {K.~G.}\ \bibnamefont
  {Chetyrkin}}\ and\ \bibinfo {author} {\bibfnamefont {A.}~\bibnamefont
  {Maier}},\ }\bibfield  {title} {\bibinfo {title} {{Massless correlators of
  vector, scalar and tensor currents in position space at orders $\alpha_s^3$
  and $\alpha_s^4$: Explicit analytical results}},\ }\href
  {https://doi.org/10.1016/j.nuclphysb.2010.11.007} {\bibfield  {journal}
  {\bibinfo  {journal} {Nucl. Phys. B}\ }\textbf {\bibinfo {volume} {844}},\
  \bibinfo {pages} {266} (\bibinfo {year} {2011})},\ \Eprint
  {https://arxiv.org/abs/1010.1145} {arXiv:1010.1145 [hep-ph]} \BibitemShut
  {NoStop}%
\bibitem [{\citenamefont {Cichy}\ \emph {et~al.}(2012)\citenamefont {Cichy},
  \citenamefont {Jansen},\ and\ \citenamefont {Korcyl}}]{Cichy:2012is}%
  \BibitemOpen
  \bibfield  {author} {\bibinfo {author} {\bibfnamefont {K.}~\bibnamefont
  {Cichy}}, \bibinfo {author} {\bibfnamefont {K.}~\bibnamefont {Jansen}},\ and\
  \bibinfo {author} {\bibfnamefont {P.}~\bibnamefont {Korcyl}},\ }\bibfield
  {title} {\bibinfo {title} {{Non-perturbative renormalization in coordinate
  space for $N_f=2$ maximally twisted mass fermions with tree-level Symanzik
  improved gauge action}},\ }\href
  {https://doi.org/10.1016/j.nuclphysb.2012.08.006} {\bibfield  {journal}
  {\bibinfo  {journal} {Nucl. Phys. B}\ }\textbf {\bibinfo {volume} {865}},\
  \bibinfo {pages} {268} (\bibinfo {year} {2012})},\ \Eprint
  {https://arxiv.org/abs/1207.0628} {arXiv:1207.0628 [hep-lat]} \BibitemShut
  {NoStop}%
\bibitem [{\citenamefont {Tomii}\ and\ \citenamefont
  {Christ}(2019)}]{Tomii:2018zix}%
  \BibitemOpen
  \bibfield  {author} {\bibinfo {author} {\bibfnamefont {M.}~\bibnamefont
  {Tomii}}\ and\ \bibinfo {author} {\bibfnamefont {N.~H.}\ \bibnamefont
  {Christ}},\ }\bibfield  {title} {\bibinfo {title} {{$O(4)$-symmetric
  position-space renormalization of lattice operators}},\ }\href
  {https://doi.org/10.1103/PhysRevD.99.014515} {\bibfield  {journal} {\bibinfo
  {journal} {Phys. Rev. D}\ }\textbf {\bibinfo {volume} {99}},\ \bibinfo
  {pages} {014515} (\bibinfo {year} {2019})},\ \Eprint
  {https://arxiv.org/abs/1811.11238} {arXiv:1811.11238 [hep-lat]} \BibitemShut
  {NoStop}%
\bibitem [{\citenamefont {Costa}\ \emph
  {et~al.}(2021{\natexlab{b}})\citenamefont {Costa}, \citenamefont
  {Panagopoulos}, \citenamefont {Panagopoulos},\ and\ \citenamefont
  {Spanoudes}}]{Costa:2021pfu}%
  \BibitemOpen
  \bibfield  {author} {\bibinfo {author} {\bibfnamefont {M.}~\bibnamefont
  {Costa}}, \bibinfo {author} {\bibfnamefont {G.}~\bibnamefont {Panagopoulos}},
  \bibinfo {author} {\bibfnamefont {H.}~\bibnamefont {Panagopoulos}},\ and\
  \bibinfo {author} {\bibfnamefont {G.}~\bibnamefont {Spanoudes}},\ }\bibfield
  {title} {\bibinfo {title} {{Gauge-invariant Renormalization of the
  Gluino-Glue operator}},\ }\href
  {https://doi.org/10.1016/j.physletb.2021.136225} {\bibfield  {journal}
  {\bibinfo  {journal} {Phys. Lett. B}\ }\textbf {\bibinfo {volume} {816}},\
  \bibinfo {pages} {136225} (\bibinfo {year} {2021}{\natexlab{b}})},\ \Eprint
  {https://arxiv.org/abs/2102.02036} {arXiv:2102.02036 [hep-lat]} \BibitemShut
  {NoStop}%
\bibitem [{\citenamefont {Bergner}\ \emph {et~al.}(2023)\citenamefont
  {Bergner}, \citenamefont {Costa}, \citenamefont {Panagopoulos}, \citenamefont
  {Piemonte}, \citenamefont {Soler~Calero},\ and\ \citenamefont
  {Spanoudes}}]{Bergner:2022see}%
  \BibitemOpen
  \bibfield  {author} {\bibinfo {author} {\bibfnamefont {G.}~\bibnamefont
  {Bergner}}, \bibinfo {author} {\bibfnamefont {M.}~\bibnamefont {Costa}},
  \bibinfo {author} {\bibfnamefont {H.}~\bibnamefont {Panagopoulos}}, \bibinfo
  {author} {\bibfnamefont {S.}~\bibnamefont {Piemonte}}, \bibinfo {author}
  {\bibfnamefont {I.}~\bibnamefont {Soler~Calero}},\ and\ \bibinfo {author}
  {\bibfnamefont {G.}~\bibnamefont {Spanoudes}},\ }\bibfield  {title} {\bibinfo
  {title} {{Nonperturbative renormalization of the supercurrent in N=1
  supersymmetric Yang-Mills theory}},\ }\href
  {https://doi.org/10.1103/PhysRevD.107.034502} {\bibfield  {journal} {\bibinfo
   {journal} {Phys. Rev. D}\ }\textbf {\bibinfo {volume} {107}},\ \bibinfo
  {pages} {034502} (\bibinfo {year} {2023})},\ \Eprint
  {https://arxiv.org/abs/2209.13934} {arXiv:2209.13934 [hep-lat]} \BibitemShut
  {NoStop}%
\bibitem [{\citenamefont {Panagopoulos}\ \emph {et~al.}(2021)\citenamefont
  {Panagopoulos}, \citenamefont {Panagopoulos},\ and\ \citenamefont
  {Spanoudes}}]{Panagopoulos:2020qcn}%
  \BibitemOpen
  \bibfield  {author} {\bibinfo {author} {\bibfnamefont {G.}~\bibnamefont
  {Panagopoulos}}, \bibinfo {author} {\bibfnamefont {H.}~\bibnamefont
  {Panagopoulos}},\ and\ \bibinfo {author} {\bibfnamefont {G.}~\bibnamefont
  {Spanoudes}},\ }\bibfield  {title} {\bibinfo {title} {{Two-loop
  renormalization and mixing of gluon and quark energy-momentum tensor
  operators}},\ }\href {https://doi.org/10.1103/PhysRevD.103.014515} {\bibfield
   {journal} {\bibinfo  {journal} {Phys. Rev. D}\ }\textbf {\bibinfo {volume}
  {103}},\ \bibinfo {pages} {014515} (\bibinfo {year} {2021})},\ \Eprint
  {https://arxiv.org/abs/2010.02062} {arXiv:2010.02062 [hep-lat]} \BibitemShut
  {NoStop}%
\bibitem [{\citenamefont {Chetyrkin}\ and\ \citenamefont
  {Tkachov}(1981)}]{Chetyrkin:1981qh}%
  \BibitemOpen
  \bibfield  {author} {\bibinfo {author} {\bibfnamefont {K.~G.}\ \bibnamefont
  {Chetyrkin}}\ and\ \bibinfo {author} {\bibfnamefont {F.~V.}\ \bibnamefont
  {Tkachov}},\ }\bibfield  {title} {\bibinfo {title} {{Integration by parts:
  The algorithm to calculate $\beta$-functions in 4 loops}},\ }\href
  {https://doi.org/10.1016/0550-3213(81)90199-1} {\bibfield  {journal}
  {\bibinfo  {journal} {Nucl. Phys. B}\ }\textbf {\bibinfo {volume} {192}},\
  \bibinfo {pages} {159} (\bibinfo {year} {1981})}\BibitemShut {NoStop}%
\bibitem [{\citenamefont {Davydychev}(1992)}]{Davydychev:1992xr}%
  \BibitemOpen
  \bibfield  {author} {\bibinfo {author} {\bibfnamefont {A.~I.}\ \bibnamefont
  {Davydychev}},\ }\bibfield  {title} {\bibinfo {title} {{Recursive algorithm
  of evaluating vertex type Feynman integrals}},\ }\href@noop {} {\bibfield
  {journal} {\bibinfo  {journal} {J. Phys. A}\ }\textbf {\bibinfo {volume}
  {25}},\ \bibinfo {pages} {5587} (\bibinfo {year} {1992})}\BibitemShut
  {NoStop}%
\bibitem [{\citenamefont {Usyukina}\ and\ \citenamefont
  {Davydychev}(1994)}]{Usyukina:1994iw}%
  \BibitemOpen
  \bibfield  {author} {\bibinfo {author} {\bibfnamefont {N.~I.}\ \bibnamefont
  {Usyukina}}\ and\ \bibinfo {author} {\bibfnamefont {A.~I.}\ \bibnamefont
  {Davydychev}},\ }\bibfield  {title} {\bibinfo {title} {{New results for two
  loop off-shell three point diagrams}},\ }\href
  {https://doi.org/10.1016/0370-2693(94)90874-5} {\bibfield  {journal}
  {\bibinfo  {journal} {Phys. Lett. B}\ }\textbf {\bibinfo {volume} {332}},\
  \bibinfo {pages} {159} (\bibinfo {year} {1994})},\ \Eprint
  {https://arxiv.org/abs/hep-ph/9402223} {arXiv:hep-ph/9402223} \BibitemShut
  {NoStop}%
\bibitem [{\citenamefont {'t~Hooft}\ and\ \citenamefont
  {Veltman}(1972)}]{tHooft:1972tcz}%
  \BibitemOpen
  \bibfield  {author} {\bibinfo {author} {\bibfnamefont {G.}~\bibnamefont
  {'t~Hooft}}\ and\ \bibinfo {author} {\bibfnamefont {M.~J.~G.}\ \bibnamefont
  {Veltman}},\ }\bibfield  {title} {\bibinfo {title} {{Regularization and
  Renormalization of Gauge Fields}},\ }\href
  {https://doi.org/10.1016/0550-3213(72)90279-9} {\bibfield  {journal}
  {\bibinfo  {journal} {Nucl. Phys. B}\ }\textbf {\bibinfo {volume} {44}},\
  \bibinfo {pages} {189} (\bibinfo {year} {1972})}\BibitemShut {NoStop}%
\bibitem [{\citenamefont {Herrlich}\ and\ \citenamefont
  {Nierste}(1995)}]{Herrlich:1994kh}%
  \BibitemOpen
  \bibfield  {author} {\bibinfo {author} {\bibfnamefont {S.}~\bibnamefont
  {Herrlich}}\ and\ \bibinfo {author} {\bibfnamefont {U.}~\bibnamefont
  {Nierste}},\ }\bibfield  {title} {\bibinfo {title} {{Evanescent operators,
  scheme dependences and double insertions}},\ }\href
  {https://doi.org/10.1016/0550-3213(95)00474-7} {\bibfield  {journal}
  {\bibinfo  {journal} {Nucl. Phys. B}\ }\textbf {\bibinfo {volume} {455}},\
  \bibinfo {pages} {39} (\bibinfo {year} {1995})},\ \Eprint
  {https://arxiv.org/abs/hep-ph/9412375} {arXiv:hep-ph/9412375} \BibitemShut
  {NoStop}%
\bibitem [{\citenamefont {Ciuchini}\ \emph {et~al.}(1995)\citenamefont
  {Ciuchini}, \citenamefont {Franco}, \citenamefont {Martinelli}, \citenamefont
  {Reina},\ and\ \citenamefont {Silvestrini}}]{Ciuchini:1995cd}%
  \BibitemOpen
  \bibfield  {author} {\bibinfo {author} {\bibfnamefont {M.}~\bibnamefont
  {Ciuchini}}, \bibinfo {author} {\bibfnamefont {E.}~\bibnamefont {Franco}},
  \bibinfo {author} {\bibfnamefont {G.}~\bibnamefont {Martinelli}}, \bibinfo
  {author} {\bibfnamefont {L.}~\bibnamefont {Reina}},\ and\ \bibinfo {author}
  {\bibfnamefont {L.}~\bibnamefont {Silvestrini}},\ }\bibfield  {title}
  {\bibinfo {title} {{An Upgraded analysis of epsilon-prime epsilon at the
  next-to-leading order}},\ }\href {https://doi.org/10.1007/BF01566672}
  {\bibfield  {journal} {\bibinfo  {journal} {Z. Phys. C}\ }\textbf {\bibinfo
  {volume} {68}},\ \bibinfo {pages} {239} (\bibinfo {year} {1995})},\ \Eprint
  {https://arxiv.org/abs/hep-ph/9501265} {arXiv:hep-ph/9501265} \BibitemShut
  {NoStop}%
\end{thebibliography}%

\end{document}